\documentclass[aps,prd,preprintnumbers,superscriptaddress,nofootinbib]{revtex4}

\pdfoutput=1

\usepackage[dvips]{graphicx}
\usepackage{feynmf}
\usepackage{color}
\usepackage{bm,latexsym,amsmath,amssymb,amsfonts,mathrsfs}
\usepackage[colorlinks=true,linkcolor=magenta,citecolor=magenta]{hyperref}
\usepackage{slashed}

\usepackage{physics}
\usepackage[amssymb,]{SIunits}
\usepackage{xcolor}
\usepackage[caption=false]{subfig}
\usepackage{multirow}

\usepackage{tensor}
\usepackage{ulem}

\newcommand*{\ba}{\begin{eqnarray}}
\newcommand*{\ea}{\end{eqnarray}}

\newcommand*{\mpl}{M_{\rm Pl}}
\newcommand{\simgt}{\lower.5ex\hbox{$\; \buildrel > \over \sim \;$}}
\newcommand{\simlt}{\lower.5ex\hbox{$\; \buildrel < \over \sim \;$}}

\newcommand*{\Lag}{{\cal L}}

\newcommand*{\p}{\partial}
\newcommand*{\phib}{\bar \phi}
\newcommand*{\Xb}{\bar X}

\newcommand*{\phid}{\dot{\phi}}
\newcommand*{\phidd}{\ddot{\phi}}

\newcommand*{\rhob}{\rho_{\rm b}}
\newcommand*{\rhoc}{\rho_{\rm c}}
\newcommand*{\rhom}{\rho_{\rm m}}
\newcommand*{\rhor}{\rho_{\rm r}}
\newcommand*{\rhobdot}{\dot{\rho}_{\rm b}}
\newcommand*{\rhocdot}{\dot{\rho}_{\rm c}}
\newcommand*{\rhordot}{\dot{\rho}_{\rm r}}

\newcommand*{\deltab}{\delta_{\rm b}}
\newcommand*{\deltac}{\delta_{\rm c}}
\newcommand*{\deltam}{\delta_{\rm m}}

\newcommand*{\wceff}{w_{\rm c}^{\rm eff}}
\newcommand*{\wphieff}{w_{\phi}^{\rm eff}}

\newcommand*{\gb}{{\bar g}}
\newcommand*{\Lphi}{{\cal L}_\phi}

\newcommand*{\Ldm}{{\cal L}_{{\rm c}}}
\newcommand*{\Tphiu}{T^{(\phi)}}
\newcommand*{\Tbu}{T^{({\rm b})}}
\newcommand*{\Tru}{T^{({\rm r})}}
\newcommand*{\Tiu}{T^{(i)}}

\newcommand*{\Tdmu}{T^{({\rm c})}}
\newcommand*{\Tphid}{T_{(\phi)}}

\newcommand*{\Tid}{T_{(i)}}
\newcommand*{\Tdmd}{T_{({\rm c})}}

\newcommand*{\fmeff}{f_{\rm m}^{\rm eff}}
\newcommand*{\fm}{f_{\rm m}}

\newcommand*{\A}{{\cal A}}

\newcommand*{\E}{{\cal E}}

\def\d{\delta}

\def\({\biggl(}
\def\){\biggr)}
\def\[{\biggl[}
\def\]{\biggr]}

\newcommand*{\parsec}{\textrm{pc}}

\begin{document}

\title{Redshift space distortions in the presence of non-minimally coupled dark matter}

\author{Fabio Chibana}
\affiliation{Department of Physics, Tokyo Institute of Technology,
	2-12-1 Ookayama, Meguro-ku, Tokyo 152-8551, Japan}

\author{Rampei Kimura}
\affiliation{Waseda Institute for Advanced Study, Waseda University,
	1-6-1 Nishi-Waseda, Shinjuku, Tokyo 169-8050, Japan}

\author{Masahide~Yamaguchi}
\affiliation{Department of Physics, Tokyo Institute of Technology,
	2-12-1 Ookayama, Meguro-ku, Tokyo 152-8551, Japan}

\author{Daisuke~Yamauchi}
\affiliation{Faculty of Engineering, Kanagawa University, 
	3-27-1 Kanagawa-ku, Yokohama, Kanagawa, 221-8686, Japan}

\author{Shuichiro~Yokoyama}
\affiliation{Kobayashi Maskawa Institute, Nagoya University, Chikusa, Aichi
  464-8602, Japan}
\affiliation{Kavli IPMU (WPI), UTIAS, The University of Tokyo,
	Kashiwa, Chiba 277-8583, Japan}

 \date{\today}

\begin{abstract}
In this paper, we fully investigate cosmological scenarios in which dark matter is non-minimally coupled to an extra scalar degree of freedom.
The interaction is realized by means of conformal and disformal terms in the transformed gravitational metric.
Considering linear perturbation theory, we show that the growth rate of dark matter differs from the uncoupled case and that the well-known Kaiser formula undergoes modification. 
As a result, redshift space distortion measurements cease to be a direct probe of the linear growth rate of total matter, since the distortion factor has an extra, coupling-dependent term.
We study the effect of the coupling in three cosmological models, two conformally and one disformally coupled, and forecast the constraints on the coupling, and other cosmological parameters, from future galaxy surveys.
\end{abstract}

\preprint{WU-AP/1906/19}

\maketitle


\section{Introduction}


Current astrophysical and cosmological observations point to the existence of dark matter (DM)~\cite{Gil-Marin:2014sta,More:2014uva, Zwicky1933,Babcock1939,Kahn1959,Clowe:2003tk} and dark energy (DE)~\cite{Perlmutter:1998hx,Riess:1998cb, Ade:2015xua}, the former behaving as a pressureless (cold) fluid called cold dark matter (CDM), while the latter being responsible for the universe's current phase of accelerated expansion. 
However, while a lot of different models have been proposed in the literature, the fundamental nature of these components remains unknown, giving rise to one of the most pressing mysteries in present-day cosmology.

Among many dark energy models, a modification of the gravitational sector has been widely discussed.
Examples of such approach include 
quintessence~\cite{Zlatev:1998tr},  
K-essence~\cite{Chiba:1999ka,ArmendarizPicon:2000dh},
$f(R)$ gravity~\cite{Carroll:2003wy,Starobinsky:2007hu,Hu:2007nk,Capozziello:2003tk}, 
Horndeski~\cite{Horndeski:1974wa,Deffayet:2011gz,Kobayashi:2011nu}, 
Gleyzes-Langlois-Piazza-Vernizzi (GLPV)~ \cite{Gleyzes:2014dya}, 
and Degenerate Higher-Order Scalar-Tensor (DHOST)~\cite{Langlois:2015cwa} theories. 
The common feature of these models is the inclusion of a scalar field $\phi$, which mimics the role of the cosmological constant at late times, and their theoretical predictions, as well as observational constraints in the cosmological context, have been extensively investigated (see~\cite{Tsujikawa:2010zza,Clifton:2011jh} for reviews).

Another crucial question is how matter (baryons and DM) is coupled to such an extra degree of freedom.  
A robust way to parameterize the coupling is introducing a new metric $\gb_{\mu\nu}$, related to the original one, $g_{\mu\nu}$, by conformal and disformal factors, which in general depend on the scalar field and its kinetic term~\cite{Bekenstein:1992pj}.  
If both baryons and DM couple to the same $\gb_{\mu\nu}$, it is known that this non-minimally coupled theory can be mapped into the minimally coupled one, resulting in a scalar-tensor theory with the minimal coupling, converted from an original one through a disformal transformation \cite{Achour:2016rkg}.
On the other hand, if baryons and DM couple to gravity with different metrics, this kind of mapping is unavailable.
Non-minimal couplings to baryons are strongly constrained by local tests of gravity both at solar-system and astrophysical scales~\cite{Uzan:2002vq, Kapner:2006si}.
Modified gravity models usually employ screening mechanisms (e.g. chameleons~\cite{Khoury:2003aq, Hui:2009kc} and Vainshtein~\cite{Dvali:2006su}) in order to explain these tight constraints~\cite{Joyce:2014kja}.
Hence, it is quite natural to assume that baryons are minimally coupled, reducing the need for screening mechanisms.
However, the situation changes when one considers non-minimally coupled DM, and an interaction between the scalar field and the DM particle cannot be excluded a priori (see \cite{Wang:2016lxa} for a recent review and the references therein).
The modifications introduced by the non-minimal coupling manifestly affect the cosmological scenario, and the resulting evolution of large-scale structures can be used to study said coupling.

One of the most reliable large-scale structure observable is the galaxy two-point correlation function (i.e. power spectrum), which can give information about the growth of density perturbations.  
Since spectroscopic surveys measure the redshift of galaxies, the observed position of each galaxy is different from the actual one, due to the combined effect of the peculiar velocity and the Hubble expansion.
In other words, the power spectrum in real space does not coincide with the one in redshift space.
This effect is called redshift space distortion (RSD) and it allows us to extract statistical information regarding the peculiar velocities of galaxies \cite{Kaiser:1987qv}.

The next generation of spectroscopic surveys, such as PFS~\cite{2014PASJ...66R...1T}, eBOSS~\cite{Delubac:2016qem}, Euclid~\footnote{http://www.euclid-ec.org/}~\cite{2016SPIE.9904E..0OR}, and WFIRST~\footnote{https://wfirst.gsfc.nasa.gov/} \cite{2015arXiv150303757S}, combined with future HI galaxy surveys such as the Square Kilometre Array (SKA)~\footnote{https://www.skatelescope.org/}~\cite{Maartens:2015mra,Bacon:2018dui,Yamauchi:2016ypt}, will improve the measurements of peculiar velocities, enabling us to test various cosmological models at an extraordinary level. 
The recent measurements of the cosmic microwave background (CMB) by Planck satellite~\cite{Ade:2015xua, Aghanim:2018eyx} have greatly improved our knowledge regarding the universe at its early stages, and a similar gain is awaited from the forthcoming galaxy surveys, which are expected to constrain DE to an unrivaled precision.  

As reviewed in section~\ref{sec:eff_f_rsd}, it was recently shown~\cite{Kimura:2017fnq} that when only DM is non-minimally coupled to the DE scalar field, the matter linear growth rate inferred by measurements of galaxies peculiar velocities is no longer defined logarithmic derivative of the linear growth function, as in the standard $\Lambda$CDM case, but rather it has an additional dependence due to the modifications to the continuity equation. 
Therefore, the relation between the real space and redshift space power spectra, the so-called Kaiser formula, must be modified, and the non-minimal coupling leads to a different interpretation of the growth function for matter perturbations. 

In the present work, we extend the analysis done in \cite{Kimura:2017fnq}, computing explicitly the difference between the effective growth rate, inferred from the peculiar velocity, and the actual one.
We consider Einstein's theory with an additional scalar degree of freedom non-minimally coupled to DM.
The coupling is introduced using a new metric, consisting of a conformal part, proportional to the gravitational metric, and another term including derivatives of the scalar field, the disformal part.
We study three models, parameterized by the conformal and disformal functions, solving both the background and linear equations. 
Additionally, we forecast the constraints to the coupling parameter from two representative future galaxy surveys, namely, Euclid and SKA.
In section~\ref{sec:set_up} we define the framework for CDM conformally/disformally coupled to a K-essence scalar field, while in section~\ref{sec:bg_and_lin_pert} we present the relevant background and linear perturbation equations. 
We also derive the evolution equations for both baryons and DM in the quasi-static limit. 
In section~\ref{sec:eff_f_rsd} we review how the modifications to the DM continuity equation, due to the non-minimal coupling, motivate the definition of the effective growth rate and reformulate the Kaiser formula.
Concrete cosmological examples are studied in section~\ref{sec:examples} choosing three different sets of conformal and disformal coupling functions.
We numerically solve both the background and linear perturbation equations and study how the interaction affects the cosmic evolution. 
In section~\ref{sec:forecasts} we perform a forecast analysis and predict the constraining power from future galaxy surveys, assuming SKA2-like and Euclid-like specifications, to our choices of coupling functions. 
Finally, in section~\ref{sec:conclusion} we conclude and summarize our results. 
In Appendix~\ref{app:metric_transformation}, details on conformal and disformal transformations are summarized. Scale dependence of the mass term and dispersion relation in the quasi-static limit are discussed in Appendices~\ref{sec:k_independence} and~\ref{sec:dispersion_relation}, respectively.


\section{Conformally/Disformally coupled dark matter \label{sec:set_up}}


In the present paper, we consider the action of Einstein's gravity with a general scalar field $\phi$ and matter fields, 
\ba
S &=&
\int {\rm d}^4 x \sqrt{-g} \left[ \frac{\mpl^2}{2} R[g] + \Lphi[g,\phi ]\right] +S_{\rm m}\,,
\label{action}
\ea
where $\mpl$ is the reduced Planck mass, $R$ is the Ricci scalar defined by the metric $g_{\mu\nu}$, $\Lphi$ is the scalar field Lagrangian, 
and $S_{\rm m}$ is the action for the (non-relativistic and relativistic) matter sector. 
We assume that the Lagrangian for baryons and radiation are defined in the usual, i.e., with these species coupled to the metric $g_{\mu\nu}$, while DM is conformally/disformally coupled to the scalar field through the metric $\gb_{\mu\nu}$, such that $S_{\rm m}$ can be written as 
\ba
S_{\rm m}=\int{\rm d}^4x\Bigl[
	\sqrt{-g}{\cal L}_{i} [g_{\mu\nu}, \psi_{i}]
	+\sqrt{-\bar g}{\cal L}_{\rm c} [\overline g_{\mu\nu},\psi_{\rm c}]
\Bigr] \,,
\ea
where the index $i = {\rm b}, {\rm r}$ stands for baryons and radiation, respectively, whereas $\psi_{\rm c}$ denotes the dark matter field.
The transformed metric is defined as
\ba
\gb_{\mu\nu}= A(\phi, X) g_{\mu\nu} + B(\phi, X) \phi_\mu \phi_\nu \,,
\ea
where $\phi_\mu = \partial_\mu \phi$.
The functions $A$ and $B$ are called conformal and disformal factors, respectively, and, in principle, both depend on the scalar field $\phi$ and its kinetic term $X := -(\p \phi)^2/2$. 
The case in which these couplings depend only on $\phi$ was investigated in \cite{Gleyzes:2015rua}.
For the scalar field, we consider the K-essence Lagrangian $\Lphi=K(\phi, X)$, which does not change a gravitational sector, i.e., $\phi$ does not couple directly to $R$, the Ricci scalar.

The variation of the action with respect to $g^{\mu\nu}$ yields the Einstein equation,
\ba
\mpl^2 G_{\mu\nu}  = \Tphiu_{\mu\nu} +  T^{\rm (m)}_{\mu\nu} + \Tru_{\mu\nu} \,,
\ea
where $\Tphiu_{\mu\nu}$, $T^{\rm (m)}_{\mu\nu} = \Tbu_{\mu\nu} + \Tdmu_{\mu\nu}$, and $\Tru_{\mu\nu}$  are the scalar field, total matter, and radiation energy-momentum tensors, respectively, defined as
\ba
 \Tphid^{\mu\nu} = \frac{2}{\sqrt{-g}} \frac{\delta (\sqrt{-g} \Lphi)}{\delta g_{\mu\nu}} 
 ,\qquad
 \Tid^{\mu\nu} = \frac{2}{\sqrt{-g}} \frac{\delta (\sqrt{-g} \Lag_{i}) }{ \delta g_{\mu\nu}}
 ,\qquad
 \Tdmd^{\mu\nu} = \frac{2 }{\sqrt{-g}} \frac{\delta (\sqrt{-\gb} \Ldm) }{\delta g_{\mu\nu}} \,.
\ea
Analogously, one could define $\bar{T}_{\rm (c)}^{\mu \nu}$, the DM energy-momentum tensor in the frame in which it is minimally coupled to gravity.
It can be shown that, for a homogeneous and isotropic background, if in one frame DM is a pressureless perfect fluid, it will also be in the other. 
This equivalence holds up to linear order (see Appendix~\ref{app:metric_transformation}).

The energy-momentum tensor of the scalar field can be directly calculated from the Lagrangian, yielding
\ba \label{eq:emt_phi}
 \Tphid^{\mu\nu} &=& K g^{\mu\nu} +K_X \phi^\mu \phi^\nu \,,
\ea
where the subscript in $K_X$ represents the derivative with respect to the kinetic term $X$.
The Bianchi identity immediately implies the conservation of the total energy-momentum tensor, $\nabla^\mu T_{\mu\nu} =\nabla^\mu T_{\mu\nu}^{(\phi)} +\nabla^\mu T_{\mu\nu}^{\rm (m)} +\nabla^\mu T_{\mu\nu}^{\rm (r)} = 0$.
Indeed, since baryons and relativistic matter are minimally coupled to the metric $g_{\mu\nu}$, their conservation equations take the usual form, 
\ba
\nabla^\mu \Tiu_{\mu\nu} =0 \,.
\label{EOMb}
\ea
On the other hand, that does not hold anymore for the scalar field or DM, individually, but rather, the sum of these components is conserved:
\ba
\nabla^\mu \Tphiu_{\mu\nu} +\nabla^\mu \Tdmu_{\mu\nu}=0 .
\label{EOMg}
\ea

In order to derive the equation of motion for $\phi$, it is useful to first calculate the following quantities\footnote{See Appendix~\ref{app:metric_transformation} for details.}
\ba
\frac{\p}{\p\phi}(\sqrt{-\gb} \Ldm) 
&=& \frac{\delta (\sqrt{-\gb} \Ldm)}{\delta g_{\mu\nu}} \frac{\delta g_{\mu\nu}}{\delta \gb_{\alpha\beta}} \frac{\delta \gb_{\alpha\beta}}{\delta \phi} 
= \sqrt{-g}\, Z, \\
\frac{\p}{\p\phi_\mu}(\sqrt{-\gb} \Ldm) 
&=& \frac{\delta (\sqrt{-\gb} \Ldm)}{\delta g_{\mu\nu}} \frac{\delta g_{\mu\nu}}{\delta \gb_{\alpha\beta}} \frac{\delta \gb_{\alpha\beta}}{\delta \phi_\mu} 
= \sqrt{-g}\, W^\mu ,
\ea
where we have defined $Z$ and $W^{\mu}$ as
\begin{align}
	Z &= \frac{1}{2A} \[A_\phi + A_X C X (A_\phi -2 B_\phi X)\] 
			\Tdmd 
		+\frac{1}{2A}\[B_\phi + B_X C X (A_\phi -2 B_\phi X)\] 
			\Tdmd^{\mu\nu} \phi_\mu \phi_\nu \,, \\
	W^\mu &= \frac{B}{A} \Tdmd^{\mu\nu} \phi_\nu
			-\frac{C}{2A}(A-2BX) \Bigl(A_X \Tdmd +B_X \Tdmd^{\alpha\beta} \phi_\alpha\phi_\beta\Bigr) \phi^\mu \,.
\end{align}
In the above equations, $\Tdmd$ is the trace of the DM energy-momentum tensor, while the function $C$ is defined as
\begin{equation}
	C = \frac{1}{A-A_X X+2B_X X^2},
\end{equation}
and the subscripts $\phi$ and $X$ represent the derivatives with respect to the scalar field and its kinetic term, respectively.
Using these results, the variation with respect to $\phi$ gives the equation of motion for the scalar field,
\ba
(K_X g^{\mu\nu} - K_{XX} \phi^\mu \phi^\nu) \,\phi_{\mu\nu} + K_\phi -2 K_{\phi X} X=Q,
\label{EOMs}
\ea
where $Q$ is the deviation from the minimally coupled DM, 
\ba
Q = \nabla_\mu W^\mu- Z \,.
\label{Q}
\ea
The combination of Eqs.~\eqref{EOMg}, \eqref{EOMs} and \eqref{Q} leads to
\ba
\nabla^\mu \Tdmu_{\mu\nu} + Q \phi_\nu = 0 \,.
\label{EOMDM}
\ea
Hence, the coupling function $Q$ determines the exchange of energy and momentum between the scalar field and the DM component.


\section{Background and linear perturbations \label{sec:bg_and_lin_pert} }


Let us consider linear perturbations in Newtonian gauge on a spatially flat Friedmann-Lema\^{i}tre-Robertson-Walker (FLRW) metric,
\ba \label{eq:line_element}
{\rm d}s^2=-[1+2\Phi(t, {\bm x})]{\rm d}t^2 + a^2(t) [1-2\Psi(t, {\bm x})]{\rm d}{\bm x}^2\,,
\ea
and define the energy-momentum tensor for each fluid as 
\begin{align} 
\label{eq:emt_def_00}
T^{\rm(I)}{}\indices{^{0}_{0}} &= -  \rho_{\rm I}(t) \left[ 1 + \delta_{\rm I}(t, {\bm x}) \right] \,, \\ 
\label{eq:emt_def_0i}
T^{\rm(I)}{}\indices{^0_{i}}  &= - \left[ \rho_{\rm I}(t) + p_{\rm I}(t) \right]  \p_i v_{\rm I}(t, {\bm x})  \,, \\ 
\label{eq:emt_def_ij}
T^{\rm(I)}{}\indices{^i_{j}}  
	&=  \left[ p_{\rm I}(t) + \delta p_{\rm I}(t, {\bm x}) \right] \delta^{i}_{j} 
		+ \left[ \Sigma_{\rm I} (t, {\bm x}) \right] \indices{ ^i_{j}}, 
	\quad \Sigma_{\rm I}{}\indices{^{i}_{i}}=0 \,.
 \end{align}
Here $\rho_{\rm I}$ is the energy density, $\delta_{\rm I}$ is the overdensity; $p_{\rm I}$ and $\d p_{\rm I}$ are the background pressure and  its perturbation, respectively;  $v_{\rm I}$ is the 3-velocity potential; and $\Sigma_{\rm I}{}\indices{^i_j} = T^{\rm(I)}{}\indices{^i_j} -  \delta^i_j T^{\rm(I)}{}\indices{^k _k} / 3$ is the traceless part of the energy-momentum tensor. Here the index ${\rm I}= {\rm b, r, c, m}$ represents baryon, radiation, DM, and the total matter, and $p_{\rm I }= \delta p_{\rm I } = \Sigma_{\rm I}= 0$ for baryon and DM (i.e. CDM) as well as the total matter.

From the above definition, the overdensity and velocity potential for the total matter fluid can be respectively expressed as
\ba
\label{eq:totalm}
\deltam &=& \omega_{\rm c} \deltac + \omega_{\rm b} \deltab \,,\\
v_{\rm m} &=& \omega_{\rm c}v_{\rm c} + \omega_{\rm b} v_{\rm b}\,,
\label{eq:totalv}
\ea
where $\omega_{\rm I} = \rho_{\rm I} / \rho_{\rm m}$ is the fractional energy density.
We also split the scalar field as $\phi(t, {\bm x})\to \phi(t) + \delta\phi(t,{\bm x})$, and introduce the convenient notation $\delta Q(t, {\bm x}) = Q (t, {\bm x})- Q_0(t)$, where $Q_0$ is a background value.


\subsection{Background equations}


At background level, Einstein equations become
\ba
&&H^2 =\frac{1}{3M_{\rm Pl}^2}\Bigl(\rho_\phi + \rhob + \rhoc +  \rhor )\,,\\
&&3 H^2+2 \dot{H}  = - \frac{1}{M_{\rm Pl}^2}\left( p_\phi + \frac{1}{3} \rhor \right) \,,
\ea
where $H = \dot{a}/a$, and the energy density and pressure for k-essence field are respectively given by
\ba
\rho_\phi &=& \phid^2 K_X-K \label{EDscalar} \, ,\\
p_\phi &=& K. \label{PRscalar}
\ea	
The equation of motion for $\phi$, Eq.~\eqref{EOMs}, yields
\ba \label{eq:eom_phi}
(K_X +K_{XX}\dot{\phi }^2  )\ddot{\phi }
+3 H \dot{\phi } K_X
+K_{\phi X}\dot{\phi }^2 
-K_{\phi }=-Q_0,
\ea
while the energy-momentum conservation for radiation, baryons and CDM leads to the background continuity equations
\ba
\rhordot + 4 H \rhor &=& 0 \,, \\
\rhobdot + 3 H \rhob &=& 0\,,\\
\label{rhobB}
\rhocdot+3 H \rhoc &=&Q_0 \phid \,.
\label{rhocB}
\ea
The background coupling $Q_0$ can be directly calculated from (\ref{Q}), resulting in
\begin{equation}  \label{eq:Q0}
	Q_0 	= \frac{\rhoc}{\phid} \dv{\Gamma}{t} \,,  
	\qquad
	\Gamma = \ln \left[  \frac{2A - A_X\phid^2+B_X\phid^4}{ 2 \sqrt{ A-B\phid^2}} \right]  \,.
\end{equation}
In general, the above expression for $Q_0$ contains $\phidd$ such that, in order to obtain the explicit expression of $Q_0$, one needs to eliminate $\phidd$ using the equation of motion for the scalar field, solving it for $Q_0$.

Using the definitions of energy density and pressure for the scalar field, Eqs.~\eqref{EDscalar} and \eqref{PRscalar}, respectively, the equation of motion~\eqref{eq:eom_phi} can be rewritten as 
\begin{equation} \label{rhophiB}
\dot{\rho}_\phi + 3H (1 + w_\phi) \rho_\phi = - Q_0 \phid,
\end{equation}
which makes, according to the Eq.~\eqref{rhocB}, the conservation of sum $\rhoc + \rho_\phi$ evident. 
The equation of state for the scalar field is defined in the usual way: $w_\phi = p_\phi/\rho_\phi$.
Furthermore, the CDM energy density can be found from Eq.~\eqref{rhocB}, which yields the solution
\begin{equation} \label{eq:rho_cdm}
	\rhoc \propto a^{-3} e^{\Gamma(\phi,X)}\,. 
\end{equation}
At the classical level, the interaction can be interpreted as a variation in the mass of the CDM particle.


\subsection{Linear perturbations}


In linear perturbation theory, the $(0,0)$, $(0, i)$, along with the trace and traceless part of $(i, j)$,  components of the Einstein equation in Fourier space read, respectively,
\begin{align}
\label{eq:einstein_00}
2\mpl^2 \left[ 3 H(H \Phi +\dot{\Psi}) + \frac{k^2}{a^2}  \Psi  \right] 
	&= -\delta\rho_{\phi} - \rhom \deltam - \rhor \delta_{\rm r} \,,\\
\label{eq:einstein_0i}
2 \mpl^2 \left[ \dot{\Psi} + H \Phi \right ] 
	&= K_X \phid \delta \phi + \rhom v_{\rm m} + \frac{4}{3} \rhor v_{\rm r}  \,, \\ 
\label{eq:einstein_ii}
2\mpl^2 \left[\ddot{\Psi} +3 H \dot{\Psi} +H \dot{\Phi}  + \left(2 \dot{H} + 3 H^2 \right) \Phi
	+ \frac{1}{3} \frac{k^2}{a^2} \left( \Psi - \Phi \right)  \right] 
	&= \delta p_\phi + \delta p_{\rm r}  \,,\\
 \label{eq:einstein_traceless}
2\mpl^2 \frac{k^2}{a^2} \left( \Psi - \Phi \right) 
	 &= 4  \rhor \sigma_{\rm r} \,,
\end{align}
where $\delta \rho_\phi$ and $\delta p_\phi$ are the scalar field energy density and pressure perturbations, respectively, while the anisotropic shear $\sigma$ is defined as 
\begin{equation}
\left( \rho  + p \right) \sigma 
	\equiv -\left( \hat{k}_{i} \hat{k}_{j} - \frac{1}{3} \delta_{i j} \right) \Sigma\indices{^{i}_{j}} \,,
\end{equation}
with $\hat k_i:=k_i/k$.
From the energy-momentum tensor, Eq.~\eqref{eq:emt_phi}, combined with the definitions~\eqref{eq:emt_def_00} and \eqref{eq:emt_def_0i}, we find 
\ba
\delta \rho_{\phi}&=&
-\left( K_{\phi }-\phid^2 K_{{\phi X}}\right)\delta \phi 
 +\left( K_X+\phid^2  K_{XX}\right)\phid \delta \phid 
-\left( K_X+\phid^2 K_{XX} \right)\phid^2 \Phi \,,\\
\delta p_{\phi}&=&
K_{\phi } \delta \phi +\phid K_X
\left(\delta\phid - \phid \Phi \right) \,.
\ea
Here, the subscripts $\phi$ and $X$ in the right-hand side represent the derivatives with respect to the background values of the scalar field and its kinetic term, respectively.
We can also compute the perturbed equation of motion for the scalar field,
\ba \label{eq:eom_delta_phi}
-\A_1 \frac{k^2}{a^2}\delta \phi- m_\phi^2 \, \delta \phi 
+3\A_1 \phid \dot{\Psi} +\A_2 \phid \dot{\Phi} +\A_3 \Phi 
-\A_2 \delta \ddot{\phi} +\A_4 \delta \phid 
=\delta Q ,
\ea
where
\ba
\A_1&=&K_X \,,\\
\A_2&=&K_X+K_{XX}\dot{\phi }^2 \,,\\
\A_3&=&6 H \dot{\phi } K_X+\dot{\phi }^2 K_{\phi X}
		+3 H \dot{\phi }^3 K_{XX}+\dot{\phi }^4 K_{\phi XX}
		+\ddot{\phi } (2 K_X+5 \dot{\phi }^2 K_{XX}+\dot{\phi }^4 K_{XXX}) \,, \\
\A_4&=&-3 H K_X-3 H \dot{\phi }^2 K_{XX}
		-\ddot{\phi } (K_{XXX} \dot{\phi }^3 +3 K_{XX}\dot{\phi })
		-\dot{\phi } K_{\phi X}-\dot{\phi }^3 K_{\phi XX} \,, \\ 
m_\phi^2&=&-K_{\phi \phi }+3 K_{\phi X} H \dot{\phi } +K_{\phi \phi X}  \dot{\phi }^2 
		+\ddot{\phi } (K_{\phi X} + K_{\phi XX}\dot{\phi }^2) \,,
\ea
and the perturbation of $Q$ is given by
\ba
&&\delta Q = 
(R_1+ R_2) \phid  \dot{\delta}_{\rm c} + Q_0 \deltac
+ R_1 \phid \frac{k^2}{a^2}v_{\rm c} + R_2 \frac{k^2}{a^2}\delta \phi
-3(R_1+R_2) \phid \dot{\Psi}  -R_3 \phid \dot{\Phi} \nonumber\\
&&~~~~~~~~~
+R_4 \phid \Phi 
+R_3 \delta \ddot{\phi} 
+R_5 \delta \phid 
+R_6 \delta \phi \,,
\label{deltaQ}
\ea
in which, for convenience, we have defined the quantities
\ba
R_1 &=& \frac{B}{A} \rhoc   \,,\label{R1}\\
R_2 &=& - \frac{C}{2A}(A- B \phid^2 ) \left( A_X - B_X \phid^2 \right) \rhoc   \,,  \label{R2}\\
R_3 &=& \rhoc \phid \omega_{X} \,, \\
R_4&=&-Q_0(\omega + \omega_{X} \phid^2)
	-\rhoc (\omega_{\phi} + 3  \omega_{X} \phidd
	+ {\dot \omega}_{X}\phid
	-\zeta_{X} \phid) \,, \\
R_5 &=& \omega_{X} Q_0  \phid^2
+\rhoc (\omega_{\phi} + \omega_{X} \phidd
+ {\dot \omega}_{X}\phid
-\zeta_{X} \phid) \,, \\
R_6 &=& \omega_{\phi} Q_0 \phid + \rhoc({\dot \omega}_{\phi}- \zeta_{\phi}) \,,
\ea
with
\ba
\omega &=& \left( R_1 + R_2 \right) \frac{\phid}{\rhoc} \,, \\
\zeta &=& -\frac{1}{2A} \[A_\phi + \frac{1}{2} C A_X \phid^2  (A_\phi - B_\phi \phid^2) \] 
			+ \frac{\phid^2}{2A} \[ B_\phi + \frac{1}{2}  C B_X \phid^2 ( A_\phi - B_\phi \phid^2 ) \] \,.
\ea
The energy-momentum conservation for baryonic matter yields the standard equation,
\ba
\dot{\delta}_{\rm b} - 3 \dot{\Psi} +\frac{k^2}{a^2} v_{\rm b} &=& 0 \,,  \\
\dot{v}_{\rm b} -\Phi &=&0 \label{eq:euler_baryons} \,,
\ea
while CDM follows 
\ba
\dot{\delta}_{\rm c} - 3 \dot{\Psi} + \frac{k^2}{a^2} v_{\rm c} 
	&=& \frac{\phid}{\rhoc}\Bigl(\delta Q-Q_0\delta_{\rm c}\Bigr)+\frac{Q_0}{\rhoc} \delta\phid\,,\\
\dot v_{\rm c} - \Phi 
	&=&\frac{Q_0}{\rhoc} \left( \delta\phi - \phid v_{\rm c}\right)\label{eq:euler_cdm}\,.
\ea
For radiation, the respective equations do not change, remaining the same as in the standard $\Lambda$CDM case.


\section{Effective growth rate and redshift space distortions \label{sec:eff_f_rsd}}
	

In this section, we derive the effective growth rate of matter perturbations for the disformally coupled CDM and show that it can significantly differ from the standard definition. 


\subsection{Quasi-static approximation}


Currently, data coming from galaxy surveys are restricted to scales much smaller than the cosmological horizon. 
Moreover, for scales well inside the sound horizon of scalar field perturbations, one can employ the quasi-static approximation \cite{Sawicki:2015zya}.
Effectively, this corresponds to dropping time derivatives, with respect to space derivatives, of metric and scalar field perturbations.
Furthermore, since we are interested in the effects of the DE scalar field at late-times, we neglect the contributions from relativistic matter (radiation) here and hereafter.
In such an approximation, Einstein equations become constraint equations for the gravitational potentials $\Phi$ and $\Psi$,
\ba \label{eq:poisson_qs}
\frac{k^2}{a^2}  \Psi = \frac{k^2}{a^2} \Phi =  -\frac{1}{2\mpl^2}
\rhom \deltam \,,
\ea
while Eq.~\eqref{eq:eom_phi} gives a constraint for the scalar perturbation,
\ba \label{eq:eom_phi_qs}
- \A \, \frac{k^2}{a^2}\delta \phi \equiv -\left(\A_1 \frac{k^2}{a^2}+m_\phi^2\right) \delta \phi 
=\delta Q \,,
\ea
where $\delta Q $ is given by
\ba \label{eq:delta_Q_qs}
\delta Q &=& 
(R_1+ R_2) \phid  \dot{\delta}_{\rm c} + Q_0 \deltac
+ R_1 \phid \frac{k^2}{a^2 }v_{\rm c}  + R_2 \frac{k^2}{a^2} \delta\phi  \,.
\ea
In Eq.~\eqref{eq:eom_phi_qs} we have defined $\A=\A(t, k)$ which in general depends on the wave-number when the Compton wavelength of the scalar field is smaller than the Hubble horizon scale. 
When the scalar field consists of a canonical kinetic term and the scalar mass scale is irrelevant within sub-horizon scales (see Appendix~\ref{sec:k_independence}), then we have $\A=1$ and recover the previous results~\cite{Kimura:2017fnq}.
The continuity and Euler equations for baryon are given by
\ba
\dot{\delta}_{\rm b} +\frac{k^2}{a^2} v_b &=& 0 \,, \label{eq:baryon_continuity_qs}\\
\dot{v}_{\rm b} -\Phi &=&0 \,,
\ea
and those for CDM are modified as follows:
\ba
\label{eq:cdm_continuity_qs}
\dot\delta_{\rm c}+\frac{k^2}{a^2}v_{\rm c }
	&=& \frac{\phid}{\rhoc} \left( \delta Q-Q_0\delta_{\rm c} \right) \,, \\
\label{eq:cdm_euler_qs}
\dot v_{\rm c}-\Phi &=&\frac{Q_0}{\rhoc}\left(\delta\phi -\dot\phi v_{\rm c}\right)\,.
\ea
Using $\d \phi$ and $\d Q$ given in Eqs.~\eqref{eq:eom_phi_qs} and \eqref{eq:delta_Q_qs} , these become
\begin{align}
 \left( 1-\Upsilon_1\right)\left(\dot{\delta}_{\rm c}+\frac{k^2}{a^2}v_{\rm c}\right)
 	&= \Upsilon_2\left(\dot\delta_{\rm c} - \epsilon \deltac\right)
 \,,\label{eq:modified continuity eq}\\
 \left( 1-\Upsilon_1\right)\frac{k^2}{a^2} \left(\dot{v}_{\rm c} - \Phi \right)
 	&= 
		\Upsilon_3
 		\left( \dot{\delta}_{\rm c } - \epsilon \deltac \right) \,,
 	\label{eq:modified euler eq}
\end{align}
where, for convenience, we introduced the quantities
\begin{equation} \label{eq:upsilon}
	\Upsilon_1 = \frac{\phid^2}{\rhoc}\frac{\A R_1}{\A+R_2}\,,
	\qquad 
	\Upsilon_2 = \frac{\phid^2}{\rhoc}\frac{\A R_2}{\A+R_2}\,,
	\qquad
	\Upsilon_3 = \frac{Q_0\dot\phi}{\rhoc}\Bigl( 1-\Upsilon_1 -\Upsilon_2\Bigr) -\epsilon\Upsilon_2 \,,
\end{equation}
with $\epsilon = Q_0 / \A \phid$.
In the derivation of Eq.~\eqref{eq:modified euler eq} from Eq.~\eqref{eq:cdm_euler_qs}, we have eliminated $v_{\rm c}$ using Eq.~\eqref{eq:modified continuity eq}.
Note that for purely conformal couplings, i.e. $A_{X}=B=0$, the functions $\Upsilon_1$ and $\Upsilon_2$ vanish.
Taking the time derivative of the continuity equations, \eqref{eq:baryon_continuity_qs} and \eqref{eq:cdm_continuity_qs}, and eliminating the metric perturbations, scalar field perturbations, and velocity terms, we obtain the following second-order differential equations for baryons and CDM, respectively,
\ba \label{eq:delta_b_eom}
&& \ddot{\d}_{\rm b} + 2H  \dot{\d}_{\rm b} - \frac{1}{2\mpl^2} \rhom \deltam = 0 \,,
\ea
\ba \label{eq:delta_c_eom}
&& (1-\Upsilon_1-\Upsilon_2) \, \ddot{\d}_{\rm c} 
+ 2H(1-\E_1) \,  \dot{\d}_{\rm c}
- \frac{1}{2\mpl^2} \[(1- \E_2)\,  \rhoc \deltac + (1-\Upsilon_1)\,  \rhob \deltab\]
=0 \,,
\ea
where we have introduced two new functions of background quantities,
\begin{align} 
\E_1 &= \Upsilon_1
			+\frac{1}{2H}
				\biggl[
					\frac{1-\Upsilon_1}{a^2}\frac{\rm d}{{\rm d}t}\left(\frac{a^2\Upsilon_2}{1-\Upsilon_1}\right) 
					-\Upsilon_3-\epsilon\Upsilon_2
				\biggr]
	\,, \label{eq:E1}\\
\E_2 
	&= \Upsilon_1
			+\frac{2\mpl^2}{\rhoc}
				\biggl[
					\frac{1-\Upsilon_1}{a^2}\frac{\rm d}{{\rm d}t}\left(\frac{a^2\epsilon\Upsilon_2}{1-\Upsilon_1}\right)
					-\epsilon\Upsilon_3
				\biggr]		
	\,. \label{eq:E2}
\end{align}
Hence, the CDM growth function experiences a different Hubble friction, $H_{\rm eff}$, and feels a different gravitational pull, $G_{\rm eff}$, depending on the couplings. 
As a result, the evolution of $\deltac$ itself is different from the standard uncoupled case.


\subsection{Linear growth rate}


For sub-horizon scales, the system formed by the second-order differential Eqs.~\eqref{eq:delta_b_eom} and \eqref{eq:delta_c_eom} does not depend on the wavenumber $k$.
Therefore, one can isolate the time dependence from the $k$~dependence of the initial conditions and express the growing solutions for the 
baryon and CDM density contrasts as
\ba \label{eq:growth_function}
\d_{\rm I}(t, {\bm k} ) = D_{\rm I} (t) \delta_0 ({\bm k})\,.
\ea
Here, $D_{\rm I}(t)$ is the $k$-independent growth factor while $\delta_0 ({\bm k})$ represents the initial density contrasts. Another quantity which is useful in describing the evolution of the perturbations is the so-called linear growth rate, defined as
\begin{equation}
	f_{\rm I}(t)\equiv \dv{ \ln D_{\rm I}}{\ln a},
\end{equation}
which quantifies the rate of growth in a Hubble time.
The continuity equations, \eqref{eq:baryon_continuity_qs} and \eqref{eq:cdm_continuity_qs}, can then be recast in terms of the respective growth factors, and the velocity potentials become  
\ba \label{eq:velocity_potential}
v_{\rm I}(t,{\bm k})  = - \frac{a^2 H}{k^2} f_{\rm I}^{\rm eff}(t, k) \delta_{\rm I}(t,{\bm k})\,.
\ea
For baryons, we immediately see that $f_{\rm b}=f_{\rm b}^{\rm eff}$. 
On the other hand, Eq.~\eqref{eq:cdm_continuity_qs} implies that the \textit{effective} linear growth rate for CDM, $f_{\rm c}^{\rm eff}$, can deviate significantly from the standard one:
\ba
f_{\rm c}^{\rm eff} 
= f_{\rm c}-\frac{\Upsilon_2}{1-\Upsilon_1}\left( f_{\rm c}-\frac{Q_0}{\A H \dot\phi}\right)
\equiv f_{\rm c}+\Delta f_{\rm c}
\,.\label{eq:f_eff_cdm}
\ea
Furthermore, employing Eq.~\eqref{eq:totalv}, the \textit{effective} linear growth rate of the total matter becomes
\begin{equation} \label{eq:f_eff_m}
	\fmeff
		= \frac{\omega_{\rm c}D_{\rm c}f_{\rm c}^{\rm eff}+\omega_{\rm b}D_{\rm b}f_{\rm b}}{\omega_{\rm c}D_{\rm c}+\omega_{\rm b}D_{\rm b}}
		\equiv f_{\rm m}+\Delta f_{\rm m} \,, 
\end{equation}
\begin{equation} \label{eq:Delta_f_m}
 \Delta f_{\rm m} 
 	= \omega_{\rm c}\frac{D_{\rm c}}{D_{\rm m}}\Delta f_{\rm c}
	-\omega_{\rm b}\frac{Q_0\dot\phi}{H\rho_{\rm m}}\frac{D_{\rm c}-D_{\rm b}}{D_{\rm m}} \,,
\end{equation}
where $D_{\rm m}=\omega_{\rm c}D_{\rm c}+\omega_{\rm b}D_{\rm b}$ is the total matter growth function.
The first term on the right-hand side of Eq.~\eqref{eq:Delta_f_m} comes from the definition of the CDM effective growth rate, Eq.~\eqref{eq:f_eff_cdm}, while the second is reflex of the non-minimal coupling on the actual growth rate
(in the minimally coupled case, the density ratios $\omega_{\rm b}$ and $\omega_{\rm c}$ are constant, while the baryon and dark matter growth function are equal;
naturally, that is no longer the case for when there is a non-minimal coupling).
Although $\fmeff$ is naturally given by the growth-factor-weighted average of the effective growth rates for CDM and baryons, the non-trivial terms in the CDM continuity equation and the background dynamics lead to a deviation from $f_{\rm m}$, the actual growth rate.


\subsection{Modified Kaiser formula \label{sec:modified_kaiser_formula}}


Spectroscopic surveys determine the distance to galaxies using their redshift.
However, this measurement includes not only the Hubble flow but also a contribution from the galaxies' peculiar velocities, which will lead to errors in the determination of distances.
On large scales, galaxies tend to follow the underlying matter distribution and fall into overdense regions.
As a consequence, in real space, the clustering of galaxies tends to not have a preferred direction.
In redshift space, however, the galaxy maps will show an anisotropy coming from the error in the determination of distances.

On linear scales, these anisotropies can be systematically taken into account.
The peculiar velocity $v_{\rm g}$ can be related to the total matter fluid velocity $v_{\rm m}$ by imposing some reasonable physical condition, such as momentum-conservation law for each galaxy~\cite{Gleyzes:2015rua}.
In the standard case, this leads to an expression for the galaxy power spectrum in redshift space, the so-called Kaiser formula~\cite{Kaiser:1987qv}
\begin{equation} \label{eq:kaiser_formula}
	P_{\rm g,s}(\bm{k}, t) = \left[ b_{\rm g}(t) + \fm (t)\, \mu^2 \right]^2 P_{\rm m}(k,t),
\end{equation}
where $b_{\rm g}$ is a linear, scale-independent bias relating the tracing population of galaxies and underlying matter field distribution, $\mu$ is the cosine of the angle between the wave vector and the line of sight, and $P_{\rm m}$ is the  matter power spectrum in real space.  
According to Eq.~\eqref{eq:growth_function}, the matter power spectrum can be written in terms of the growth factor as $P_{\rm m}(k,t)=D_{\rm m}^2(t)P_{\rm m,0}(k)$, where $P_{\rm m,0}$ is the initial matter power spectrum.
Eq.~\eqref{eq:kaiser_formula} assumes that, since galaxies move according to a common gravitational field, its velocities are not biased with respect to the matter velocity field. 
This formula holds not only for $\Lambda$CDM, but also for quintessence and modified gravity models with a minimal coupling.

As previously seen, when there is an interaction in the dark sector, the velocity potential, Eq.~\eqref{eq:velocity_potential}, is related not to the growth rate itself but to the effective growth rate. 
Hence, we extend Eq.~\eqref{eq:kaiser_formula} to a modified Kaiser formula to take into account the coupling as
\begin{equation}
	P_{\rm g,s}(\bm{k},t) = \left[ b_{\rm g}(t) + \fmeff(t,k) \, \mu^2 \right]^2 P_{\rm m}(k,t),
\end{equation}
with $\fmeff$ is given in Eq.~\eqref{eq:f_eff_m}.
For simplicity, as suggested by the $\Lambda$CDM case~\cite{Elia:2011ds}, we assume that the relation $v_{\rm g} = v_{\rm m}$ holds even when CDM is non-minimally coupled to the scalar field.
In fact, as long as the galaxy peculiar velocity has some dependence on the dark matter velocity field (i.e. $v_{\rm g} \neq v_{\rm b}$), the coupling effect on RSD measurements of the growth rate will be present.
The distribution of galaxies in redshift space depends on the effective growth rate and, in principle, an interaction between dark matter and the scalar field can be probed using clustering measurements. 

One way to obtain the actual growth rate is to observe the time-evolution of large-scale structures using a direct probe, e.g. tomographic weak lensing. 
In such a case, the measurement of the gravitational potential gives clean information about the total matter density, via the Poisson equation~\eqref{eq:poisson_qs}. 
On the other hand, to probe the coupling between dark matter and the scalar field, the linear galaxy bias can be determined at each redshift by the cross-correlation between weak lensing and galaxy clustering \cite{Hashimoto:2015tnv} and one can then unambiguously measure the coupling using RSD measurements. 

In terms of the real space and redshift space power spectra for galaxy distribution, $P_{\rm g} \equiv b_{\rm g}^2P_{\rm m}$ and $P_{\rm g, s}$ respectively, the Kaiser formula, Eq.~\eqref{eq:kaiser_formula}, reads
\begin{equation}
	P_{\rm g, s}({\bm k},t) = \left[ 1 + \beta(t)\, \mu^2 \right]^2 P_{\rm g} (k,t),
\end{equation}
where $\beta \equiv \fm / b_{\rm g}$, which encodes the corrections due to redshift space distortions, is the redshift-distortion factor.
In our setting, this quantity becomes
\begin{equation}
	\beta^{\rm eff} = \frac{\fmeff}{b_{\rm g}} = \beta + \Delta \beta \,,
\end{equation}
with two contributions: the usual one, due to the peculiar velocity of galaxies, and also an additional term encompassing the modifications of the CDM continuity and Euler equations due to the non-minimal coupling.
For a given choice of coupling, the latter can be characterized by
\begin{equation}
	\frac{\Delta \beta}{\beta} = \frac{\Delta \fm }{ \fm} \,,
\end{equation}
where $\Delta \fm$ was defined in Eq.~\eqref{eq:f_eff_m}.
For positive values of $\Delta \fm / \fm$, the difference between the real space and redshift power spectra will be further increased, 
while for negative values, the coupling leads to a suppression of the redshift-distortion factor.


\section{Concrete examples \label{sec:examples}}


In this section we study three cosmological models, choosing the scalar field Lagrangian, along with the conformal and disformal factors, and discuss how the strength of the coupling changes the evolution of background and perturbed quantities.
To simplify the notation, throughout this section we set the reduced Planck mass, $\mpl$, to unity.

In what follows, we consider coupled quintessence models, with the canonical scalar field Lagrangian given by 
\begin{equation} \label{eq:lagrangian_quintessence}
    K(\phi, X) = X  - V(\phi),
\end{equation} 
with an inverse-power-law scalar potential \cite{Ratra:1987rm, Caldwell:1997ii}, 
\begin{equation} \label{eq:scalar_potential}
    V(\phi) = M^{2} \phi^{-n},
\end{equation}
where the slope of the potential $n$ is a positive integer and $M$ is a constant.
The potential~\eqref{eq:scalar_potential} belongs to the class of 
freezing models \cite{Caldwell:2005tm}, in which the field rolls
down the potential in the past, slowing down at the onset of the cosmic
acceleration.  In the uncoupled case, this model is known to give rise
to tracking solutions \cite{Zlatev:1998tr, Steinhardt:1999nw}: an
attractor regime in which the energy density of the scalar field is
subdominant, and both its equation of state and energy density parameter
are approximately constant.  For a wide range of initial conditions, the
evolution of the scalar field will approach the previously mentioned
attractor solution, with its equation of state, for the potential
\eqref{eq:scalar_potential}, being given by
\begin{equation} \label{eq:eos_tracking_uncoupled}
    w_\phi \approx \frac{n w_{\rm M}-2}{n+2} \,,
\end{equation}
where $w_{\rm M} = p_{\rm M} / \rho_{\rm M}$ is the equation of state parameter of the dominant fluid  (matter or radiation).
Cosmological models that admit tracking solutions are interesting because such behavior lessens the dependence of the scalar field dynamics on the initial condition.
In fact, a well-known result is that in the uncoupled case it is possible to solve the equation of motion for $\phi$ analytically (see, for instance, \cite{weinberg2008cosmology}).

As for the conformal and disformal factors, we consider the following one-parameter models:

\paragraph*{Conformal model I:}
\begin{equation} \label{eq:coupling_1}
    A(\phi,X) = e^{-2 \alpha \phi },  \qquad
    B(\phi,X) = 0.
\end{equation}

\paragraph*{Conformal model II:}
\begin{equation} \label{eq:coupling_2}
    A(\phi,X) = e^{-\alpha \phi^2 },  \qquad
    B(\phi,X) = 0.
\end{equation}

\paragraph*{Disformal model III:}
\begin{equation} \label{eq:coupling_3}
    A(\phi, X) = 1, \qquad
    B(\phi, X) = \alpha/X.
\end{equation}
In the above expressions, $\alpha$ is the dimensionless coupling constant that controls the strength of the DE/CDM interaction and must be constrained by observations.
For the conformal models~I and~II, the continuity equation in the quasi-static limit, Eq.~\eqref{eq:modified continuity eq} is the standard one, with no explicit dependence on the coupling, whereas for model~III the background coupling $Q_0$ vanishes.
As will be discussed in the following sections, the most important difference between model~I and model~II is that, even for small values of the coupling constant, the former will deviate from the tracker solution, whilst the latter still hold, by construction, the tracker properties.

In the following subsections, we present the numerical solutions for the models considered in this work.
The background and linear perturbation equations for each model were implemented in the publicly available CLASS code~\cite{Blas:2011rf}.
We assume that at early times the effect of the coupling is negligible and set the initial conditions for the scalar field using the analytic expression for the tracker solution.
All plots were made using the cosmological parameters presented on TABLE~\ref{tab:fid_params}.
Additionally, the value of $M$ in Eq.~\eqref{eq:scalar_potential} is found using a shooting method such that we get the desired value of $\Omega_\phi$, the current value of the scalar field density parameter, whereas the slope parameter $n$ is set to $0.5$.
The initial energy density of cold dark matter is also found by employing the shooting method.


\subsection{Background evolution}


In order to study the background evolution, it is useful to rewrite the continuity equations for CDM and the scalar field, Eqs.~\eqref{rhocB} and \eqref{rhophiB}, respectively, as
\begin{align}
	& \rhocdot+3 H (1+ \wceff ) \rhoc = 0,\\
	& \dot{\rho}_{\phi} +3 H ( 1 + \wphieff ) \rho_{\phi} = 0, 
\end{align}
where we have defined the expressions for the effective equation of state as
\begin{align}
	\wceff &= - \frac{Q_0 \phid}{3 H \rhoc} \label{eq:w_c_eff} \, ,\\
	\wphieff &= w_\phi + \frac{Q_0 \phid}{3 H \rho_\phi} \label{eq:w_phi_eff}\,.
\end{align}
As long as $\wceff$ is small, much less than unity, it is expected that the evolution of the CDM energy density is not affected drastically by the interaction. 
Indeed, that is the case, for small values of the coupling constant, if the kinetic energy density of the scalar field is much smaller than the CDM energy density.
Furthermore, when $\wphieff$ is approximately constant the evolution of $\phi$ has a tracking behavior, even in the coupled case.

Additionally, for the scalar field Lagrangian~\eqref{eq:lagrangian_quintessence}, the equation of motion for $\phi$ at background level, Eq.~\eqref{eq:eom_phi}, becomes
\begin{equation} \label{eq:eom_phi_quintessence}
	\phidd + 3 H \phid + V_\phi = -Q_0.
\end{equation}

\subsubsection{Conformal model I}

The conformal model I, characterized by the coupling functions~\eqref{eq:coupling_1}, is an example of purely conformal metric transformation.
This kind of interaction was first proposed in the context of coupled quintessence in \cite{Amendola:1999er} and has been further investigated in \cite{Amendola:2000ub,Zhang:2005rj,Amendola:2007yx}. 
Imposing that both terms in the right hand side of Eq.~\eqref{eq:w_phi_eff} are constant leads to a scaling behavior and constrains the potential to be the exponential one.

According to Eq.~\eqref{eq:Q0}, the coupling at background level is described by
\begin{equation}
    Q_0 = - \alpha \rhoc , \quad \Gamma = - \alpha \phi,
\end{equation}
such that, from Eq.~\eqref{eq:rho_cdm}, the CDM energy density evolves as
\begin{equation}
    \rhoc  \propto a^{-3} e^{- \alpha \phi },
\end{equation}
while the equation of motion for the scalar field, Eq.~\eqref{eq:eom_phi_quintessence}, gives
\begin{equation}
	\phidd + 3 H \phid + V_\phi = \alpha \rhoc.
\end{equation}

\begin{figure}[!ht]
	\subfloat[\label{fig:ratio_rho_cdm_m1}]{
		\includegraphics[width=0.4\textwidth]{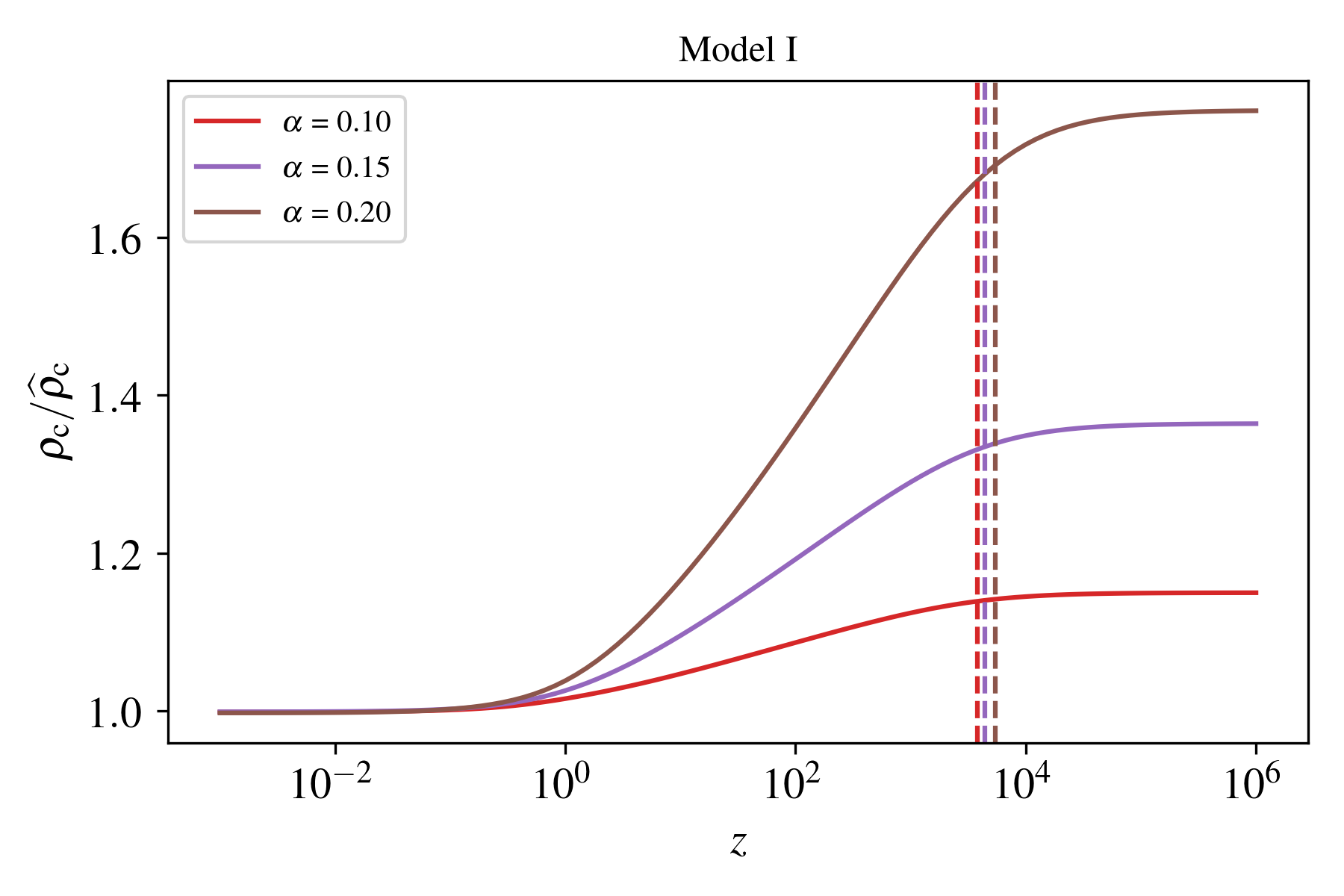}
	}
	\quad
	\subfloat[\label{fig:w_eff_cdm_m1}]{
		\includegraphics[width=0.4\textwidth]{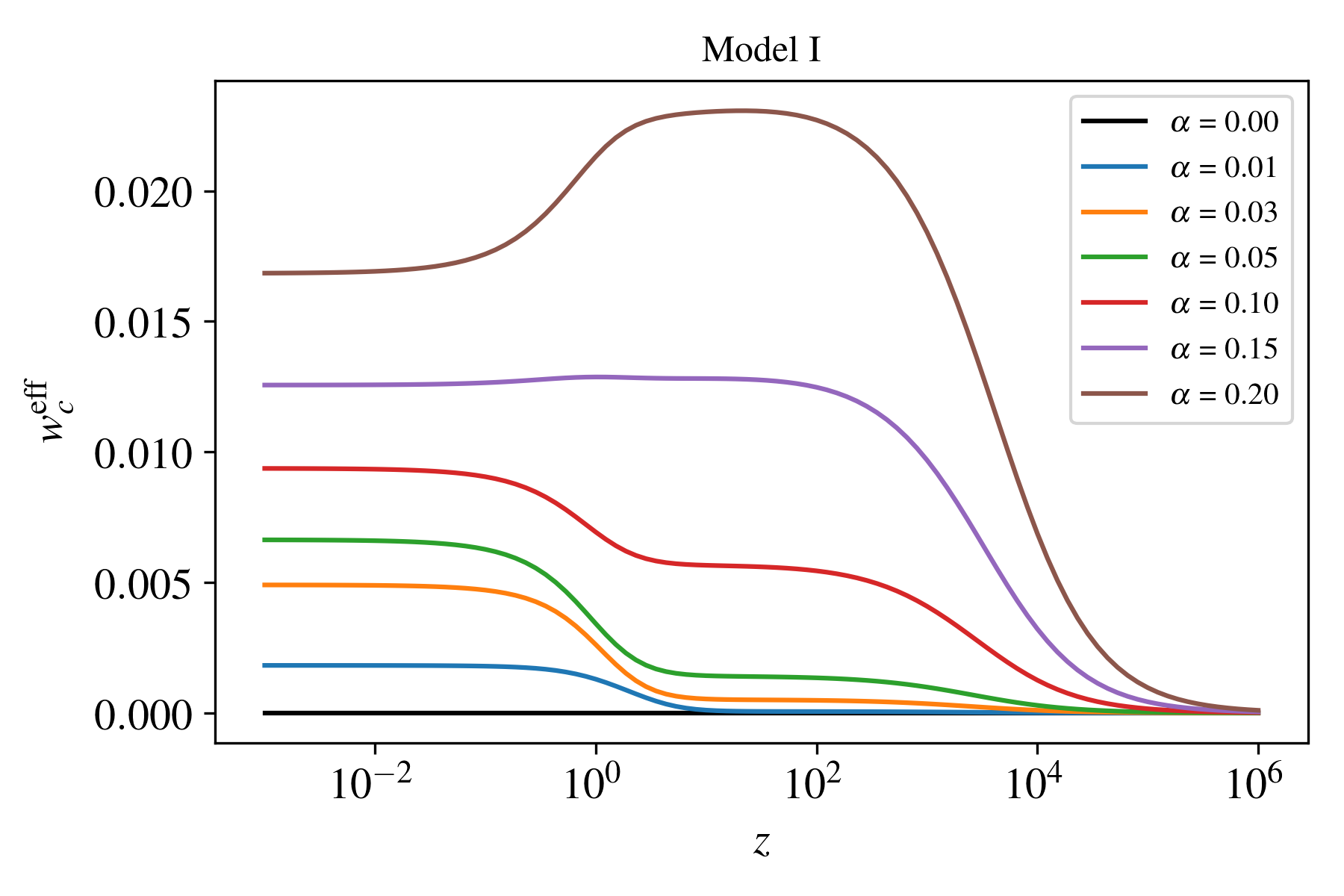}
	}
 	\caption{\label{fig:bg_evo_cdm_m1} 
	Left panel (a): Evolution of the ratio between the coupled and uncoupled CDM energy density, $\rhoc$ and $\widehat{\rho}_{\rm c}$, respectively.
 	The dashed lines correspond to the redshift at matter-radiation equality.
 	Right panel (b): Evolution of the effective equation of state of dark matter, Eq.~\eqref{eq:w_c_eff}.
 	Both plots show the respective quantities as function of the redshift for the conformal model~I, Eq.~\eqref{eq:coupling_1}, 
 	for different values of the coupling constant.}
\end{figure}

When $\alpha>0$, since $\phi$ is also positive, $\rhoc$ dilutes faster than in the uncoupled case, and energy is transferred from dark matter to the scalar field, starting around the transition from radiation to matter epoch~(Fig.~\ref{fig:ratio_rho_cdm_m1}).
As a result, for a given value of $\Omega_{\rm c}$ today, the CDM energy density in the past will be higher in coupled models than it would be in the uncoupled case.
This leads to a shift in the time of matter-radiation equality, which will take place earlier, at a higher redshift.
Also, for very small values of $\alpha$ (e.g. $\alpha \lesssim 0.15$ in Fig. \ref{fig:w_eff_cdm_m1}) the decay of $\rhoc$ is more important during DE epoch, whereas for slight larger values, the energy transfer is more salient during matter epoch.

\begin{figure}[!ht]
	\subfloat[\label{fig:w_scf_m1}]{
		\includegraphics[width=0.4\textwidth]{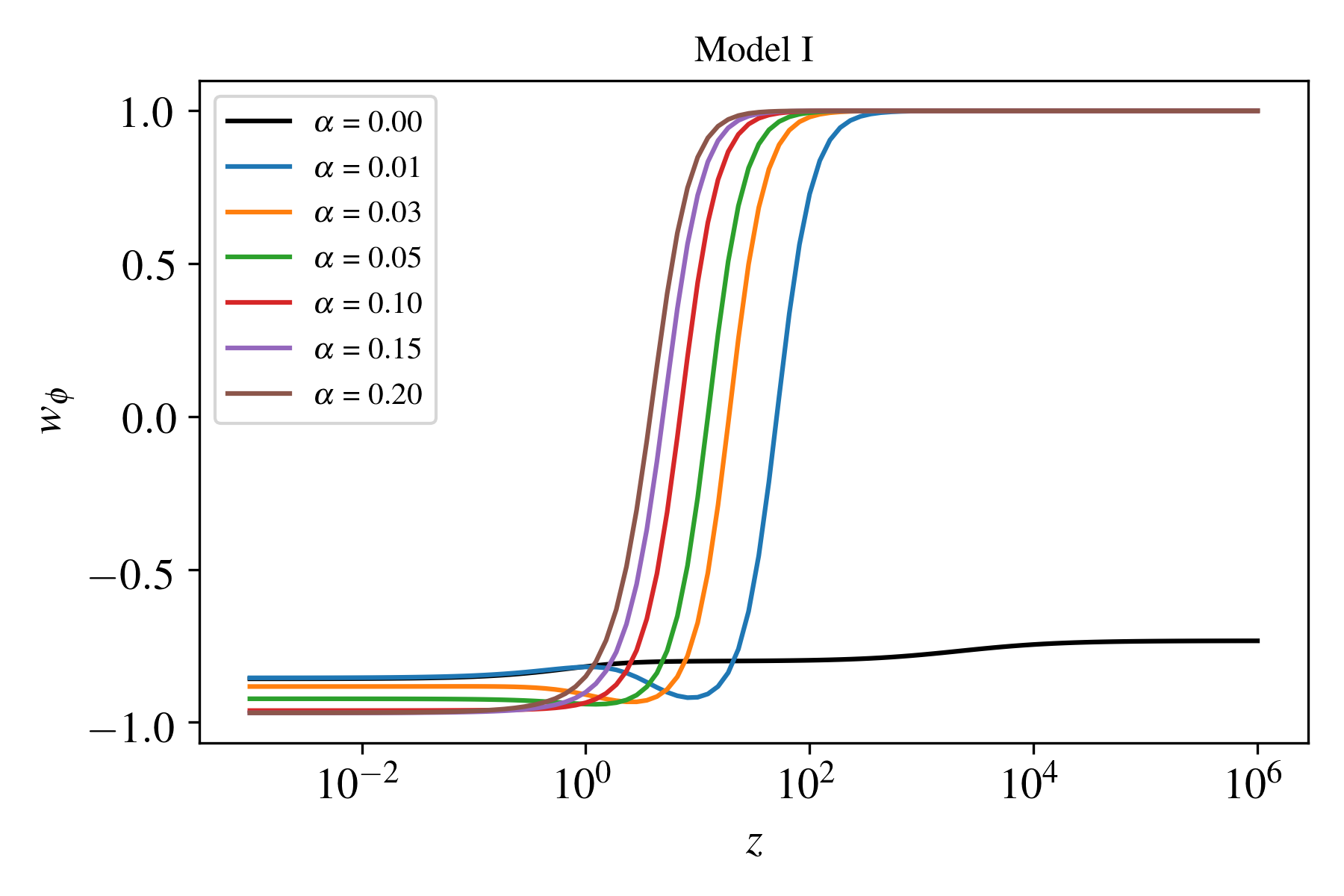}
	}
	\qquad
	\subfloat[\label{fig:w_eff_scf_m1}]{
		\includegraphics[width=0.4\textwidth]{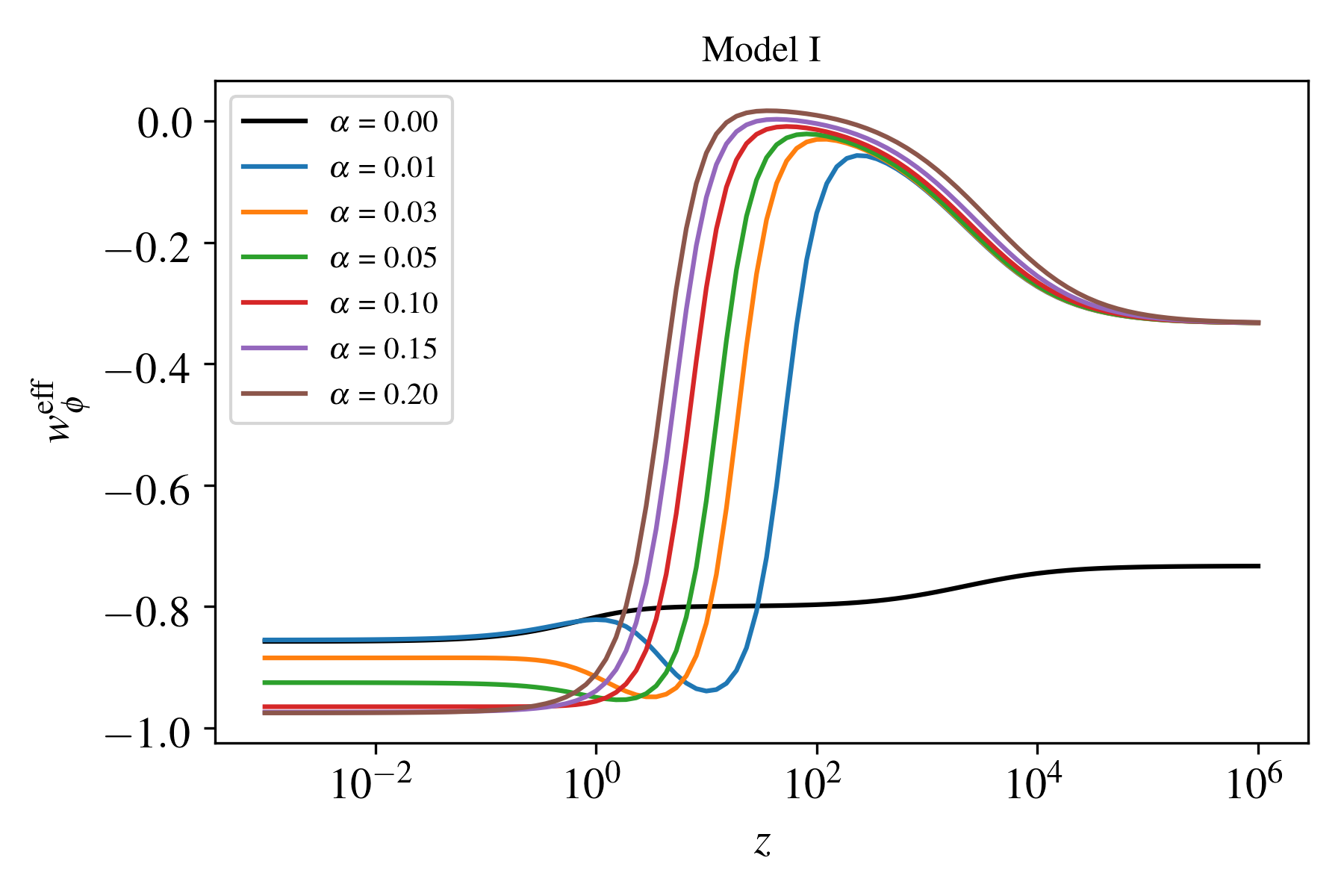}
	}
 	\caption{\label{fig:eos_m1} 
 	Left panel (a): Evolution of the scalar field equation of state, $w_\phi = p_\phi/\rho_\phi$.
 	Right panel (b): Evolution of the effective equation of state, Eq.~\eqref{eq:w_phi_eff}.
 	Both plots show the respective quantities as function of the redshift for the conformal model~I, Eq.~\eqref{eq:coupling_1}, 
 	for different values of the coupling constant.}
\end{figure}

In Fig.~\ref{fig:eos_m1} we present the evolution of the scalar field equation of state as a function of the redshift. 
In the uncoupled case, $\phi$ follows the tracker solution with, according to Eq.~\eqref{eq:eos_tracking_uncoupled}, $w_\phi \approx -0.73$ during radiation epoch, and $w_\phi \approx -0.8$ during matter epoch.
Even though the initial conditions for $\phi$ and $\phid$ in the coupled case were set using the tracker solution, the evolution of the scalar field during radiation epoch is such that its kinetic energy dominates over the potential and $w_\phi \approx 1$, departing from the tracker behavior.
For small values of the coupling constant (e.g. $\alpha \lesssim 0.01$), the field catches up with the tracker solution around the time of the onset of the accelerated expansion.
However, for larger values of $\alpha$, the evolution can deviate significantly from the uncoupled tracker solution, getting asymptotically closer to the limit $w_\phi = -1$ at the present time.


\subsubsection{Conformal model II}


As previously discussed, cosmological scenarios that admit scalar field tracking solutions are interesting.
In the uncoupled case, the equation of state of the scalar field follows the behavior of the dominant background fluid according to Eq.~\eqref{eq:eos_tracking_uncoupled}.
Hence, during the matter-dominated epoch, for small deviation from standard uncoupled quintessence, the equation of state of the background fluid $w_{\rm m}^{\rm eff} \simeq \wceff$ would still be constant, very close to zero.
Then, it is expected that Eq.~\eqref{eq:eos_tracking_uncoupled} still holds, with $\wphieff \simeq w_\phi$ also being approximately constant.
With that in mind, by imposing a few conditions on the evolution of the scalar field, we can derive an interaction term that preserves the tracker behavior. 
Suppose that, during the matter domination epoch, the scalar field not only admits a tracking solution, with both terms on the right-hand side of Eq.~\eqref{eq:w_phi_eff} being constant, but also the ratio $\rhoc / \rho_{\phi}^{p}$, for some integer $p > 1$, is constant. 
The exponent $p$ guarantees that the energy density of the scalar field decreases slower than the CDM energy density, eventually becoming dominant. 
In such case, we have $\phi \propto \phid^{1-p}$.
Since constant $w_\phi$ implies that $V \propto \phid^{2}$, the scalar potential must be $V\propto \phi^{2/(1-p)}$.
Our parameterization corresponds to $p=5$, namely $\rhoc\propto\rho_\phi^5$.
Finally, if for simplicity we assume that the conformal function $A(\phi)$ depends only on $\phi$ and that the disformal one is $B=0$, the requirement that the second term on the right-hand side of Eq.~\eqref{eq:w_phi_eff} is constant, leads to the following expression for the background coupling:
\begin{equation} \label{eq:bg_coupling_1}
    Q_0 = - \alpha \rhoc  \phi  \, , \quad \Gamma = -\frac{\alpha}{2} \phi^2.
\end{equation}
Then, from Eq.~\eqref{eq:rho_cdm}, the CDM energy density evolves as,
\begin{equation}
    \rhoc \propto a^{-3} \exp( -\frac{\alpha}{2} \phi^2) \,,
\end{equation}
and the equation of motion for the scalar field, Eq.~\eqref{eq:eom_phi_quintessence}, becomes
\begin{equation}
	\phidd + 3 H \phid + V_\phi = \alpha \rhoc  \phi.
\end{equation}

\begin{figure}[!ht]
	\subfloat[\label{fig:ratio_rho_cdm_m2}]{
		\includegraphics[width=0.4\textwidth]{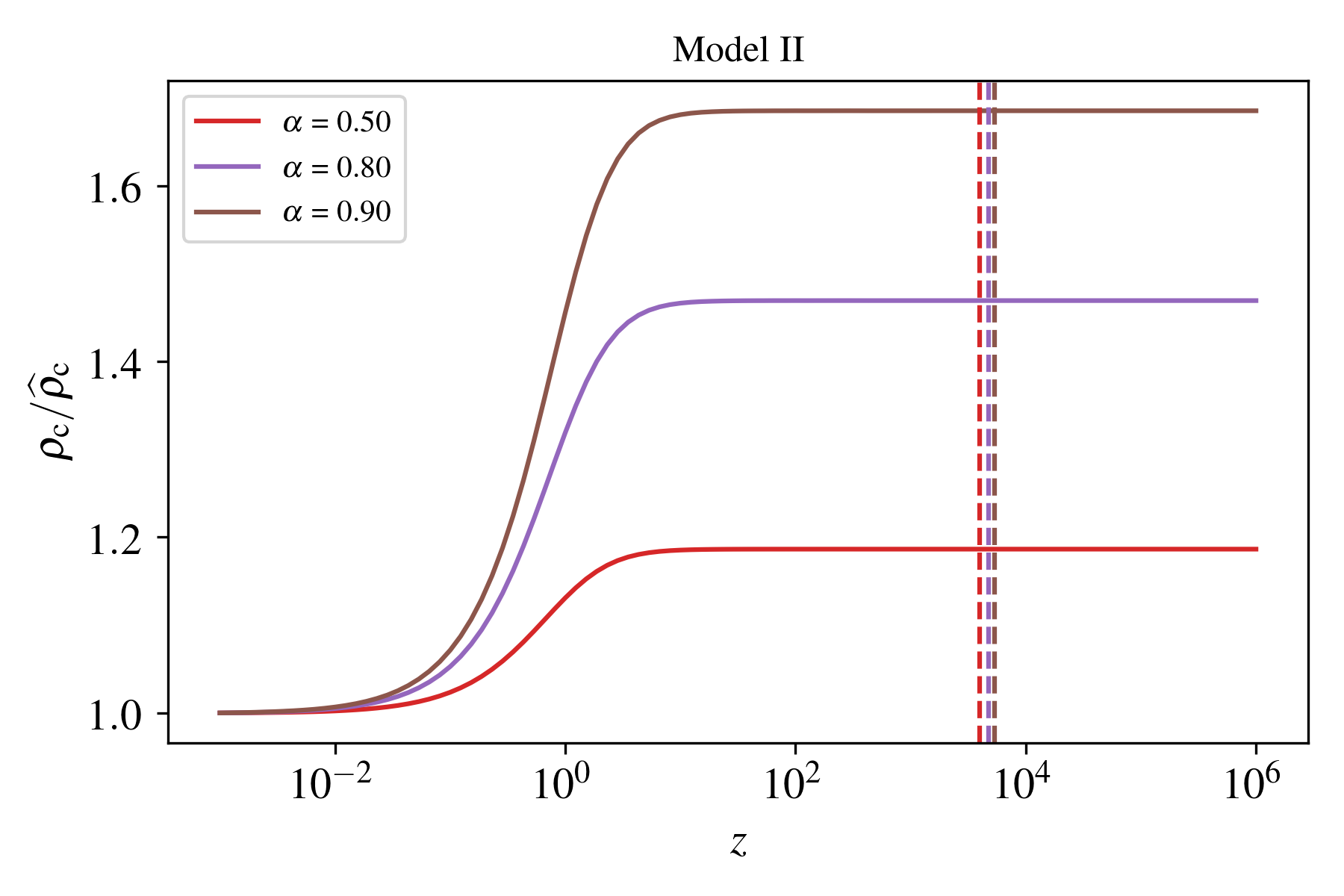}
	}
	\quad
	\subfloat[\label{fig:w_eff_cdm_m2}]{
	    \includegraphics[width=0.4\textwidth]{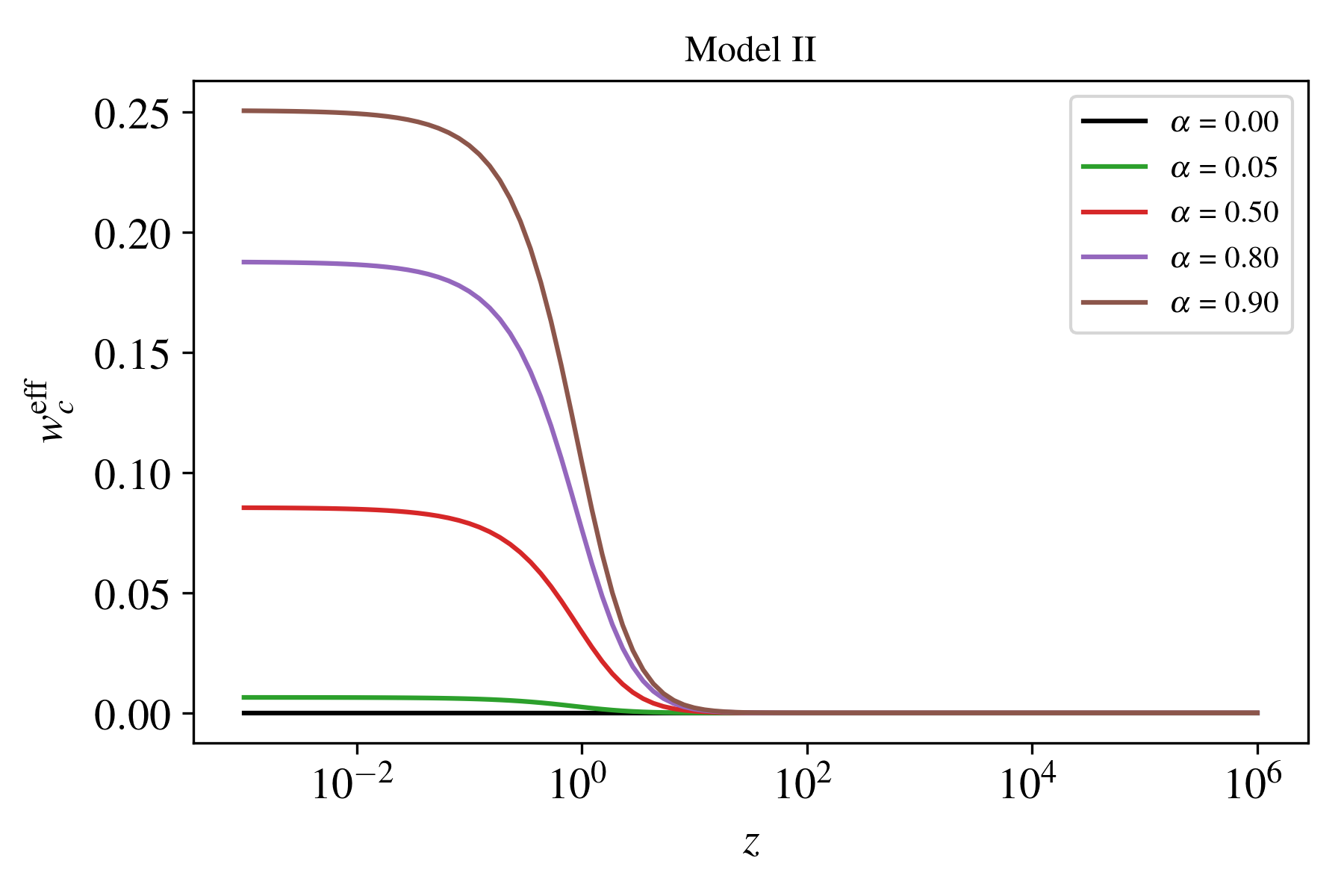}
	}
 	\caption{ \label{fig:bg_evo_cdm_m2}
 	Left panel (a): Evolution of the ratio between the coupled and uncoupled CDM energy density, $\rhoc$ and $\widehat{\rho}_{\rm c}$, respectively.
 	The dashed lines correspond to the redshift at matter-radiation equality.
 	Right panel (b): Evolution of the effective equation of state of dark matter, Eq.~\eqref{eq:w_c_eff}.
 	Both plots show the respective quantities as function of the redshift for the conformal model~II, Eq.~\eqref{eq:coupling_2}, 
 	for different values of the coupling constant.}
\end{figure}

Similarly to model~I, for positive values of the coupling constant, $\rhoc$ dilutes faster than in the uncoupled case, and there is an injection of energy from CDM to the scalar field.
However, note that the conformal factor~\eqref{eq:coupling_2} has an additional $\phi$ dependence on the exponential term, when compared to model~I.
Typically, during the cosmic evolution, the dynamics are such that $0 < \phi(t) < 1$, in reduced Planck mass units.
Hence, in comparison to model~I, the interaction in the dark sector for model~II becomes relevant at later times, around the transition from matter to DE domination epochs, when the scalar field becomes large enough to flatten the potential (Fig.~\ref{fig:ratio_rho_cdm_m2}).
This is reflection of the fact that, by construction, model~II effectively preserves the tracker properties up to the transition from matter to scalar field domination epochs.
To get the same energy density at earlier times as in model~I, the corresponding coupling constant must be significantly larger, and the ratio $\rhoc / \widehat{\rho}_{\rm c}$ increases steeply (as function of redshift) at late times.
The effective equation of state for CDM remains close to zero during most part of the cosmic expansion (Fig.~\ref{fig:w_eff_cdm_m2}) such that the evolution $\rhoc$ does not deviate much from the uncouple behavior, except at the low-$z$ region.

\begin{figure}[!ht]
	\subfloat[ \label{fig:w_scf_m2}]{
		\includegraphics[width=0.4\textwidth]{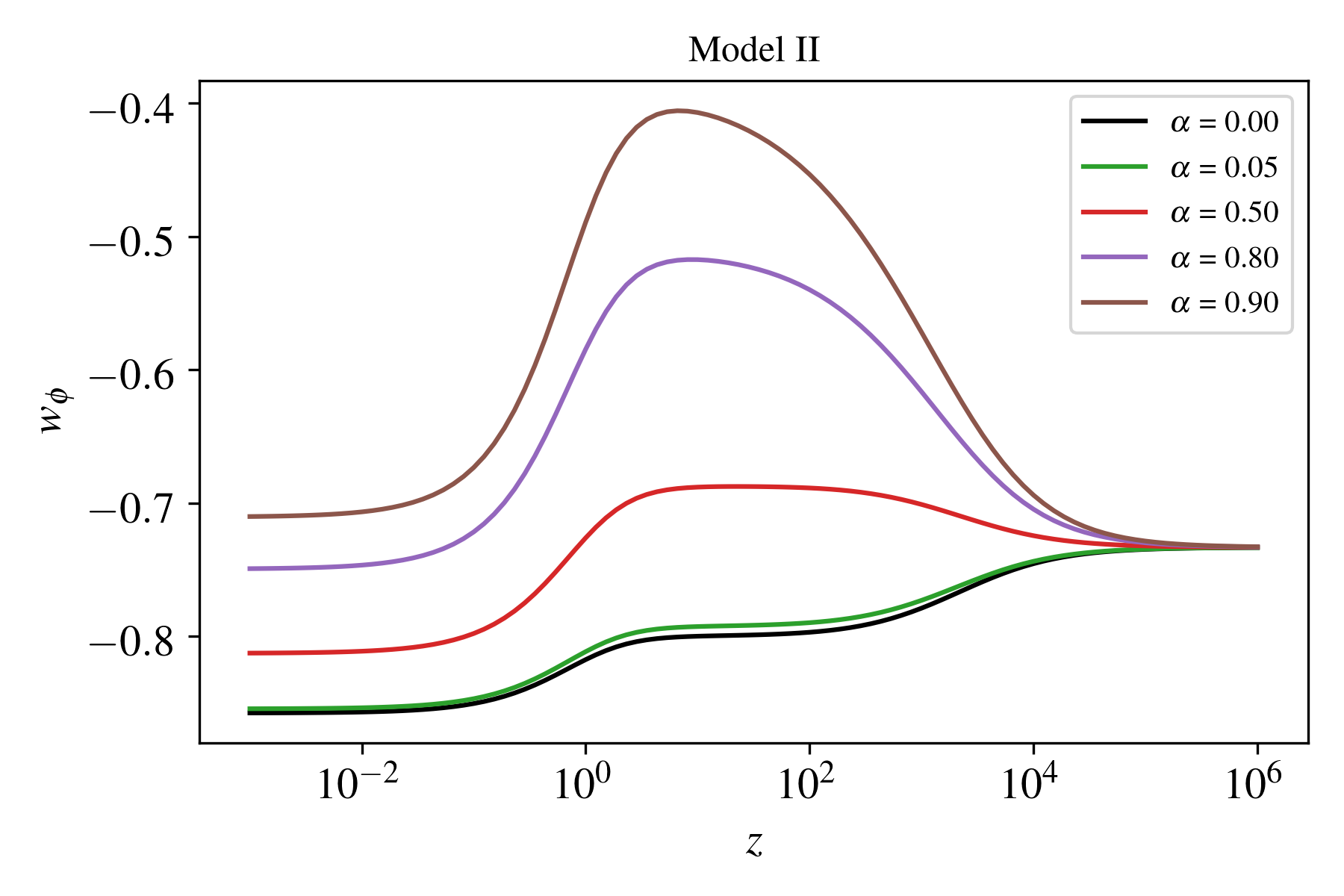}
	}
 	\qquad
 	\subfloat[\label{fig:w_eff_scf_m2}]{
 		\includegraphics[width=0.4\textwidth]{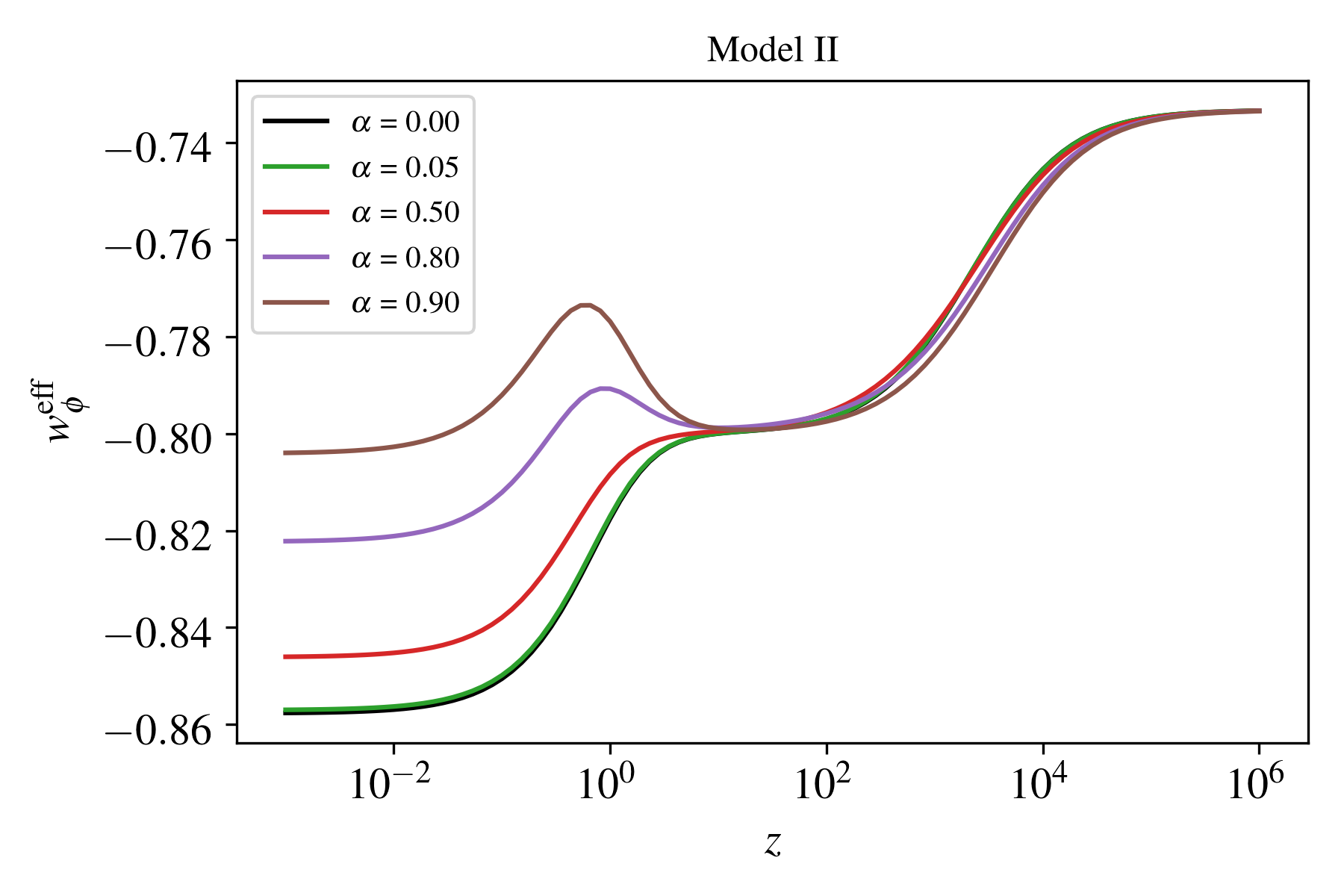}
 	}
 	\caption{\label{fig:eos_m2} 
 	Left panel (a): Evolution of the scalar field equation of state, $w_\phi = p_\phi/\rho_\phi$.
 	Right panel (b): Evolution of the effective equation of state, Eq.~\eqref{eq:w_phi_eff}.
 	Both plots show the respective quantities as function of the redshift for the conformal model~II, Eq.~\eqref{eq:coupling_2}, 
 	for different values of the coupling constant.}
\end{figure}

In this model the evolution of the coupled scalar field indeed exhibits a tracker behavior up to matter domination epoch, 
when the scalar field equation of state is roughly constant, Fig.~\ref{fig:w_scf_m2}, but the coupling slightly changes the value of  $w_\phi$.
However, this change is compensated by the $Q_0$ term in Eq.~\eqref{eq:w_phi_eff} such that the effective equation of state in the coupled case does not change with the interaction, remaining close to the uncoupled one, Fig.~\ref{fig:w_eff_scf_m2}.
In contrast to model~I, for larger values of the coupling constant, the kinetic energy of the scalar field becomes more important when compared to the scalar potential, and $w_\phi$ is less negative at late times. 


\subsubsection{Disformal model III}


As an example of purely disformal coupling, we consider model~III, characterized by the coupling functions~\eqref{eq:coupling_2}.
In this case, there is no interaction at the background level, $Q_0=0$.
Hence, the energy density of cold dark matter evolves the usual way, $\rhoc \propto a^{-3}$, and the equation of motion for the scalar field is
\begin{equation}
	\phidd + 3 H \phid + V_\phi = 0.
\end{equation}
From the observational point of view, the lack of modifications at background level with respect to the standard quintessence scenario is interesting, given the current constraints on the properties of the homogeneous universe. The effects induced by the DE/CDM interaction are limited to change the evolution of linear perturbations only.


\subsection{Perturbations}


In this subsection, we present the evolution of linear perturbations and study each model in the quasi-static limit.
Even though all the numerical results presented in this work were computed using the full set of $k$-dependent linear perturbation equations shown in section \ref{sec:bg_and_lin_pert}, it is elucidating to study the evolution in the quasi-static limit, since in said limit, one can write the equations of motion for the linear density in a simple way, allowing a straightforward discussion of the effects due to the coupling.
Furthermore, as previously discussed, for cosmological scales probed by galaxy surveys, said approximation is valid and there is no loss of physical information.
In what follow, we present the evolution of perturbations for modes with $k = 1 h \, \mega\parsec^{-1}$.
We numerically checked that these modes are sufficiently deep inside the sound horizon of scalar perturbation (see Appendix~\ref{sec:dispersion_relation}) and that the quasi-static approximation is indeed valid.


\subsubsection{Conformal model I}

\begin{figure}[!ht]
	\subfloat[\label{fig:ratio_f_m_m1}]{
		\includegraphics[width=0.4\textwidth]{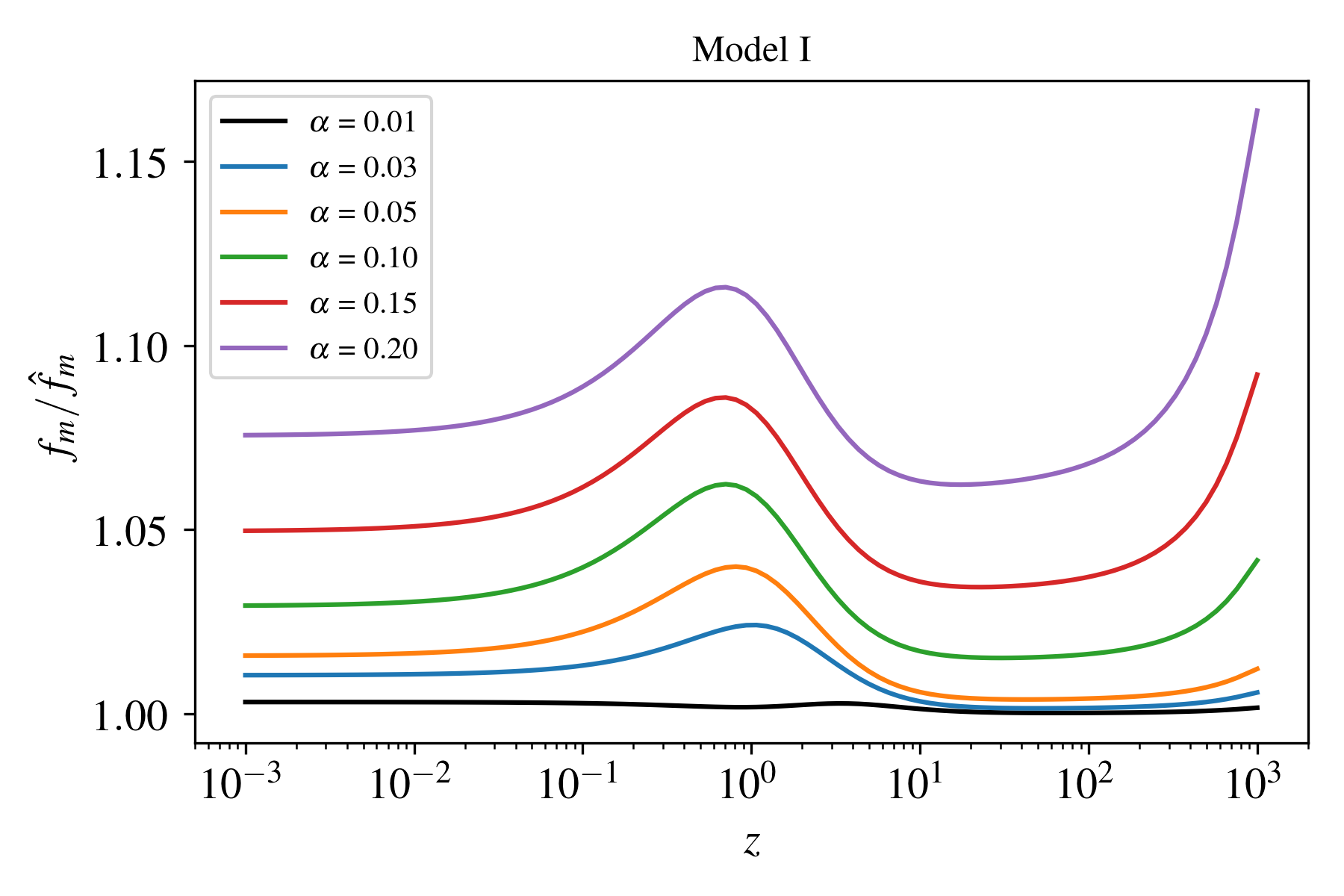}
		}
	\qquad
	\subfloat[\label{fig:Delta_f_m_m1}]{
		\includegraphics[width=0.4\textwidth]{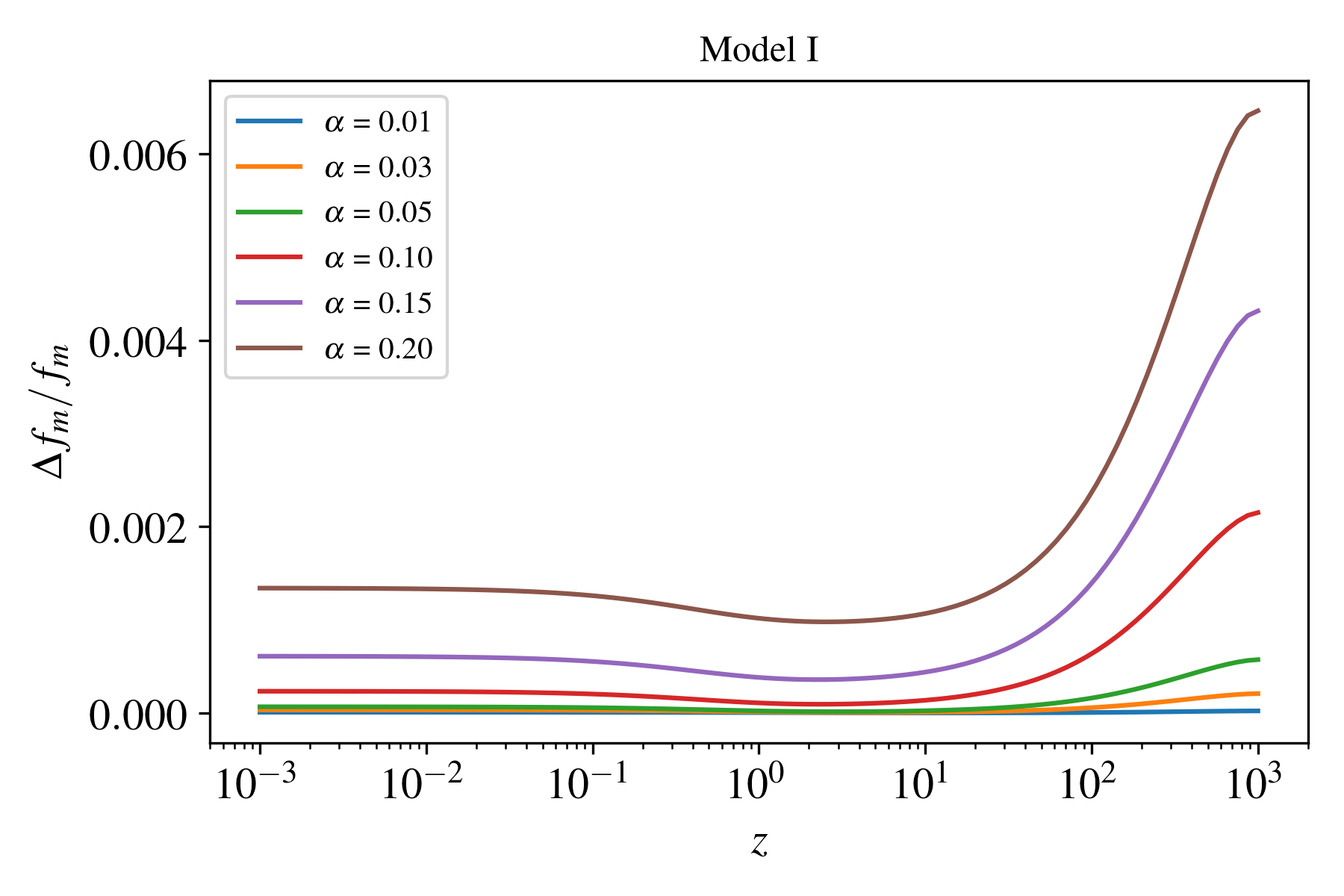}
		}
 	\caption{ \label{fig:growth_function_m1}
 		Right panel (a): ratio $\fm / \hat{f}_{\rm m}$ where $\hat{f}_{\rm m}$ is the growth rate in the uncoupled case and $\fm$ is the one with non-zero coupling. 
 		Left panel (b): difference between the effective and actual growth rates $\Delta \fm / \fm$, Eq.~\eqref{eq:f_eff_m}.
 		Both figures are shown as function of the redshift for the conformal model~I, Eq.~\eqref{eq:coupling_1}, 
 		for different values of the coupling constant.}
\end{figure}

In model~I, since the coupling is purely conformal, both $\Upsilon_1$ and $\Upsilon_2$, given by Eq.~\eqref{eq:upsilon}, vanish and the equation of motion for the CDM growth function, Eq.~\eqref{eq:cdm_continuity_qs}, becomes
\begin{equation}
	\ddot{D}_{\rm c} + 2H(1 - \E_1) \dot{D}_{\rm c} - \frac{1}{2} \left[ (1 - \E_2) \rhoc D_{\rm c} + \rhob D_{\rm b} \right] = 0 \,,
\end{equation}
with the the following contributions to the Hubble friction and gravitational constant terms, respectively:
\begin{equation} \label{eq:E1_E2_m1}
		\E_1 = \frac{\alpha}{2} \frac{\phid}{H} \,, \qquad 
		\E_2 = - 2 \alpha^2 \,.
\end{equation}
First, by inspecting Eqs.~\eqref{eq:w_phi_eff} and \eqref{eq:bg_coupling_1}, note that $\E_1 = 3 \wceff / 2$,
being proportional to the square root of the fractional kinetic energy of the scalar field such that, for small values of the coupling constant $\alpha$, the usual Hubble friction term is recovered at early times.
On the other hand, since $\E_2$ gives a constant contribution, the conformal coupling~\eqref{eq:coupling_1} changes the gravitational constant term by a factor~$\sim \mathcal{O}(\alpha^2)$ at all epochs.
For $\alpha > 0$, the interaction suppresses the friction term and enhances the gravitational pull, leading to more growth (Fig.~\ref{fig:ratio_f_m_m1}).
This is a general feature of models with non-minimal couplings, where the interaction between matter particles is mediated by a scalar field, leading to an attractive fifth-force.
As it will be shown, this behavior is also present in models II and III.

Besides, since $\Upsilon_2$ vanishes, according to Eq.~\eqref{eq:f_eff_cdm} the effective and actual growth rates for CDM coincide. 
Even though $\Delta f_{\rm c} = 0$, the effective linear growth rate for total matter, Eq.~\eqref{eq:f_eff_m}, will differ from the actual growth rate, by a factor $\sim \mathcal{O}(\alpha^2)$, due to the combined effect of the background coupling $Q_0$ and the difference between the baryon and CDM growth functions (Fig.~\ref{fig:Delta_f_m_m1}).
Hence, the non-minimal coupling reinforces the redshift-distortion factor.

\begin{figure}[!ht]
	\includegraphics[width=0.4\textwidth]{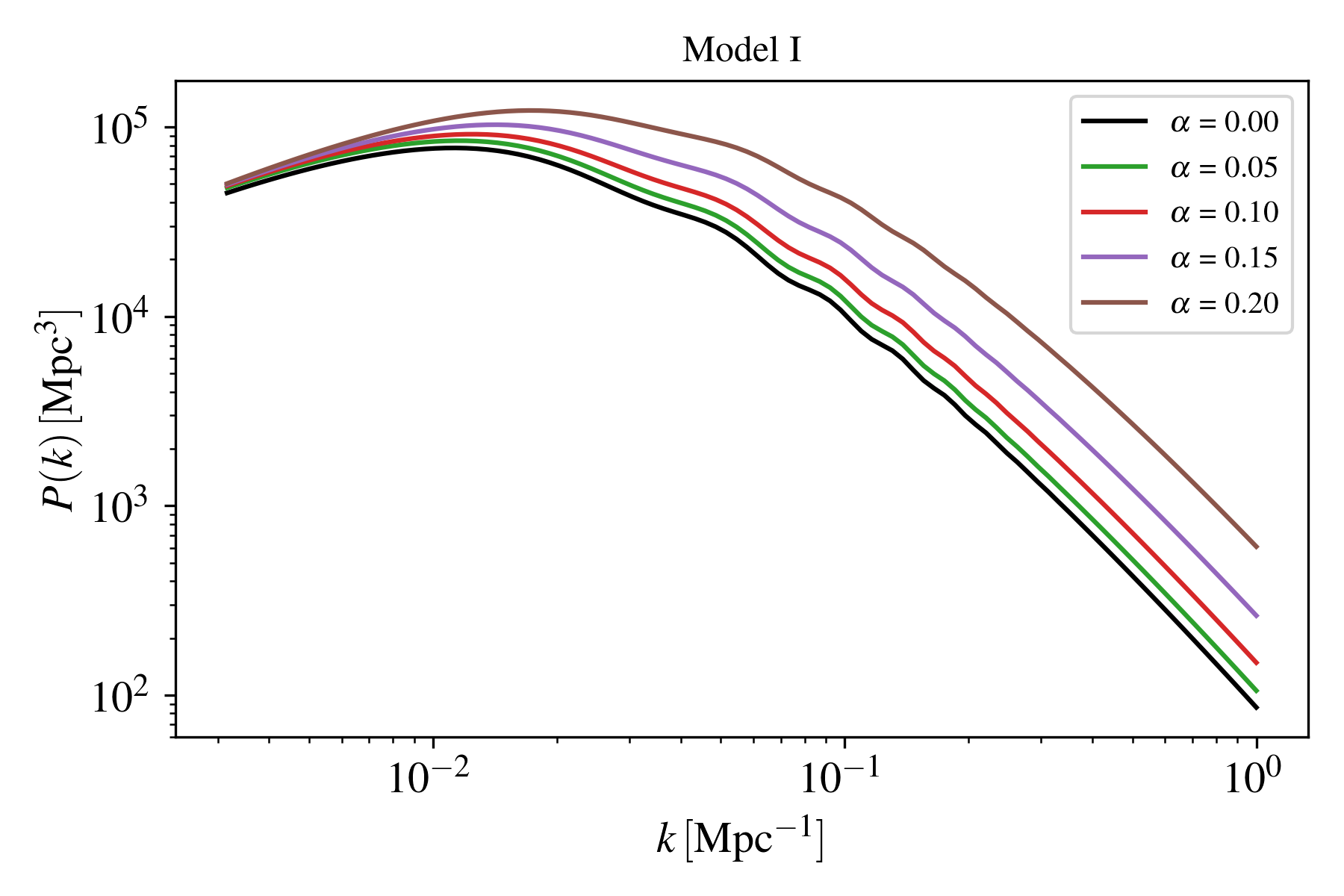}
	\caption{ \label{fig:pk_qs_m1}
 	Linear matter power spectrum for conformal model~I, Eq.~\eqref{eq:coupling_1}, 
 	for different values of the coupling constant.}
\end{figure}

Let us now turn to the matter power spectrum (Fig.~\ref{fig:pk_qs_m1}).
On small scales, i.e. in the quasi-static limit, since the scalar interaction induces more growth on the CDM linear density, the power spectrum on those scales will be equally augmented.
Also, since larger $\alpha$ makes the matter-radiation equality to take place earlier, when the Hubble horizon was smaller, the wavelength of modes that entered the horizon at the equality will be smaller.
Hence, the whole power spectrum will be shifted towards larger values of $k$.
Another effect related to the background dynamics is that, since the ratio $\rhob / \rhom$ at early times decreases for larger values of the coupling constant, the amplitude of the baryon acoustic oscillations also decreases.


\subsubsection{Conformal model II}

\begin{figure}[!ht]
    \centering
    \subfloat[\label{fig:E1_m2}]{
            \includegraphics[width=0.4\textwidth]{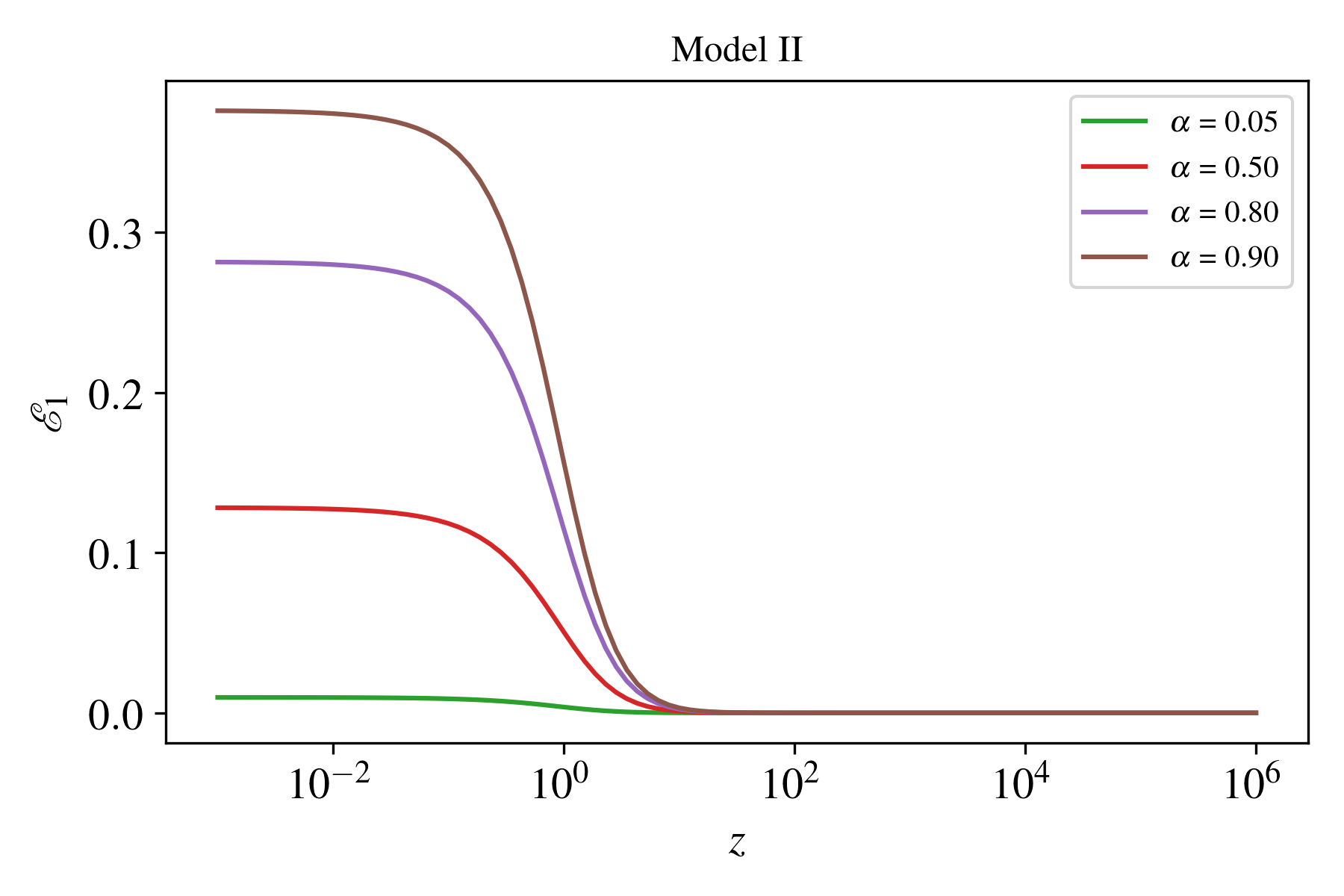}
    }
    \qquad
    \subfloat[\label{fig:E2_m2}]{
        \includegraphics[width=0.4\textwidth]{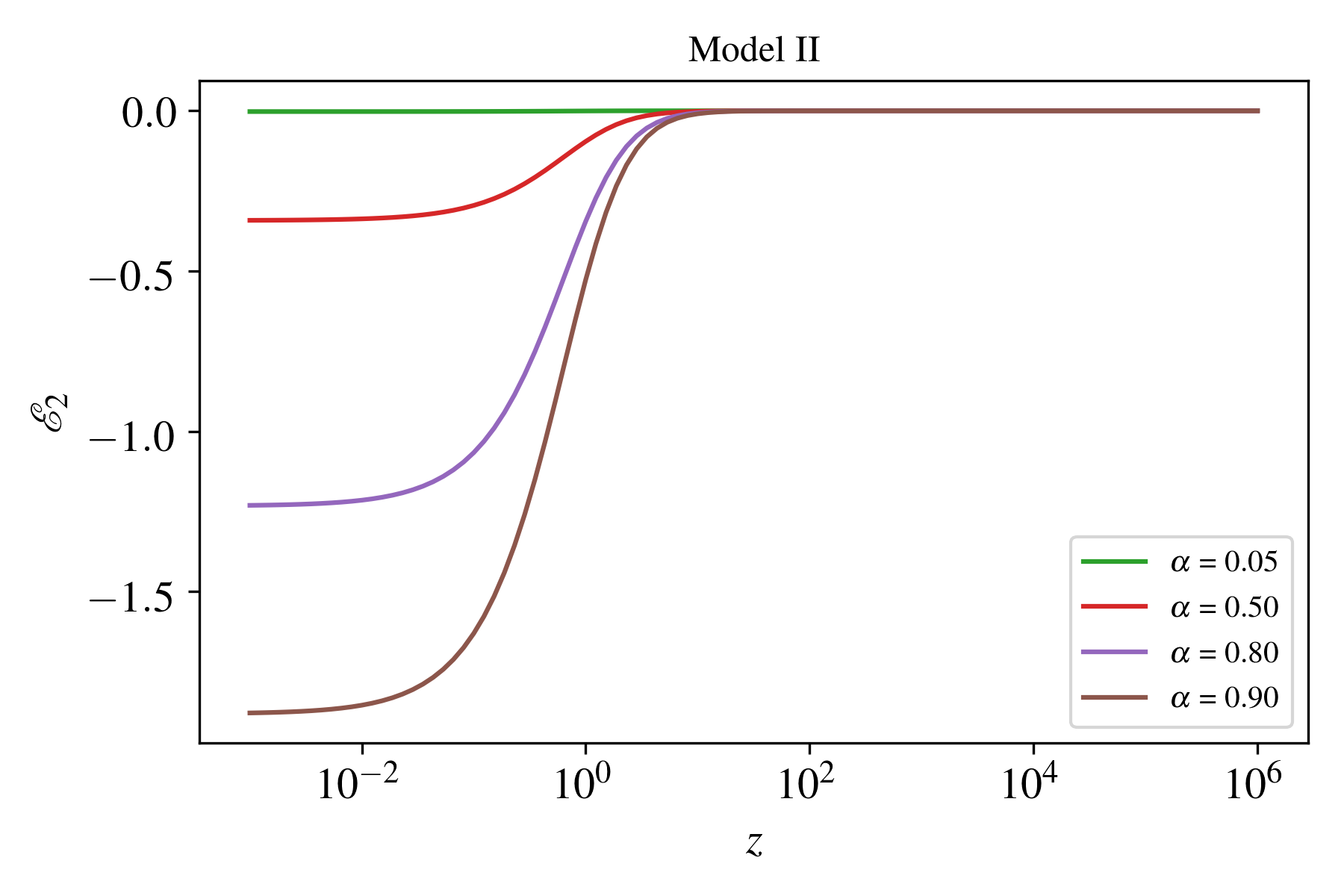}
    }
    \caption{\label{fig:E1_E2_m2}
    Evolution of the additional Hubble friction $\E_1$ (left panel (a)) and gravitational constant $\E_2$ (right panel (b)), Eq.~\eqref{eq:E1_E2_m2},  as function of redshift for the conformal model~II, Eq.~\eqref{eq:coupling_2}, for different values of the coupling constant.}
\end{figure}
\begin{figure}[!ht]
	\subfloat[\label{fig:ratio_f_m_m2}]{
		\includegraphics[width=0.4\textwidth]{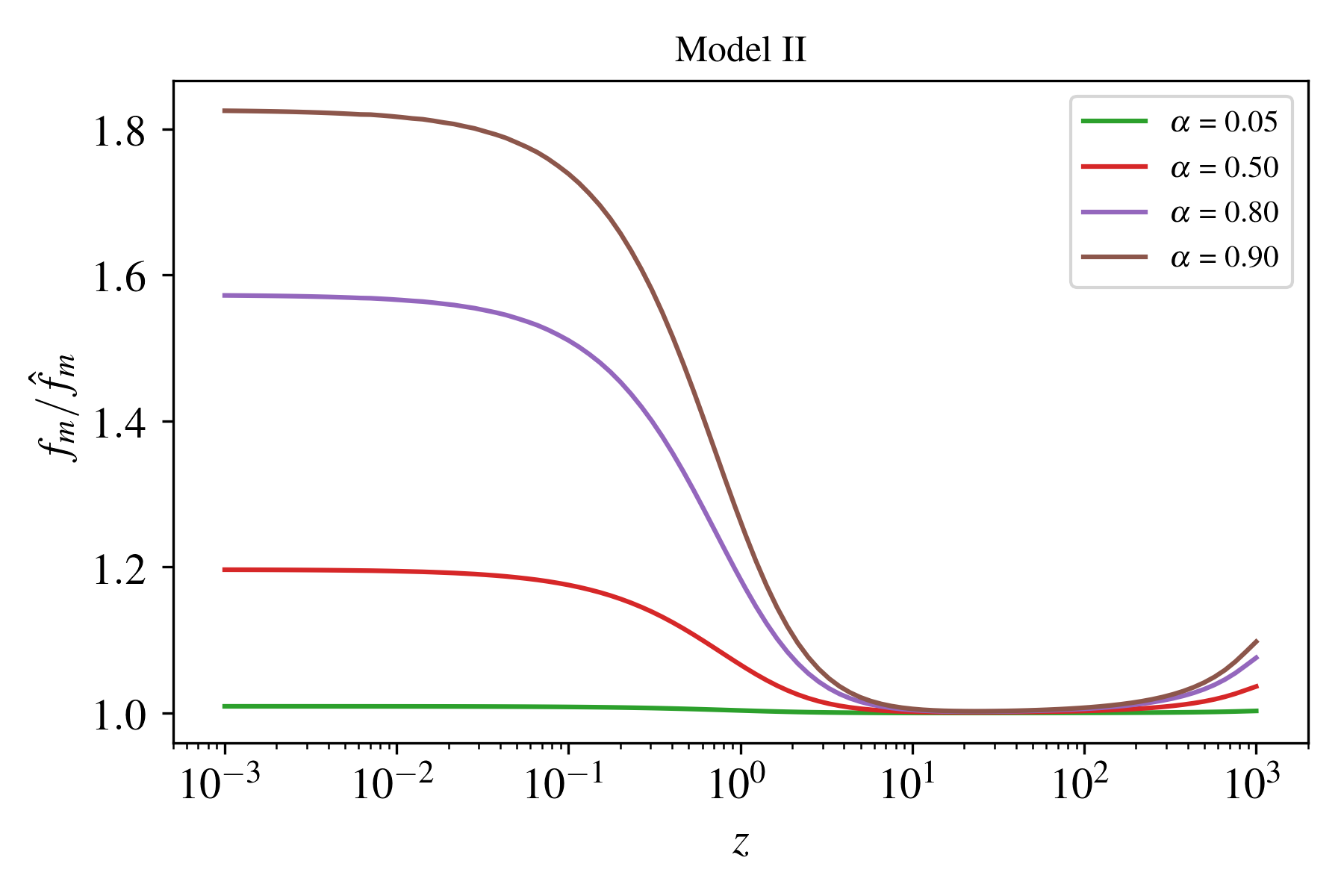}
		}
	\qquad
	\subfloat[\label{fig:Delta_f_m_m2}]{
		\includegraphics[width=0.4\textwidth]{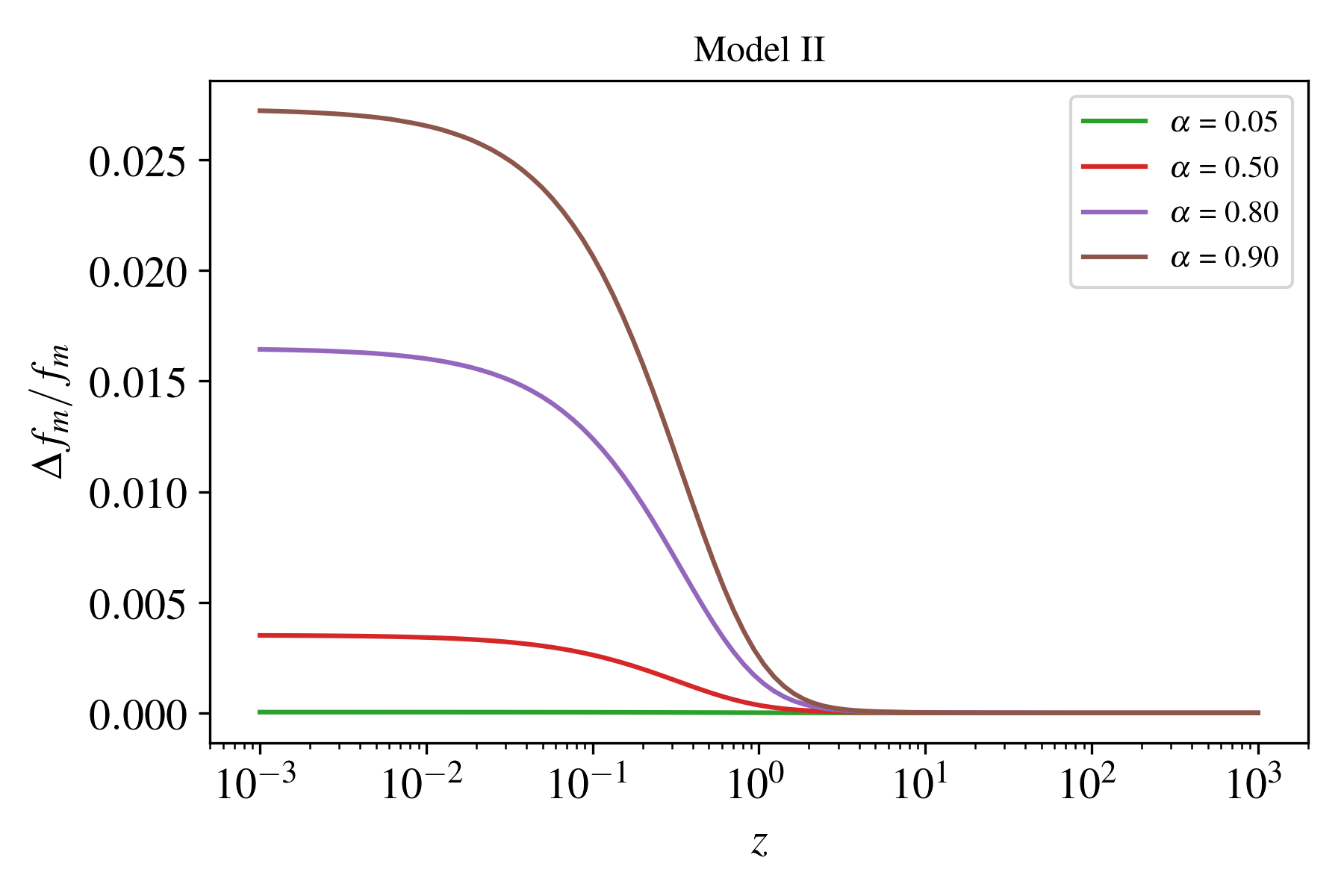}
		}
 	\caption{ \label{fig:growth_function_m2} 
 		Left panel (a): ratio $\fm / \hat{f}_{\rm m}$ where $\hat{f}_{\rm m}$ is the growth rate in the uncoupled, case and $ \fm$ is the one with non-zero coupling. 
 	    Right panel (b): difference between the effective and actual growth rates $\Delta \fm / \fm$, Eq.~\eqref{eq:f_eff_m}.
 		Both figures are shown as function of the redshift for the conformal model~II, Eq.~\eqref{eq:coupling_2}, for different values of the coupling constant.
 		}
 		\end{figure}

In model II, once again, both $\Upsilon_1$ and $\Upsilon_2$ vanish and the modifications to the evolution of dark matter growth function appear through  
\begin{equation}
    \ddot{D}_{\rm c} + 2H(1 - \E_1) \dot{D}_{\rm c} - \frac{1}{2} \left[ (1 - \E_2) \rhoc D_{\rm c} + \rhob D_{\rm b} \right] = 0 \,,
\end{equation}
where
\begin{equation} \label{eq:E1_E2_m2}
    \E_1 = \frac{\alpha \, \phi \, \phid}{ 2 \, H}, \qquad 
    \E_2 = -2 \alpha^2 \phi^2.
\end{equation}
As in conformal model~I, the modification to the Hubble friction term is related to the effective CDM equation of state by $\E_1 = 3 \wceff / 2$.
Once again, the correction to the Hubble friction and gravitational constant terms are $\mathcal{O}(\alpha)$ and $\mathcal{O}(\alpha^2)$, respectively, and the scalar force leads to more clustering. 
Similarly to the behavior at background level, since both $\E_1$ and $\E_2$ are $\phi$-suppressed, the growth function during matter domination epoch does not differ significantly from the uncoupled one during matter epoch, and all the coupling-induced growth happens at late times (Fig.~\ref{fig:E2_m2}).
Hence, in order to get a deviation from the uncoupled growth function comparable to the one in model~I, the value of the coupling constant must be considerably larger (Fig.~\ref{fig:ratio_f_m_m2}).

As in model I, $\Delta f_{\rm c}$ is zero and the deviation between effective and actual matter growth rate, also by a factor~$\sim \mathcal{O}(\alpha^2)$, is due to the background coupling and the difference between the baryon and CDM growth function~(Fig.~\ref{fig:Delta_f_m_m2}). 
However, since in model II the quantity $\E_2$ is negligible at early times, the difference $\Delta \fm$ is present only at later epochs.
Since $\Delta \fm$ is positive, on top of the enhancement of the actual growth rate due to the coupling, the effective growth will be further enhanced.

\begin{figure}[!ht]
	\includegraphics[width=0.4\textwidth]{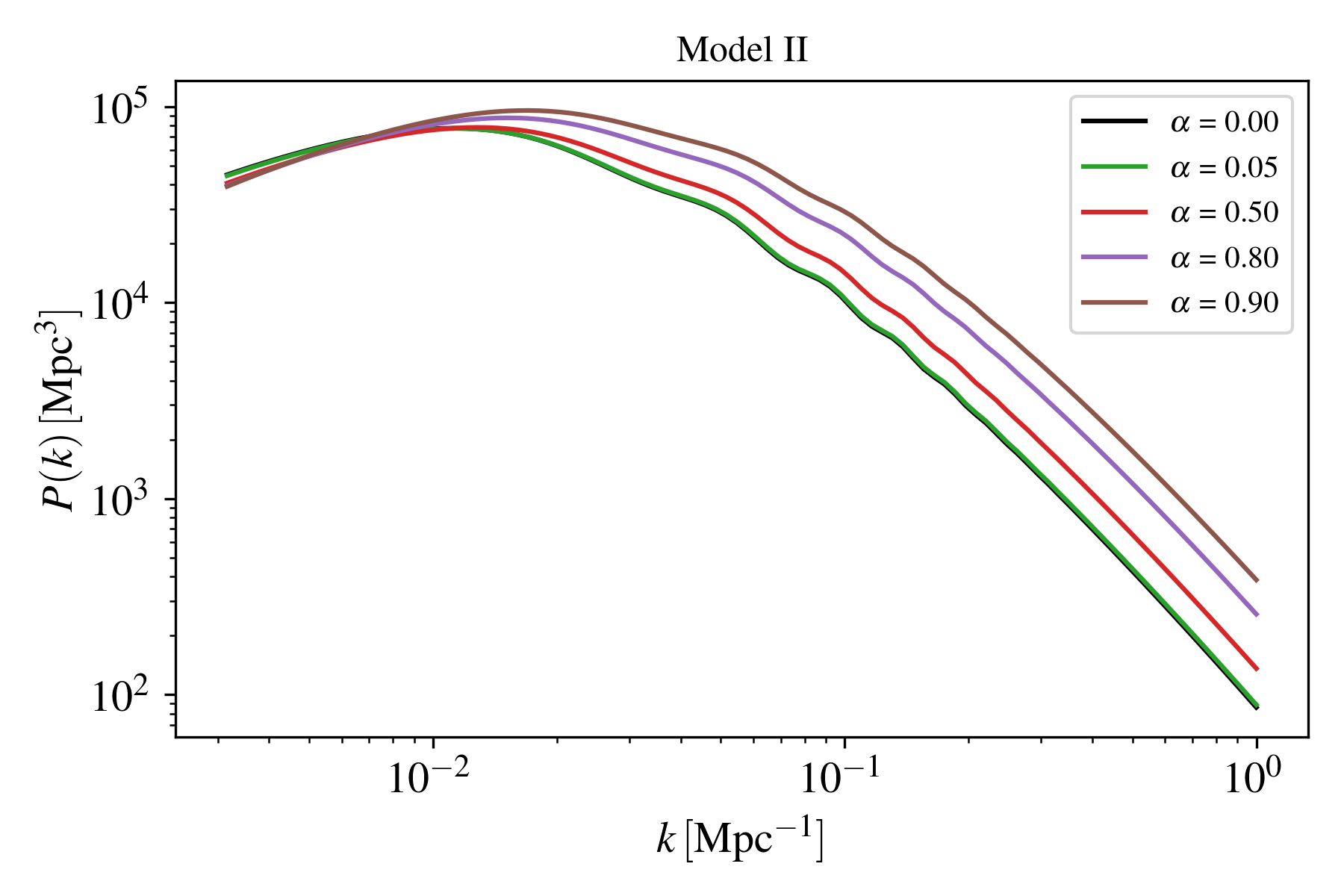}
 	\caption{\label{fig:pk_qs_m2}
	Linear matter power spectrum for conformal model~II, Eq.~\eqref{eq:coupling_2}, 
	for different values of the coupling constant.}
\end{figure}

The behavior of the matter power spectrum of linear perturbations in model~II (Fig.~\ref{fig:pk_qs_m2}) resembles that of model~I.
For larger values of the coupling constant, because of the clustering induced by the coupling, it is enhanced at small scales, whereas the change in the time of matter-radiation equality shifts the whole spectrum towards larger values of $k$.
The amplitude of the baryonic feature also decreases.
However, since the interaction is effective only during the scalar-field-dominated epoch, to get a modification of the same effect as in model~I, the coupling constant must be roughly one order of magnitude larger.


\subsubsection{Disformal model III \label{sec:ex_pert_m3}}

\begin{figure}[!ht]
	\subfloat[\label{fig:E1_m3}]{
		\includegraphics[width=0.4\textwidth]{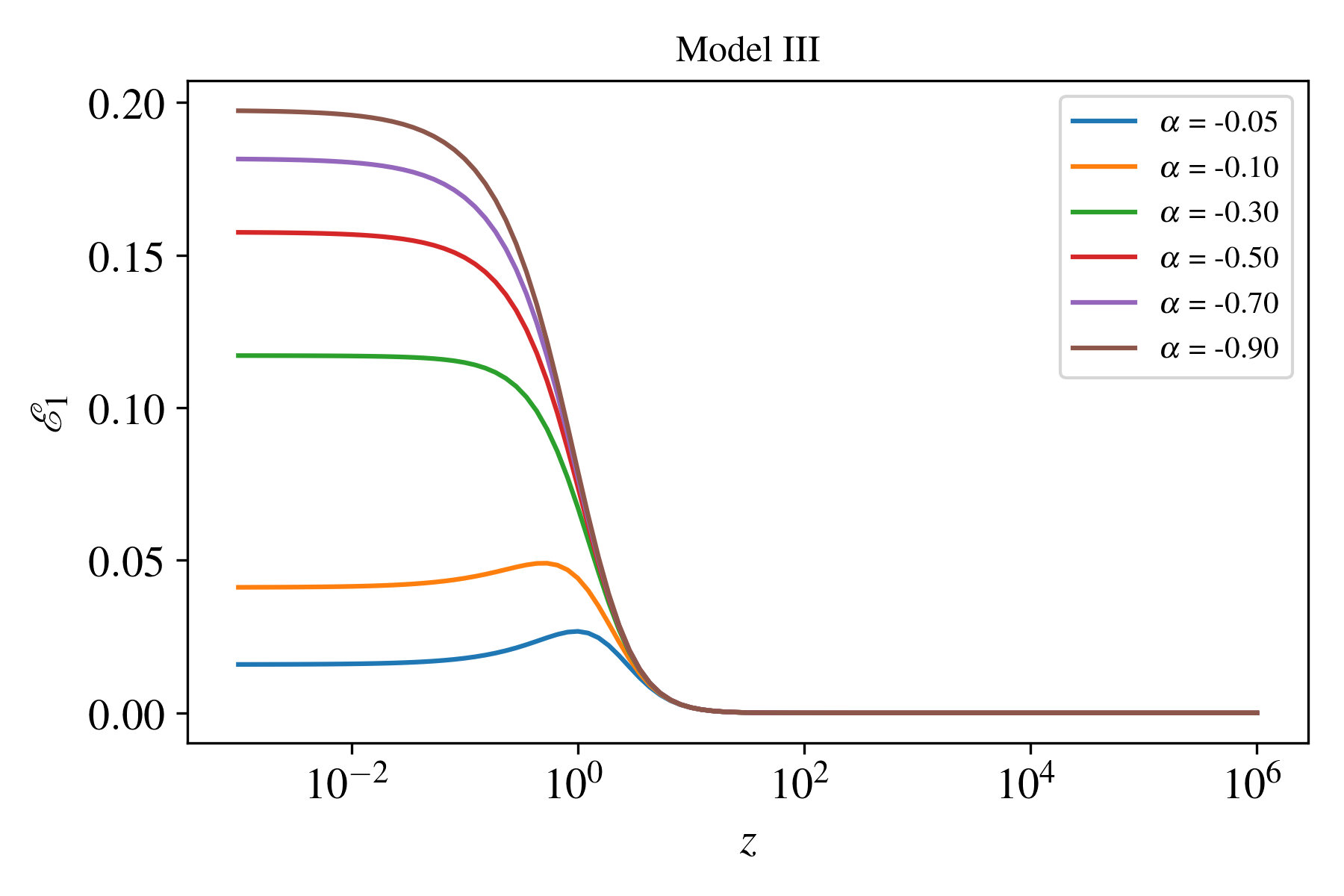}
	}
	\qquad
	\subfloat[\label{fig:E2_m3}]{
		\includegraphics[width=0.4\textwidth]{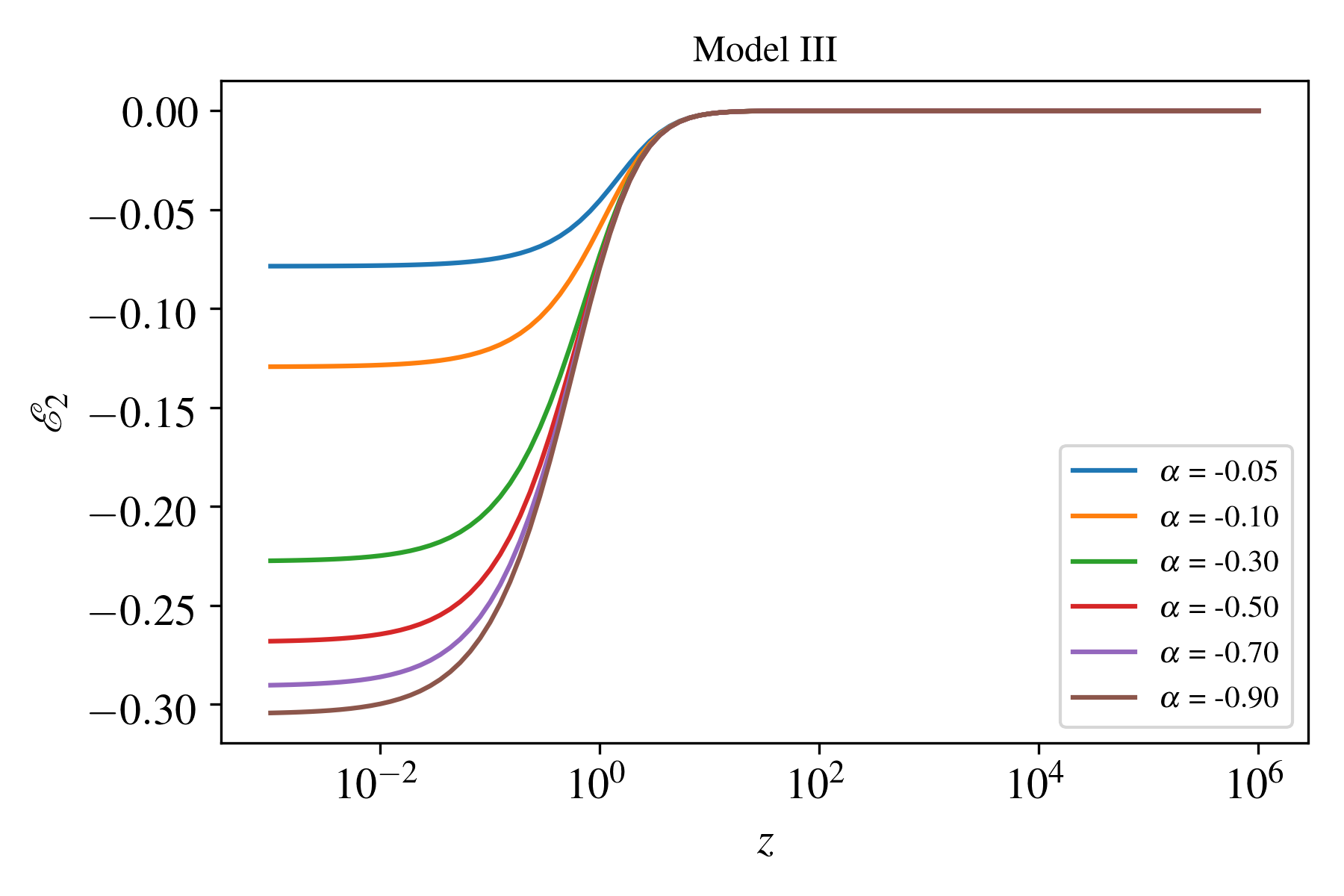}
	}
	\caption{\label{fig:E1_E2_m3}
	Evolution of the additional Hubble friction $\E_1$ (left panel (a)) and gravitational constant $\E_2$ (right panel (b)), Eq.~\eqref{eq:E1_E2_m3},  
	as function of redshift for the disformal model~III, Eq.~\eqref{eq:coupling_3}, for different values of the coupling constant.}
	\end{figure}

For model~III, with conformal and disformal factors defined in Eq.~\eqref{eq:coupling_3}, one finds $\Upsilon_1 = -\Upsilon_2= \E_2$.
Hence, the equation of motion for the CDM growth function, Eq.~\eqref{eq:cdm_continuity_qs}, becomes
\begin{equation}
	\ddot{D}_{\rm c} + 2H(1 - \E_1) \dot{D}_{\rm c} - \frac{1}{2} \left(1 - \E_2 \right) \left(\rhoc D_{\rm c} + \rhob D_{\rm b} \right) = 0 \,,
\end{equation}
where, 
\begin{equation} \label{eq:E1_E2_m3}
	\E_1 = -{
		\frac{ 2 \alpha ^2 \rhoc \phid \left( 3 H \phid + 2 V_\phi  \right)} 
		{H \left( \phid^2 - 2 \alpha \rhoc \right) \left[ (1-2 \alpha) \phid^2 - 2 \alpha  \rhoc \right]}} \, , \qquad
	\E_2 = \frac{2 \alpha \phid^2}{\phid^2 - 2 \alpha \rhoc} \, .
\end{equation}
Note that, in principle, when $\phid^2 = 2 \alpha \rhoc$ both $\E_1$ and $\E_2$ functions diverge, whereas when $(1-2 \alpha) \phid^2 - 2 \alpha  \rhoc=0$ only $\E_1$ has problems. 
The  requirement that the disformally transformed metric $\bar{g}_{\mu\nu}$ preserves the Lorentzian signature for this choice of the functions $A$ and $B$ translates into $2\alpha < 1$ (see Appendix~\ref{app:metric_transformation}), which by itself does not guarantee that the functions $\E_1$ and $\E_2$ are well defined.
However, the disformal coupling modifies the sound speed of the scalar field, not being equal to unity as it is the case for standard quintessence. 
The requirement that $c_{\rm s}^{2} > 0$ leads to the constraint $\alpha < \phid^2/ (2\rhoc)$, the ratio between the scalar field kinetic energy and the CDM energy density (see Appendix~\ref{sec:dispersion_relation}). 
Since this relation must be satisfied throughout the cosmic evolution, we consider only negative values for the coupling constant.
Similarly to model~II, at earlier times the functions $\E_1$ and $\E_2$ are negligibly small, so the growth rate is insensitive to the interaction, 
and most of the effects due to the coupling become important during onset of the DE scalar field domination epoch (Fig.~\ref{fig:E1_E2_m3}).
That is a consequence of the fact that both functions depend not only on the coupling constant, but also on the ratio between the kinetic energy of the scalar field, $\phid^2/2$, and the CDM energy density.
During matter domination epoch, this ratio is negligibly small, such that both $\E_1$ and $\E_2$ remain close to zero, regardless of the coupling constant value.
However, as the ratio gets closer to unity at late times, the coupling becomes significant, with $\E_1$ and $\E_2$ being roughly proportional to $\alpha^2$ and $\alpha$, respectively.

\begin{figure}[!ht]
	\subfloat[\label{fig:ratio_f_m_m3}]{
		\includegraphics[width=0.4\textwidth]{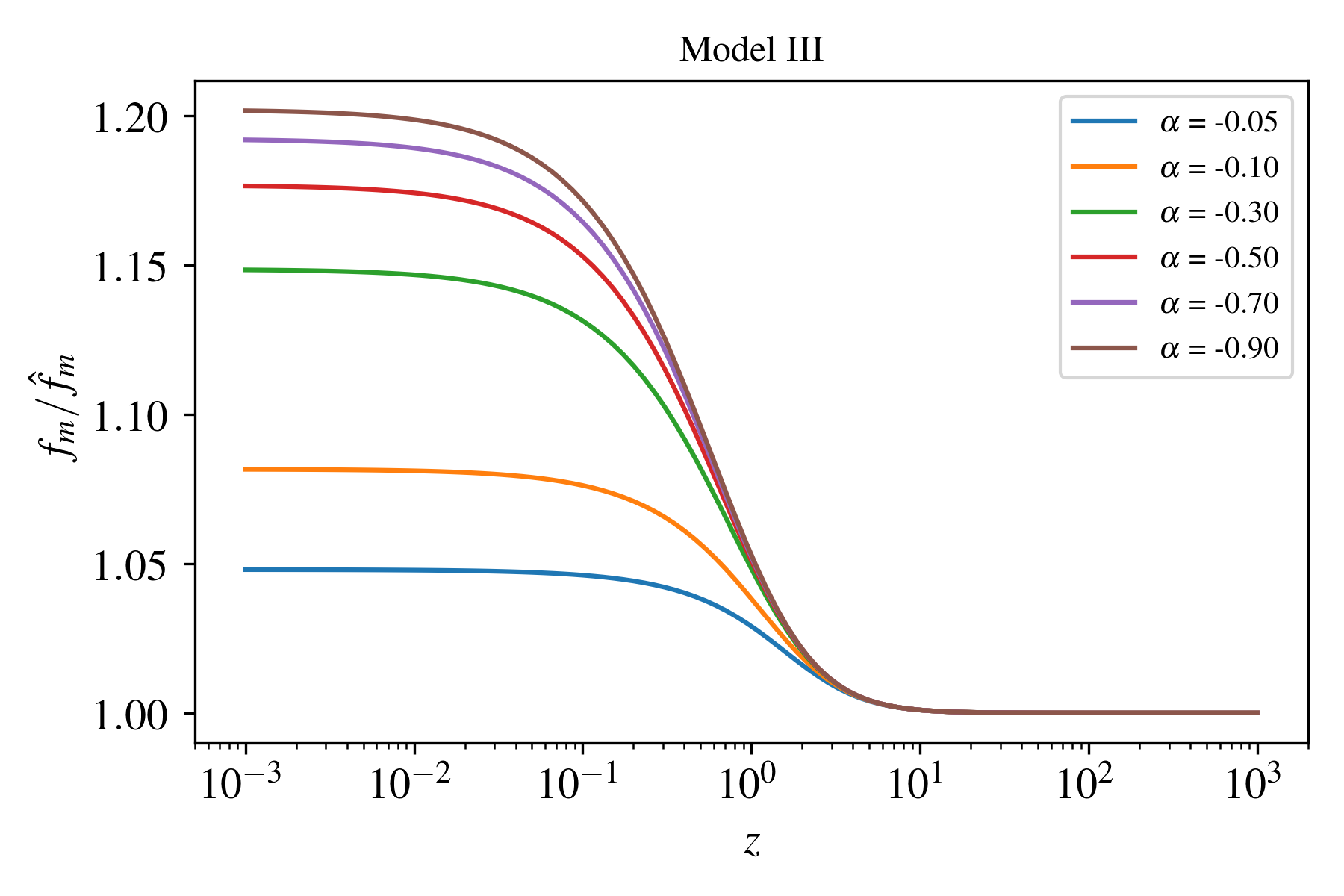}
		}
	\qquad
	\subfloat[\label{fig:Delta_f_m_m3}]{
		\includegraphics[width=0.4\textwidth]{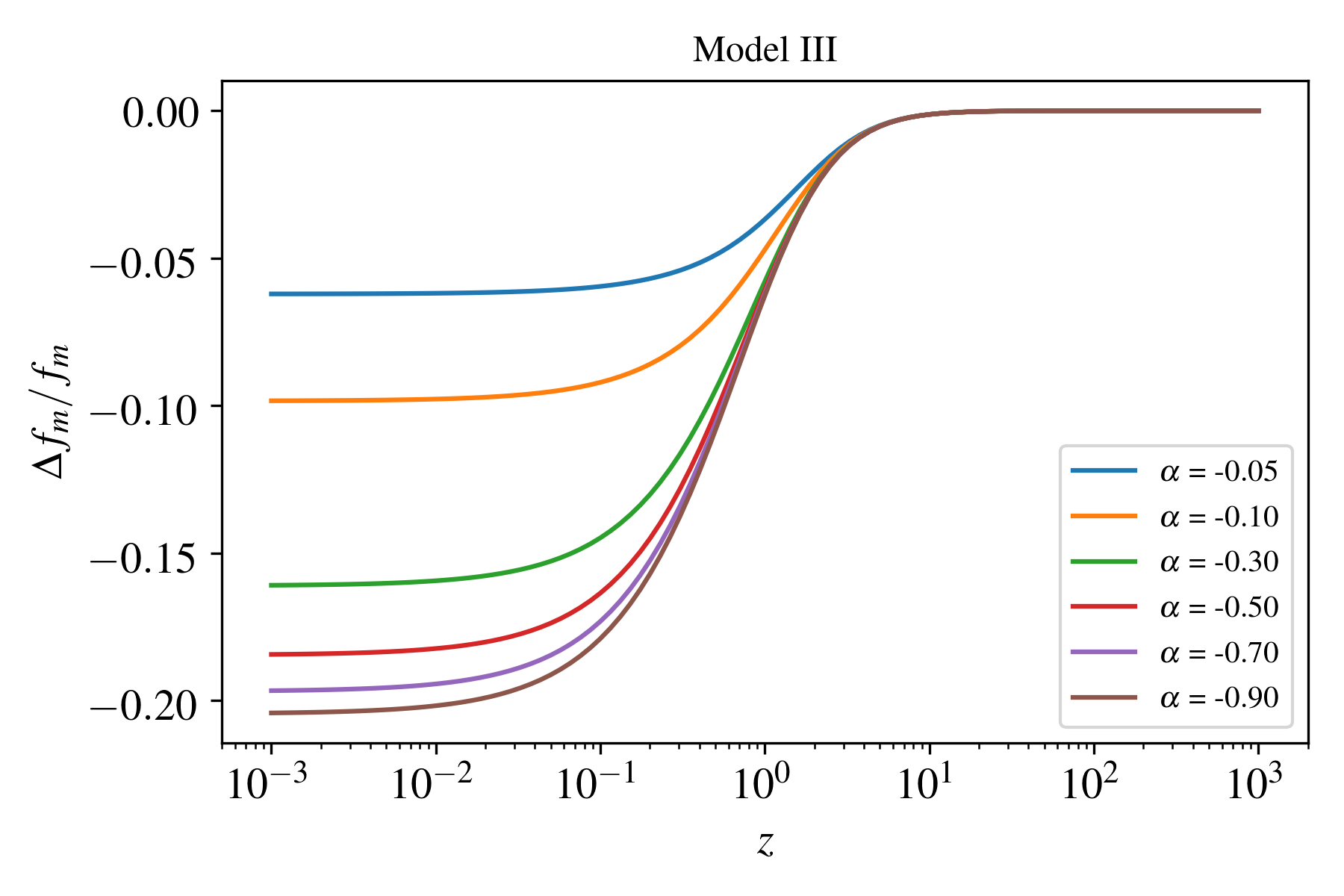}
		}
 	\caption{\label{fig:growth_function_m3}
 		Left panel (a): ratio $\fm/\hat{f}_{\rm m}$ where $\hat{f}_{\rm m}$ is the growth rate in the uncoupled, case and $\fm$ is the one with non-zero coupling. 
 		Right panel (b): difference between the effective and actual growth rates $\Delta \fm / \fm$, Eq.~\eqref{eq:f_eff_m}.
 		Both figures are shown as function of the redshift for the disformal model~III, Eq.~\eqref{eq:coupling_3}, for different values of the coupling constant.}
 		\end{figure}

Naturally, this choice of disformal factor also leads to more growth, since the Hubble friction is suppressed, while the gravitational constant term is enhanced (Fig.~\ref{fig:ratio_f_m_m3}). 
However, in contrast with the conformal models~I and II, increasing the value of $|\alpha|$ indefinitely does not lead to more growth.
Indeed, taking the limit $\alpha \to -\infty$, one finds
\begin{equation} 
	\E_1 \to -{
		\frac{ \phid \left( 3 H \phid + 2 V_\phi  \right)} 
		{2H \left( \phid^2 + \rhoc \right) }} \, , \qquad
	\E_2 \to - \frac{\phid^2}{\rhoc} \, ,
\end{equation}
which means that the modifications due to the coupling saturate.

The lack of modification at the background level implies that the deviation of $\fmeff$ from $\fm$ is due only to the term proportional to $\Delta f_{\rm c}$ in equation~\eqref{eq:f_eff_m}.
At late times, when $\phid^2 / (2 \rhoc) \sim 1$, it becomes $\Delta f_{\rm c} \approx 2 \alpha f_{\rm c} / (1 - 3\alpha)$.
In contrast to the conformally coupled models, $\Delta \fm$ for model~III is negative (Fig.~\ref{fig:Delta_f_m_m3}), being roughly proportional to $\alpha$ for small values of the coupling constant.
Therefore, the enhancement of the actual growth rate due to the coupling will be suppressed.
However, as in the case of the growth function, $|\Delta \fm|$ does not increase indefinitely when one consider larger and larger values of $|\alpha|$.

\begin{figure}[!ht]
	\subfloat[\label{fig:pk_qs_m3}]{
		\includegraphics[width=0.4\textwidth]{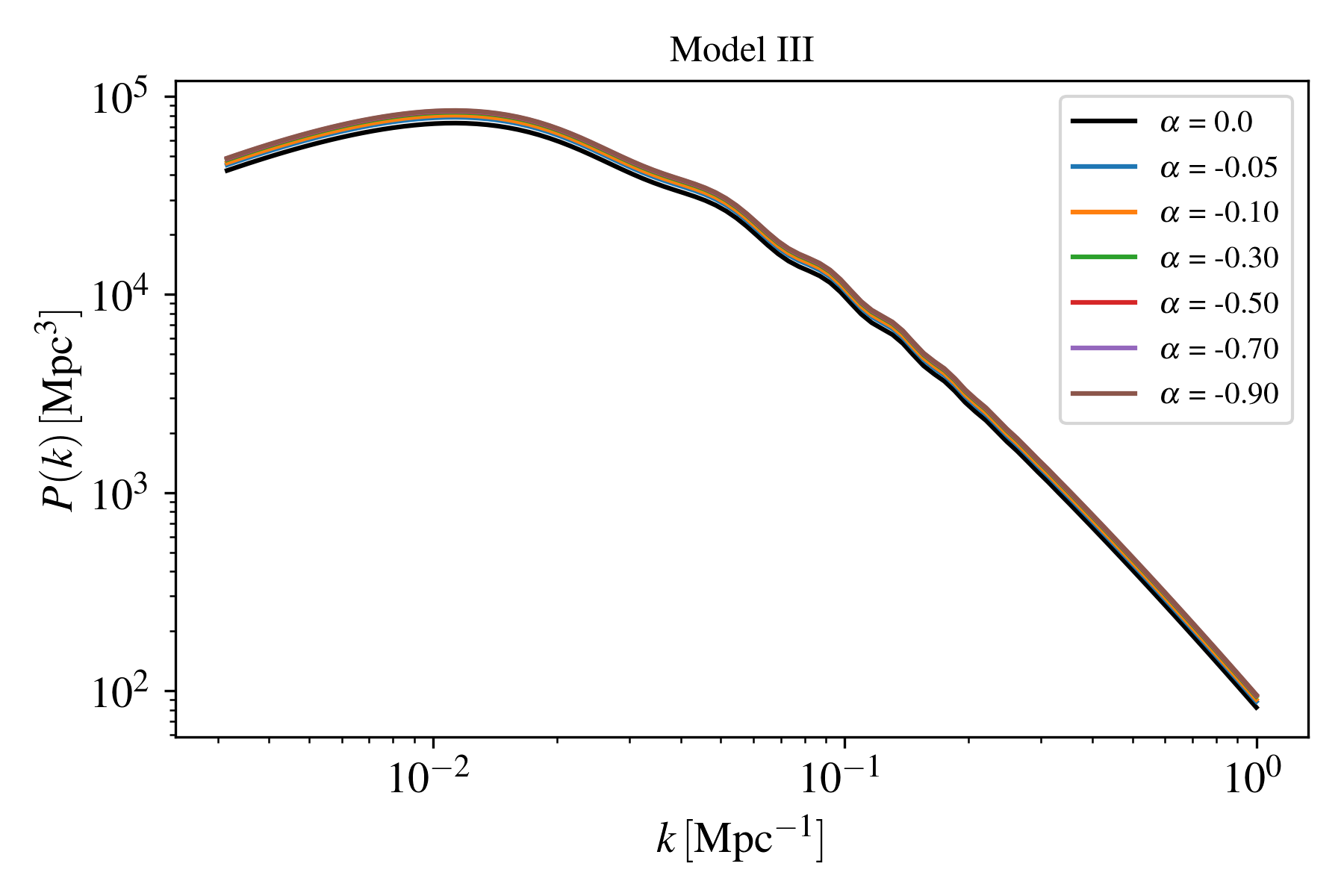}
		}
	\qquad
	\subfloat[\label{fig:ratio_pk_qs_m3}]{
		\includegraphics[width=0.4\textwidth]{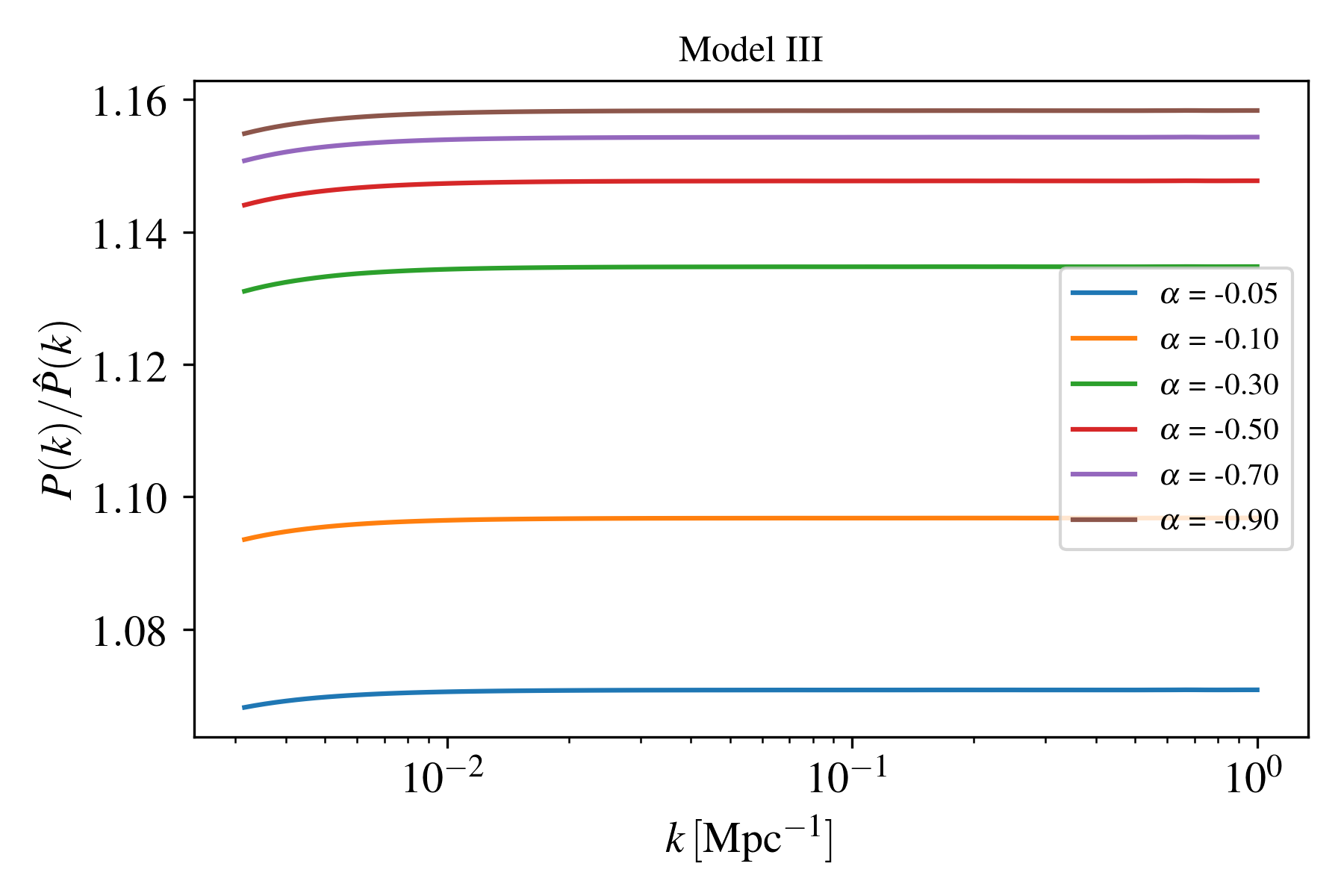}
		}
 	\caption{ \label{fig:power_spectrum_m3}
 		Left panel (a): linear matter power spectrum.
 		Right panel (b): ratio $P(k)/\hat{P}(k)$ where $\hat{P}(k)$ is the power spectrum in the uncoupled case, and $P(k)$ is the one with non-zero coupling. 
 		Both spectra are computed at the redshift $z = 0$ for the disformal model~III, Eq.~\eqref{eq:coupling_3}, for different values of the coupling constant.}
 		\end{figure}

Once again, because of the lack of modifications at the background level, the amplitude of the power spectrum simply increases, due to the coupling effect on the growth function, as shown in Fig. \ref{fig:power_spectrum_m3}.

\section{Forecasts \label{sec:forecasts}}


In this section, we introduce the approach followed to obtain the constraining power of future galaxy surveys on cosmological parameters.


\subsection{Method}


In order to forecast how well the future galaxy surveys will be able to constraint the DE/CDM coupling, we employ the Fisher matrix analysis to the galaxy power spectrum sliced over several redshift bins.
The technique consists of approximating the full likelihood for the cosmological parameters as a multivariate Gaussian distribution. 
After Taylor expanding the likelihood function $\mathcal{L}$, the Fisher matrix is defined by
\begin{equation}
	F_{\alpha \beta}= - \evaluated{ \expval{\pdv{\ln \mathcal{L}(\vb{\theta})}{\theta^\alpha}{\theta^\beta}}}_{\hat{\vb{\theta}}},
\end{equation}
where $\hat{\bm{\theta}}$ is the vector containing the maximum likelihood estimators. 
The Fisher matrix estimates the best possible precision obtainable by a given experiment.
Under the assumption that the parameters follow a Gaussian distribution, the Fisher matrix is the inverse of the covariance matrix.
Generally, that is the case near the peak of the likelihood, and the Fisher matrix provides a good approximation to the uncertainties.
The marginalized constraint on a given parameter $\theta^\alpha$ is simply $(F^{-1})_{\alpha \alpha}^{1/2}$, while the unmarginalized one, when all the other parameters are fixed, is given by $(F_{\alpha \alpha})^{-1/2}$. 
As long as the parameters are not extremely degenerated, the Fisher matrix method is an efficient tool in estimating the best possible constraints.

When applied to galaxy power spectrum observations, sliced over several redshift bins, the Fisher matrix takes the form~\cite{Tegmark:1997rp,Seo:2003pu},
\begin{equation} \label{eq:fm_ps}
	F_{\alpha \beta} = \sum_{z_i} \int_{k_{\rm min}}^{k_{\rm max}} \frac{\dd^3 {\bm k}}{(2 \pi)^3}
	V_{\rm eff}({\bm k},z_i) \pdv{\ln P_{\rm obs}({\bm k},z_i)}{\theta^\alpha}
	\pdv{\ln P_{\rm obs}({\bm k},z_i)}{\theta^\beta} \,,
\end{equation}
where the effective volume $V_{\rm eff}$ is defined as
\begin{equation}
	V_{\rm eff}({\bm k},z_i) = \left( \frac{n_{\rm g}(z_i) P_{\rm obs}({\bm k},z_i)}{n_{\rm g}(z_i) P_{\rm obs}({\bm k},z_i) + 1} \right)^2 V_{\rm survey}(z_i) \,,
\end{equation}
which comes from the fact that, since galaxies follow a Poissonian distribution, the observed power spectrum has a shot-noise correction $P_{\rm shot}=n_{\rm g}^{-1}$, with $n_{\rm g}(z_i)$ being the number density of galaxies in a redshift bin. 
When the sampling is good enough, $n_{\rm g}P_{\rm obs} \gg 1$ and both effective and actual survey volumes coincide.
Note that the total Fisher matrix is given by the summation over all redshift bins of the integrals over ${\bm k}$.

To compute the observed galaxy power spectrum, in addition to the modified Kaiser effect discussed in section~\eqref{sec:modified_kaiser_formula}, a few other effects must be taken into account, 
and the general expression reads~\cite{Seo:2003pu}
\begin{equation}
	P_{\rm obs}({\bm k},z) =  \mathcal{N}_{\rm AP}(z) \left[ b_{\rm g}(z) + \fmeff(z) \mu^2 \right]^2P_{\rm m}(k,z)
e^{-k^2 \mu^2 \sigma_{\rm NL}^{2}}.
\end{equation}
The Alcock-Packzynski factor~\cite{Alcock:1979mp}, defined as 
\begin{equation}
 \mathcal{N}_{\rm AP}(z) = \left( \frac{D_{A}^{\rm ref}(z)}{ D_{A}(z)}\right)^2 \frac{H(z)}{H^{\rm ref}(z) } ,
\end{equation}
where $D_A$ is the angular diameter distance and $H$ is the Hubble function, is a correction due to the fact that when converting angles and redshifts into distances and wavenumbers one needs to, incorrectly, assume a reference cosmology.
On the other hand, the exponential factor corresponds to errors in the observed redshift due to a line-of-sight smearing of the structure~\cite{White:2008jy}, which becomes relevant on small scales, and is characterized by the nuisance parameter $\sigma_{\rm NL}$. 
Finally, we can compute the derivative of the observed galaxy power spectrum with respect to the cosmological parameters, resulting in~\cite{White:2008jy}
\begin{align}
	\dv{\ln P_{\rm obs}}{\theta^{\alpha}} 
	&= \pdv{\ln P_{\rm m}}{\theta^{\alpha}}
	+ \frac{2}{b_{\rm g}+\fmeff \mu^2} \left( \pdv{ b_{\rm g} }{\theta^\alpha} + \mu^2 \pdv{\fmeff}{\theta^\alpha} \right)
	- k^2 \mu^2 \pdv{\sigma_{\rm NL}^2}{\theta^\alpha} \nonumber \\
	&+ \left[	1 + \frac{4 \fmeff \mu^2}{ b_{\rm g} + \fmeff \mu^2} (1 - \mu^2)
		+ \pdv{\ln P_{\rm m} }{\ln k} \mu^2 - 2k^2 \mu^2 \sigma_{\rm NL}^2 \right] \dv{\ln H}{\theta^\alpha} \nonumber \\
	&+ \left[	-2 + \frac{4 \fmeff \mu^2}{ b_{\rm g} + \fmeff \mu^2} (1 - \mu^2)
		- \pdv{\ln P_{\rm m} }{\ln k} (1-\mu^2) \right] \dv{\ln D_A}{\theta^\alpha}.
\end{align}

In this work, we perform the forecast for the set of cosmological parameters
\begin{equation} \label{eq:forecast_parameters}
	\bm{\theta} 
	= \{ \alpha, h, \Omega_{\rm b}h^2, \Omega_{\phi}, 10^9 A_s, n_s \}
	    + \{\sigma_{\rm NL}, b_{\rm g}(z_i)\} \,,
\end{equation}
where 
$\alpha$ is the coupling constant for each model considered here, 
$h$ is the dimensionless Hubble parameter, defined as $H_0 = 100h \, \kilo\meter\per\second\per\mega\parsec$, 
$\Omega_{\rm b}$ and $\Omega_{\phi}$ are, respectively, the baryon and scalar field current density parameters, 
$A_s$ is the amplitude of the primordial power spectrum at $k = 0.05 \mega\parsec^{-1}$, 
and $n_s$ is the spectral index of primordial perturbations.
The error in the observed redshift, $\sigma_{\rm NL}$, and the galaxy bias at each redshift bin, $b(z_i)$, are taken to be nuisance parameters and are marginalized over. 
Further, we consider a flat geometry.

For each model, the fiducial values for the coupling constant is chosen to be
\begin{itemize}
\item Conformal model I:   $\alpha = 0.05$.
\item Conformal model II:  $\alpha = 0.1$.
\item Disformal model III: $\alpha = -0.05$.
\end{itemize}
This choice for the conformal model~I is consistent with current constraints on coupled quintessence~\citep{Ade:2015rim}.
Even though we are interested in cosmological scenarios that differ from the standard one, we set the fiducial cosmology using current constraints on $\Lambda$CDM~\cite{Aghanim:2018eyx}. 
These values are shown in TABLE~\ref{tab:fid_params}.
We choose the value $\sigma_{\rm NL} = 7 \, \mega\parsec$, corresponding to a peculiar velocity dispersion of roughly $400 \kilo\meter\per\second$~\cite{Li:2007rpa, Bull:2014rha}.
Moreover, we assume that the slope of the scalar field potential $n$ is fixed, and set its value to  $n=0.5$, which is consistent with current constraints on quintessence~\citep{Ade:2015rim}. 

\begin{table}[h!]
\caption{\label{tab:fid_params}
	Fiducial cosmological parameters consistent with Planck 2018 results for $\Lambda$CDM~\cite{Aghanim:2018eyx}}
\begin{tabular}{l c c c c c}
\hline\hline
Parameter 	& $h$ & 	$\Omega_{\rm b}h^{2}$		&	$\Omega_\phi$		&	$10^9 A_s$	&	$n_s$ \\
\hline
Value			&	0.674	&	0.224								&	0.685					&	2.10				&	0.965 \\
\hline\hline
\end{tabular} 
\end{table}

Our goal is estimating the precision with which future galaxy redshift surveys will be able to measure the set of cosmological parameters~\eqref{eq:forecast_parameters}, especially the coupling constant $\alpha$.

In this work, we consider experiments with Euclid-like \cite{Amendola:2016saw} and SKA-like (Phase 2)~\cite{Yahya:2014yva}  configurations.
For the Euclid-like survey, we consider the range $0.7 < z < 2$, with a sky coverage of 15,000 squared degrees, and bias function $b_{\rm g} = \sqrt{1 + z}$,
while for SKA2 we take $ 0.1 < z < 2$, a sky coverage of 30,000 squared degrees, and bias function $b_{\rm g} = c_1 \exp(c_2 z)$, with constants $c_1$ and $c_2$ given in \cite{Yahya:2014yva}.
We adopt the predicted number density of galaxies as a function of redshift given in \cite{Amendola:2016saw} and \cite{Yahya:2014yva}.
We divide the survey volume into redshift bins, with $\Delta z = 0.1$, and consider the expect galaxy number density $n_{\rm g}(z_i)$  and galaxy bias $b_{\rm g}(z_i)$ at each redshift slice $z_i$.
 
At each redshift bin, the limits of integration in Eq.~\eqref{eq:fm_ps} are taken to be
\begin{equation}
	k_{\rm min}(z_i) = \left( \frac{2 \pi}{V_{\rm survey}(z_i)}	 \right)^{1/3}, \qquad
	k_{\rm max}(z_i) = 0.2h (1 + z_i)^{2/(2 + n_s)} \,,
\end{equation}
for both surveys. 
The maximum wavenumber $k_{\rm max}$ corresponds to a conservative cut-off of non-linear scales~\cite{Smith:2002dz,Sprenger:2018tdb}. 
 
The derivative with respect to a given cosmological parameter $\theta^\alpha$ in the Fisher matrix are performed numerically, with a symmetric difference quotient, evaluating the observed power spectrum at $\theta^\alpha(1 \pm 10^{-2})$.


\subsection{Forecast results}

\begin{figure}[!ht]
	\subfloat{
		\includegraphics[width=0.3\textwidth]{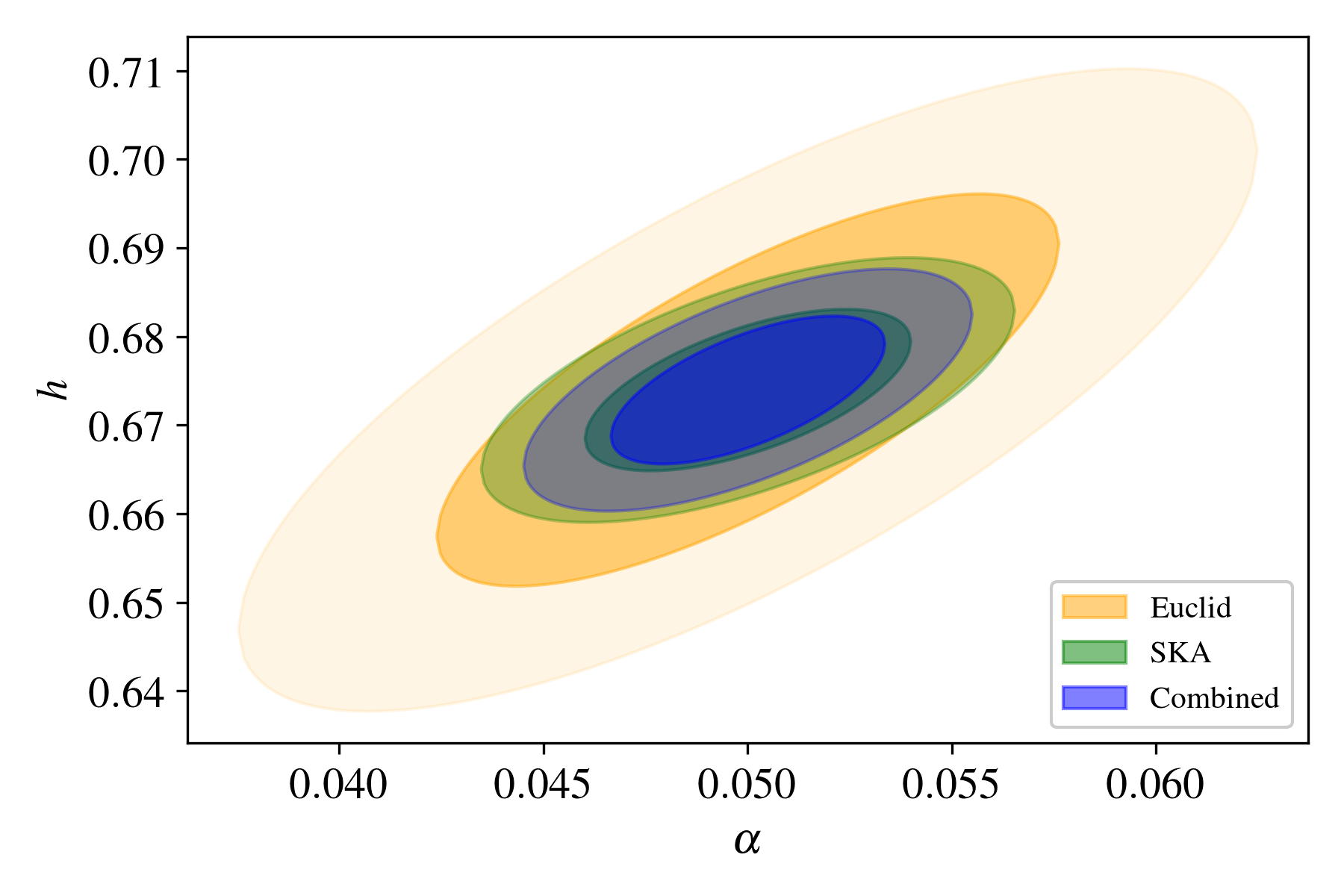}
		}
	\quad
	\subfloat{
		\includegraphics[width=0.3\textwidth]{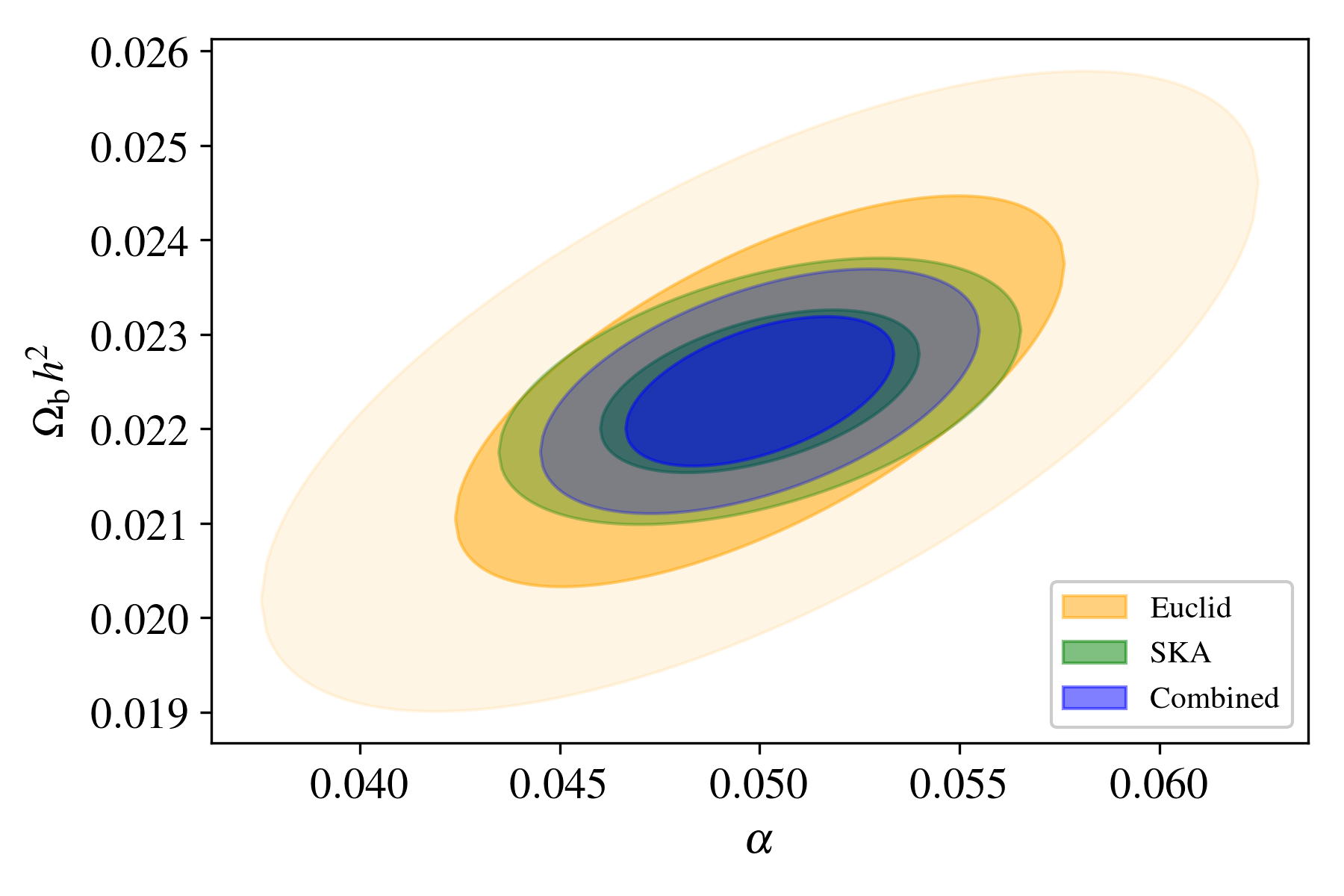}
		}
	\quad
	\subfloat{
		\includegraphics[width=0.3\textwidth]{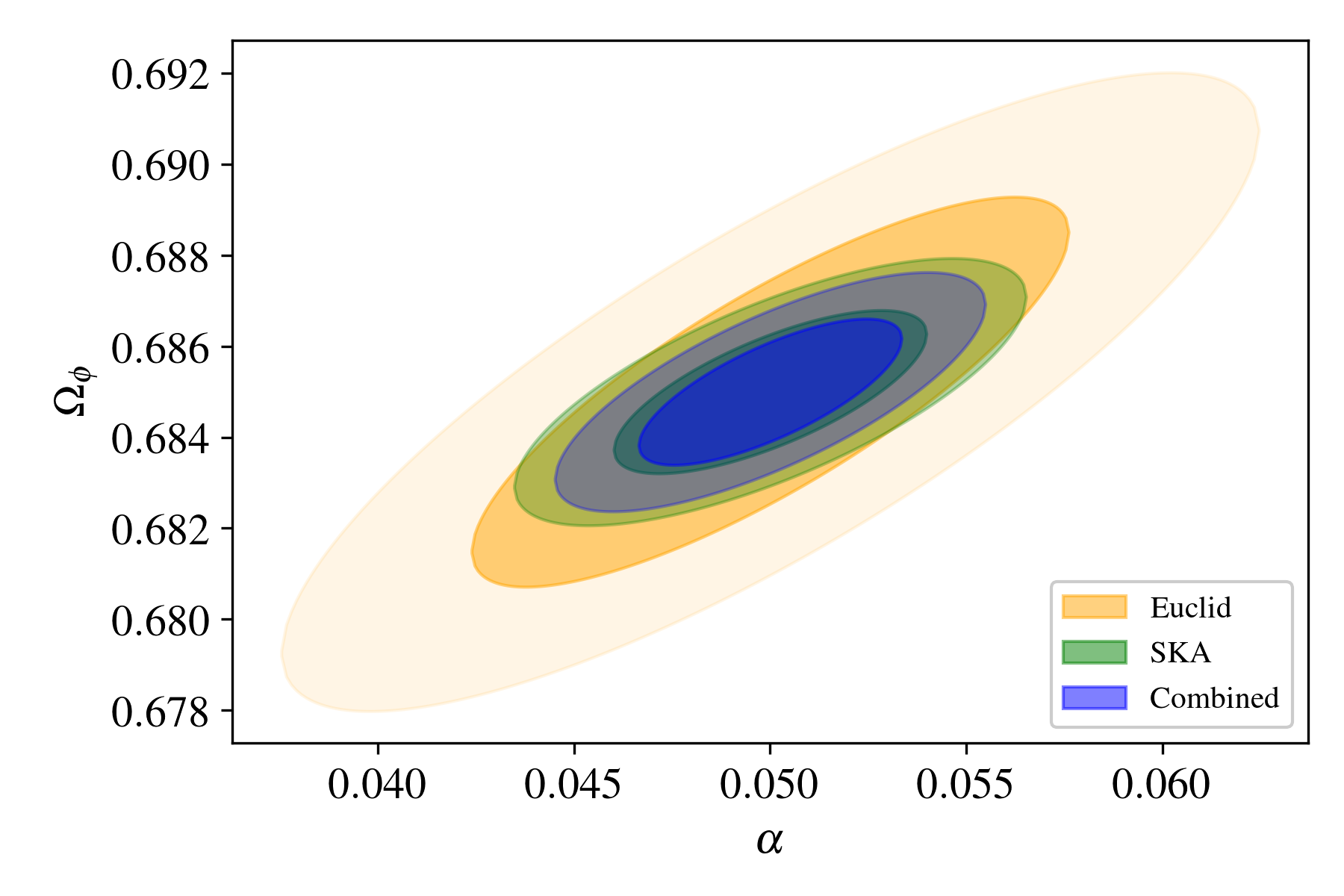}
		}
	\quad
	\subfloat{
		\includegraphics[width=0.3\textwidth]{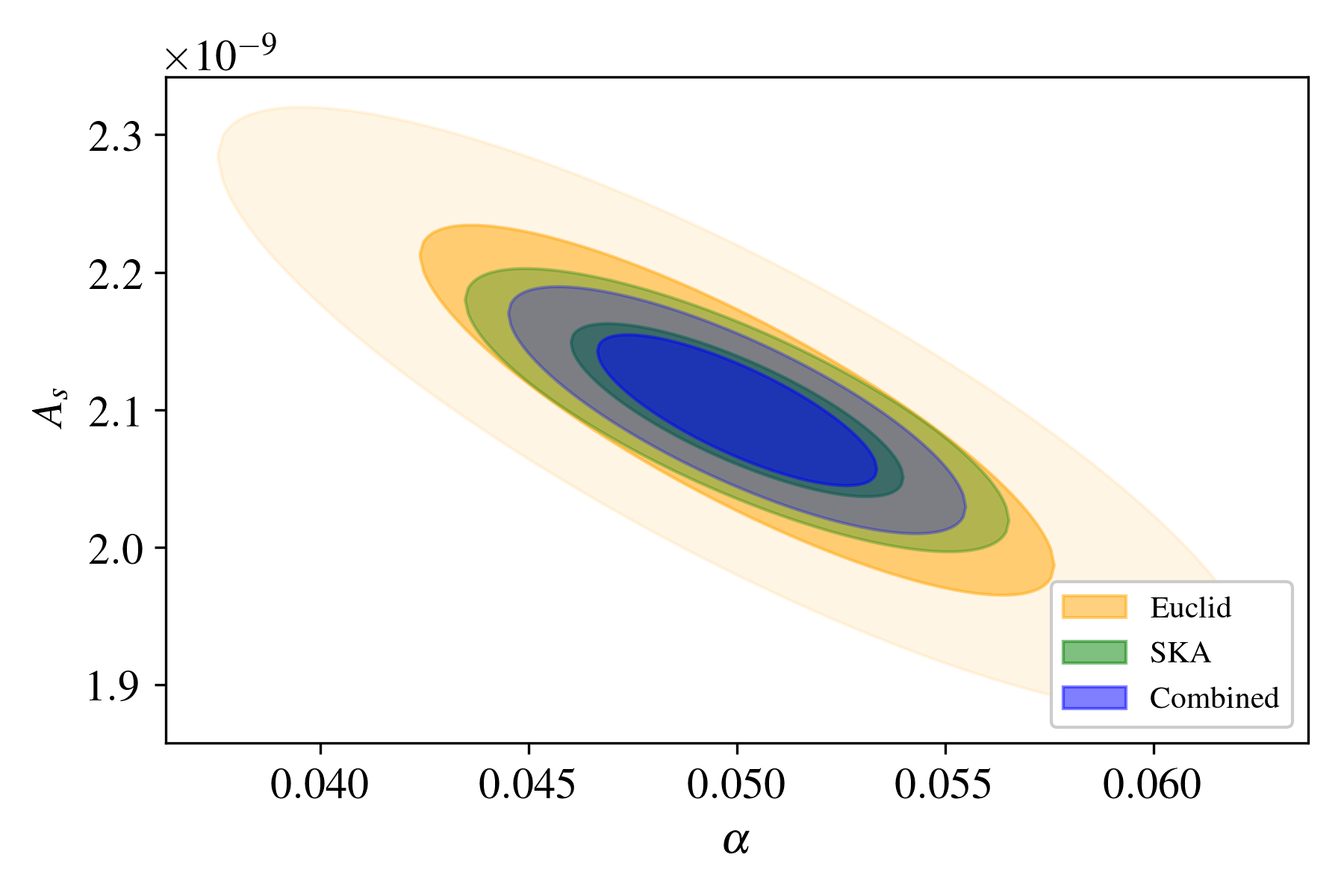}
		}
	\quad
	\subfloat{
		\includegraphics[width=0.3\textwidth]{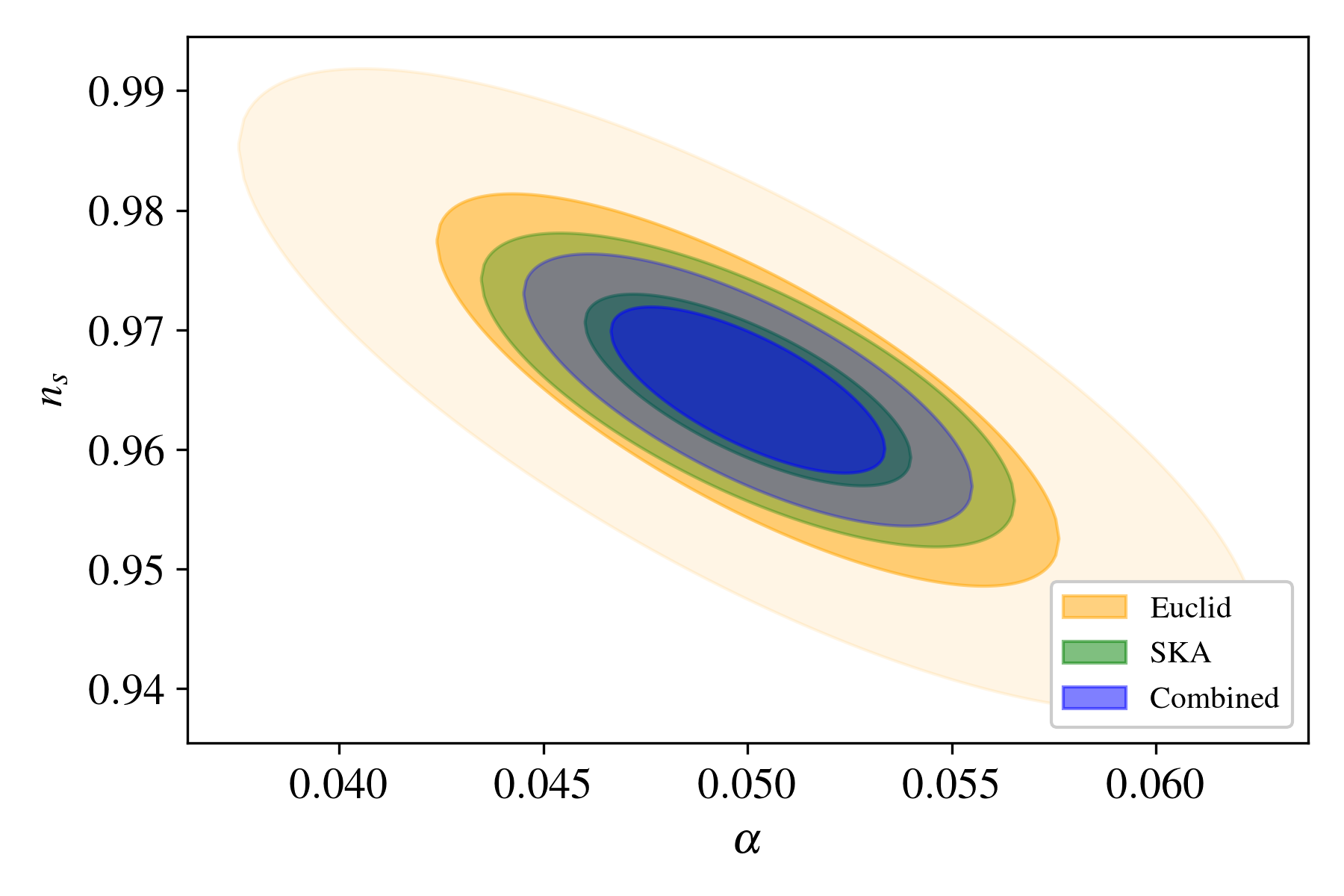}
		}
	\caption{\label{fig:m1_fm_3x}
	Confidence regions, $1\sigma$ and $2\sigma$, for selected parameters vs $\alpha$ for the conformal model~I. 
	We present the independent SKA and Euclid results, and also the combined one.
	The contours are obtained marginalizing over all the parameters except the ones being plotted.}
	\end{figure}
\begin{figure}[!ht]
	\subfloat{
		\includegraphics[width=0.3\textwidth]{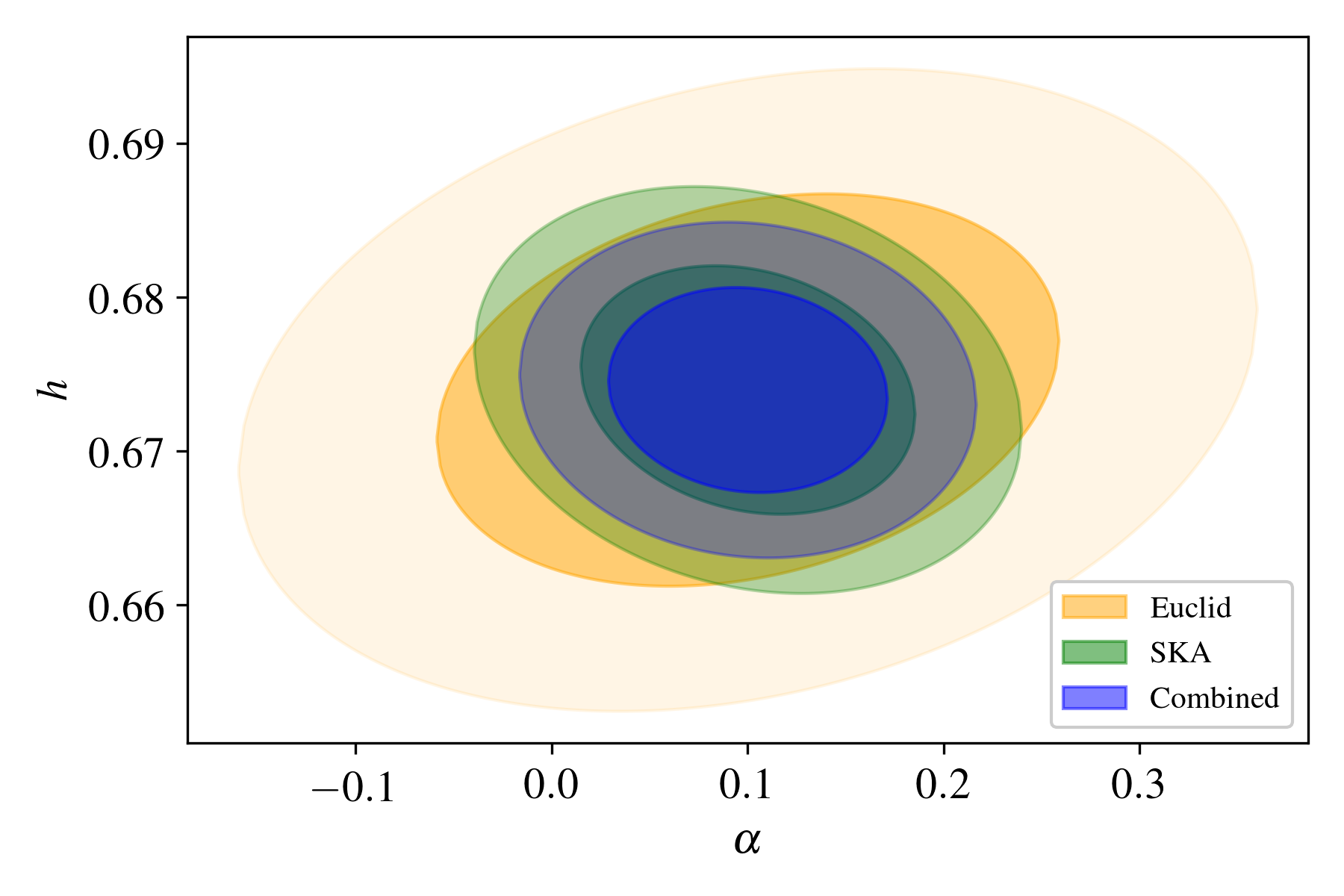}
		}
	\quad
	\subfloat{
		\includegraphics[width=0.3\textwidth]{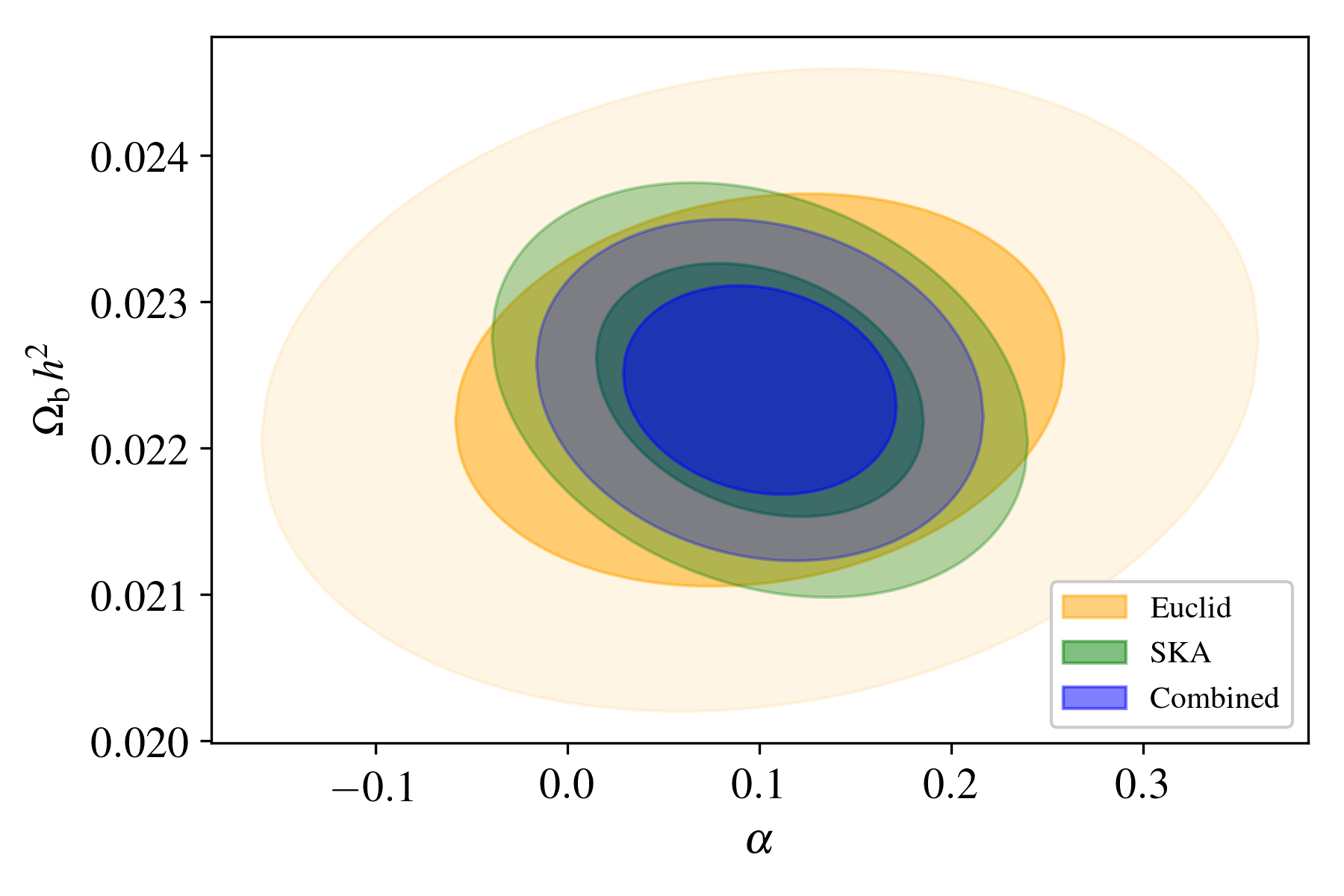}
		}
	\quad
	\subfloat{
		\includegraphics[width=0.3\textwidth]{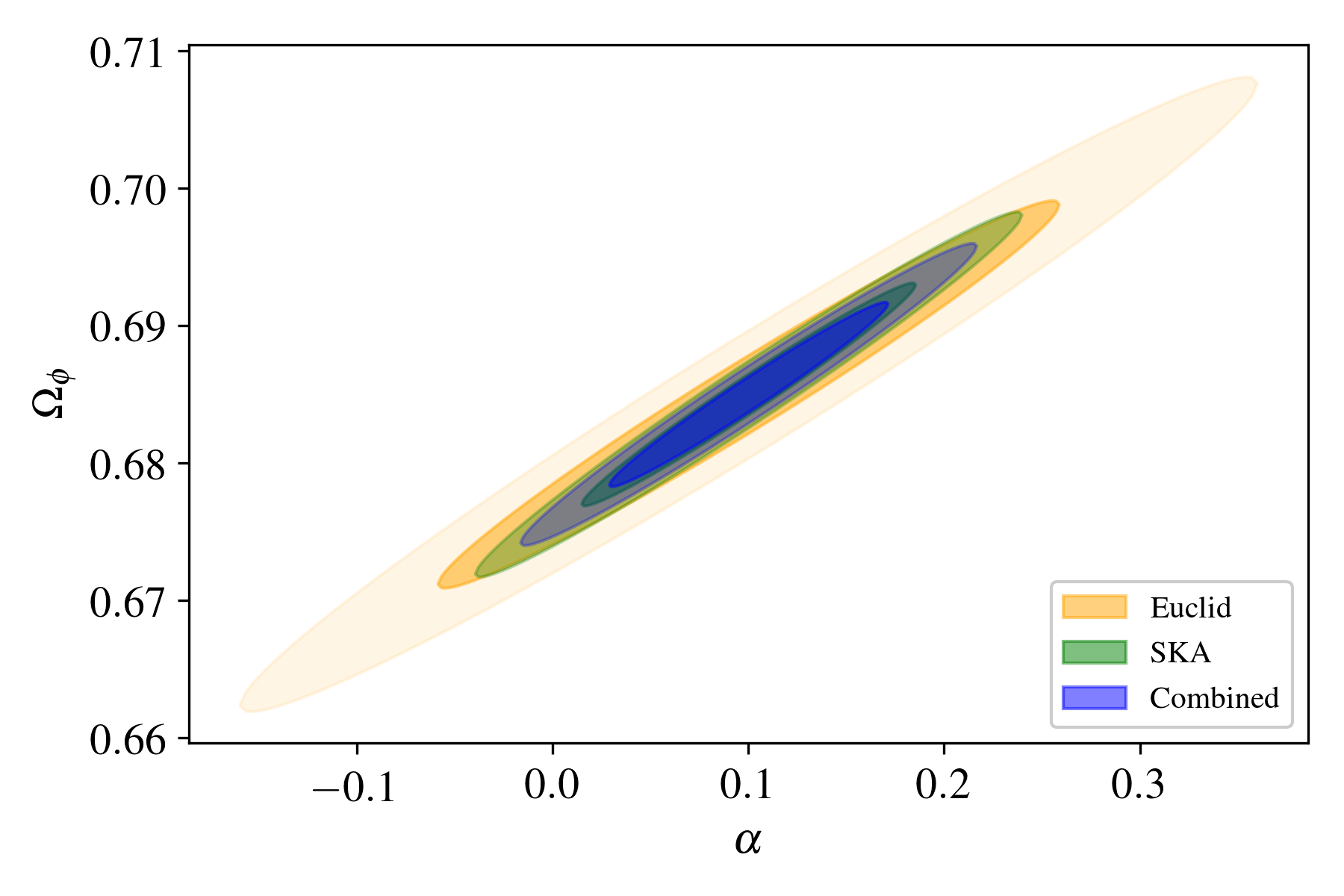}
		}
	\quad
	\subfloat{
		\includegraphics[width=0.3\textwidth]{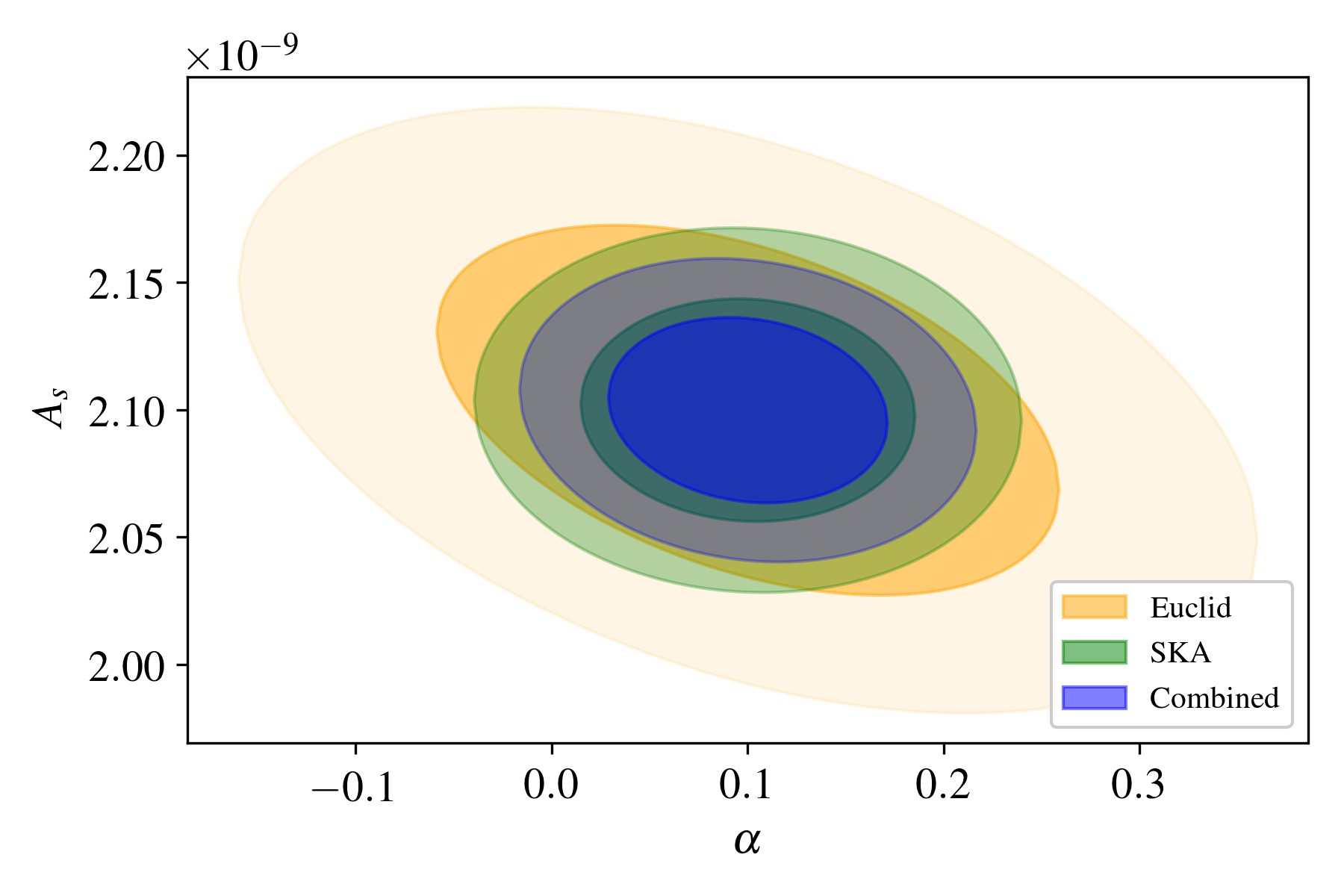}
		}
	\quad
	\subfloat{
		\includegraphics[width=0.3\textwidth]{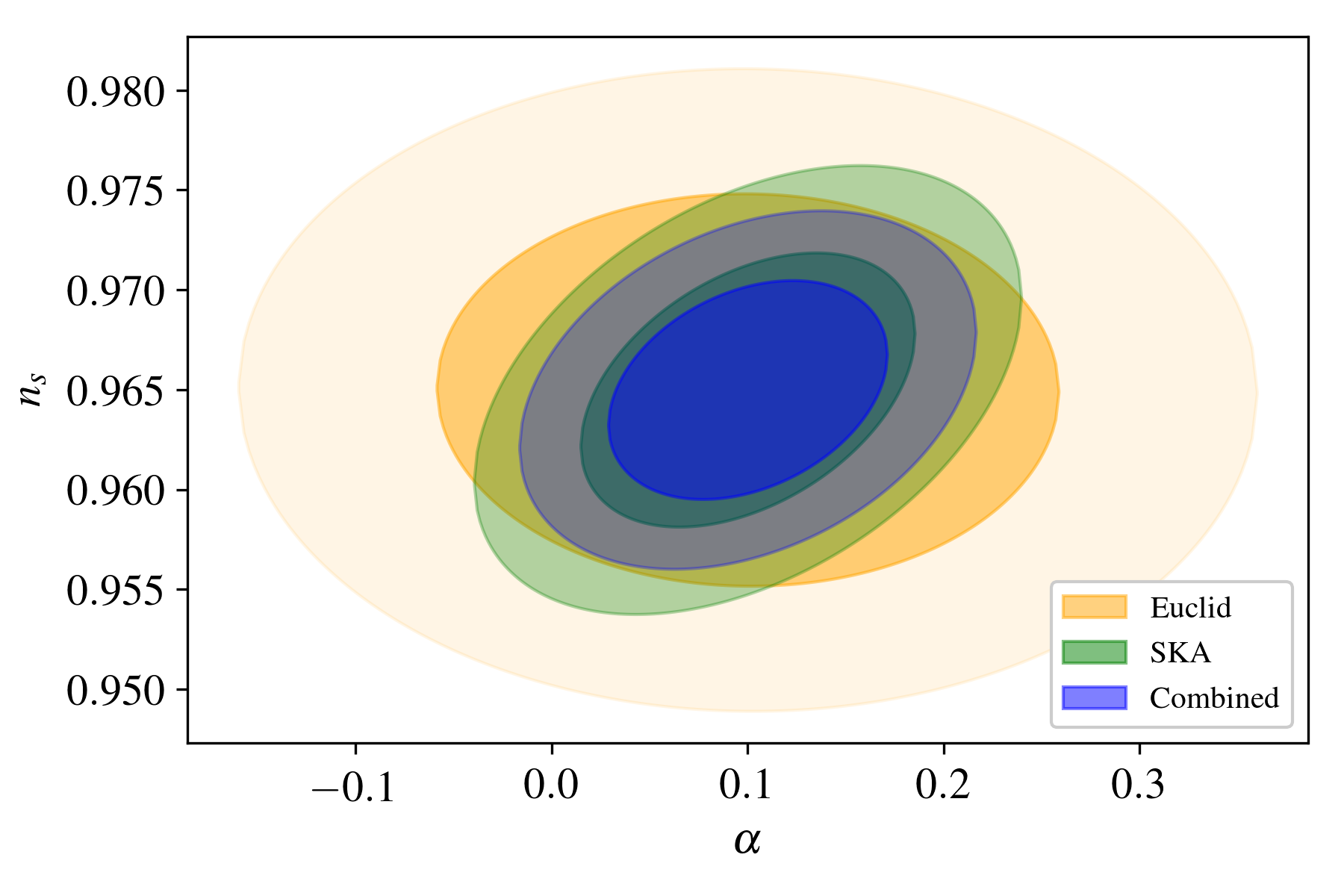}
		}
	\caption{\label{fig:m2_fm_3x_v2}
	Confidence regions, $1\sigma$ and $2\sigma$, for selected parameters vs $\alpha$ for the conformal model~II
	assuming a fiducial coupling constant $\alpha = 0.1$.
	We present the independent SKA and Euclid results, and also the combined one.
	The contours are obtained marginalizing over all the parameters except the ones being plotted.}
\end{figure}
\begin{figure}[!ht]
	\subfloat{
		\includegraphics[width=0.3\textwidth]{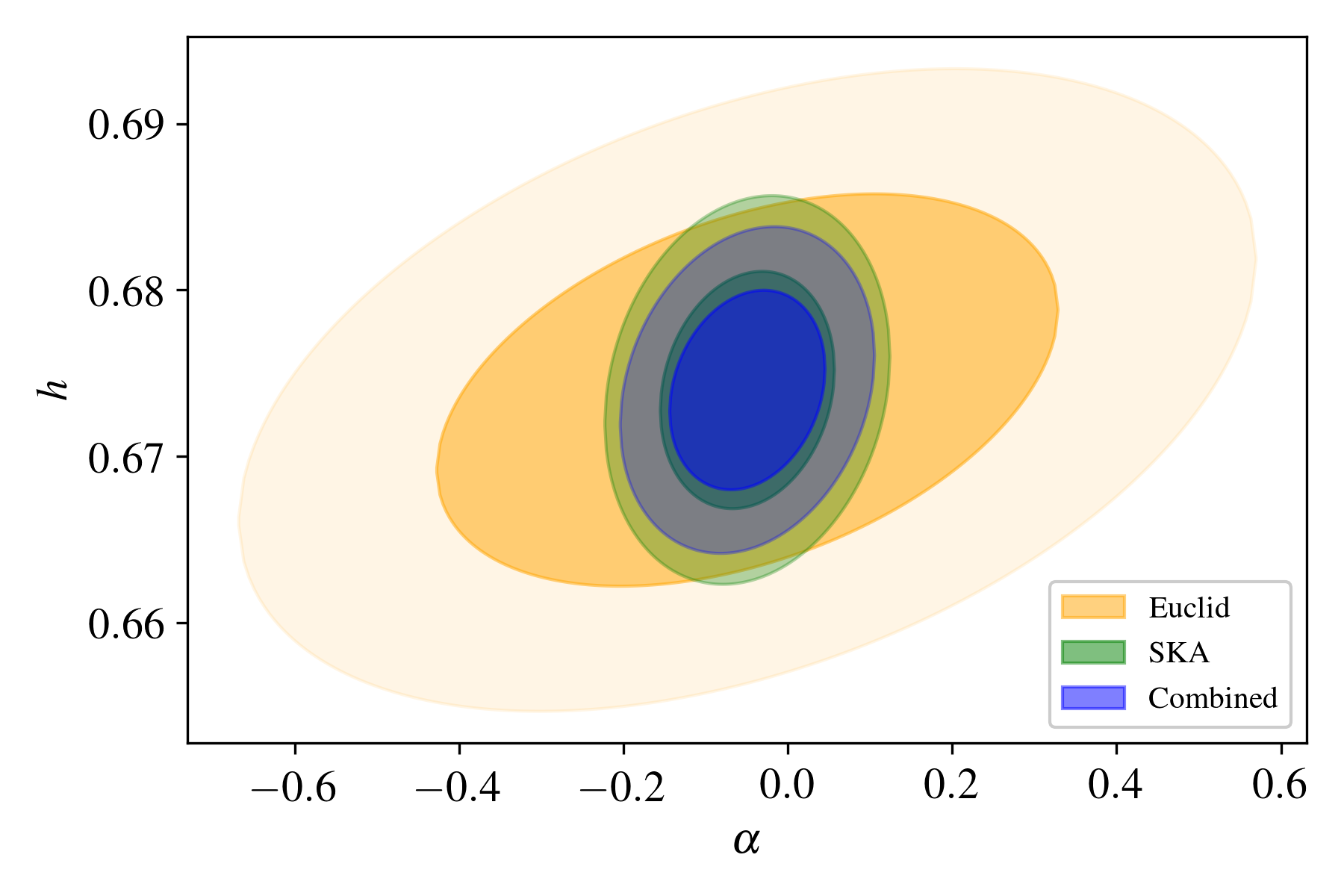}
		}
	\quad
	\subfloat{
		\includegraphics[width=0.3\textwidth]{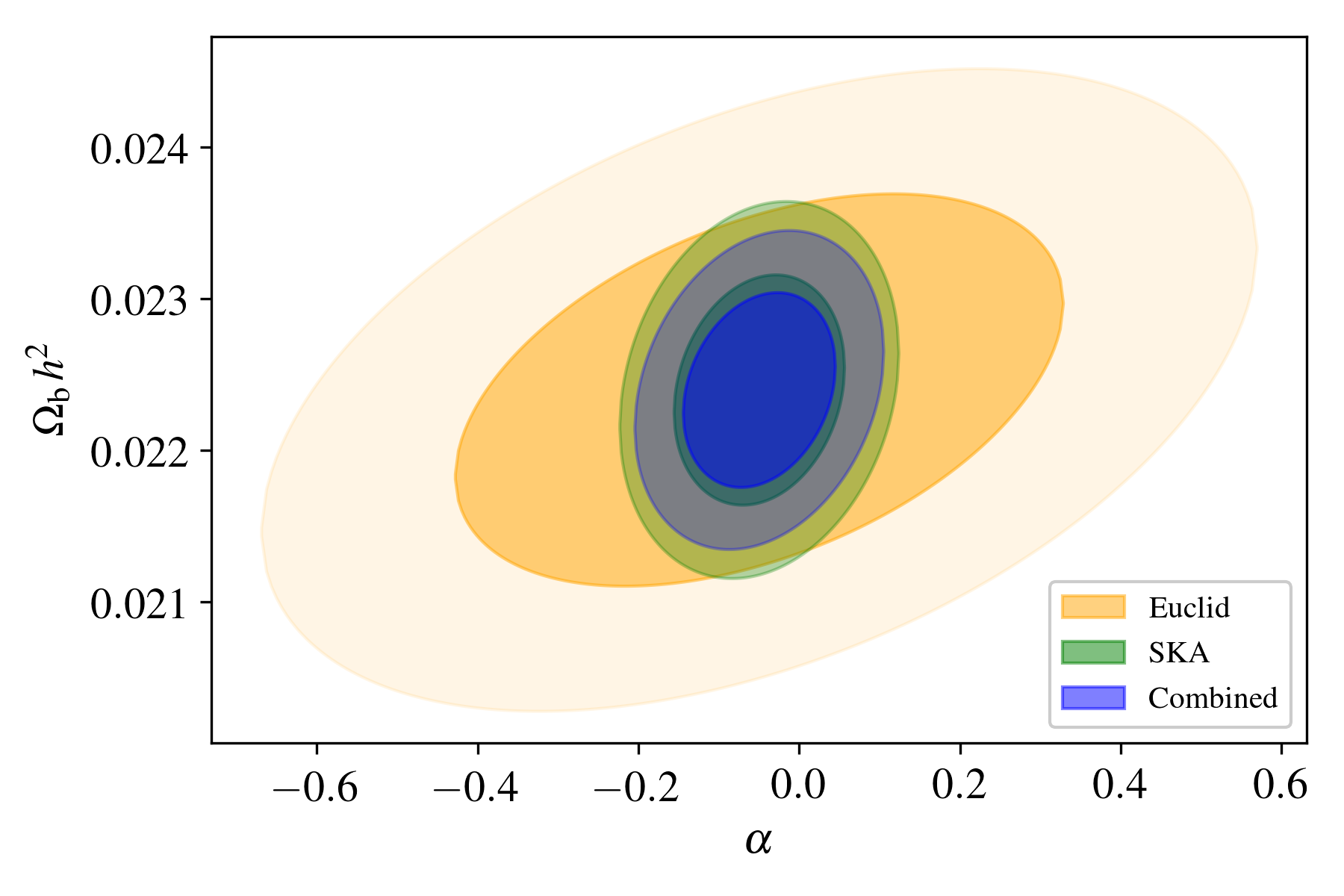}
		}
	\quad
	\subfloat{
		\includegraphics[width=0.3\textwidth]{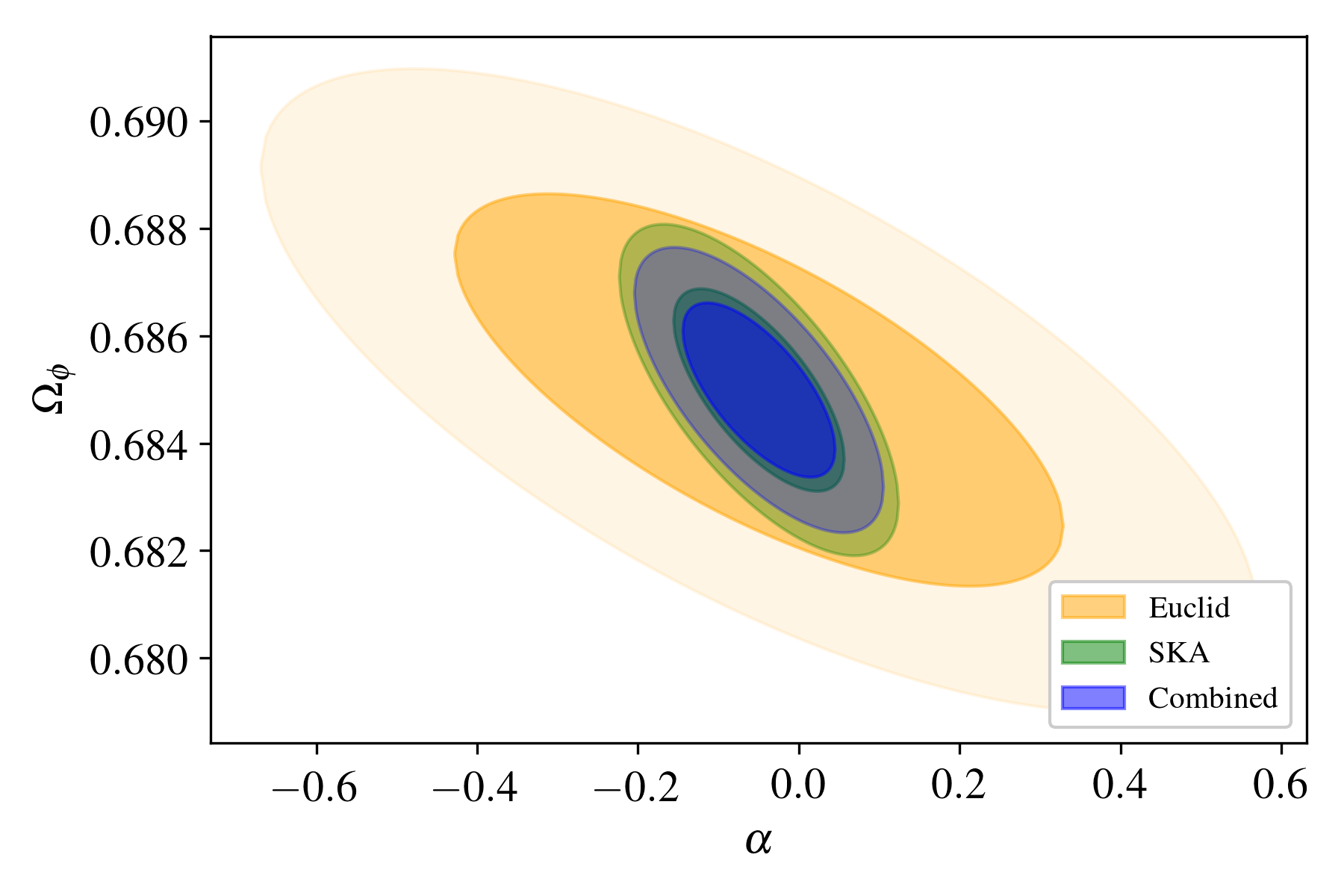}
		}
	\quad
	\subfloat{
		\includegraphics[width=0.3\textwidth]{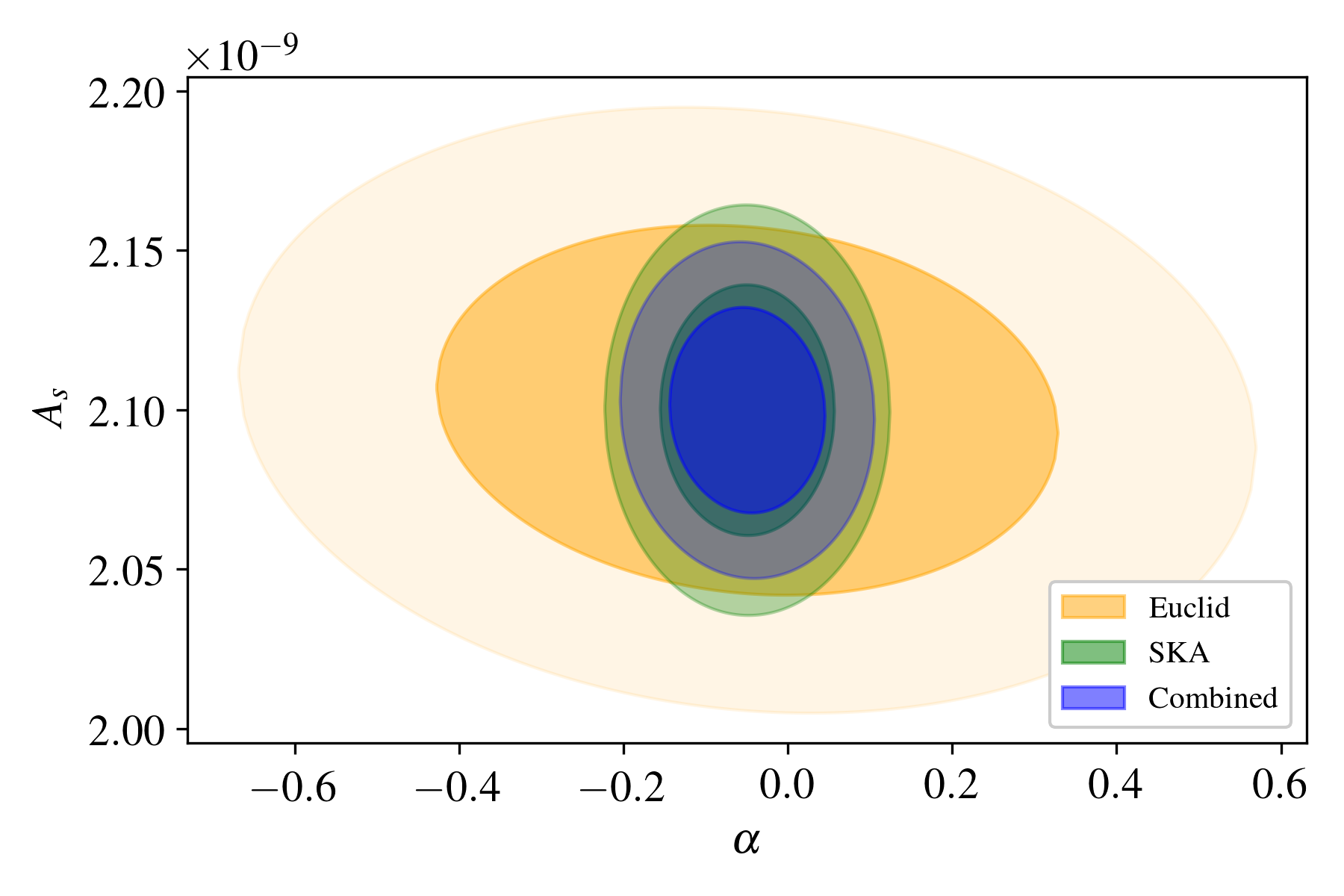}
		}
	\quad
	\subfloat{
		\includegraphics[width=0.3\textwidth]{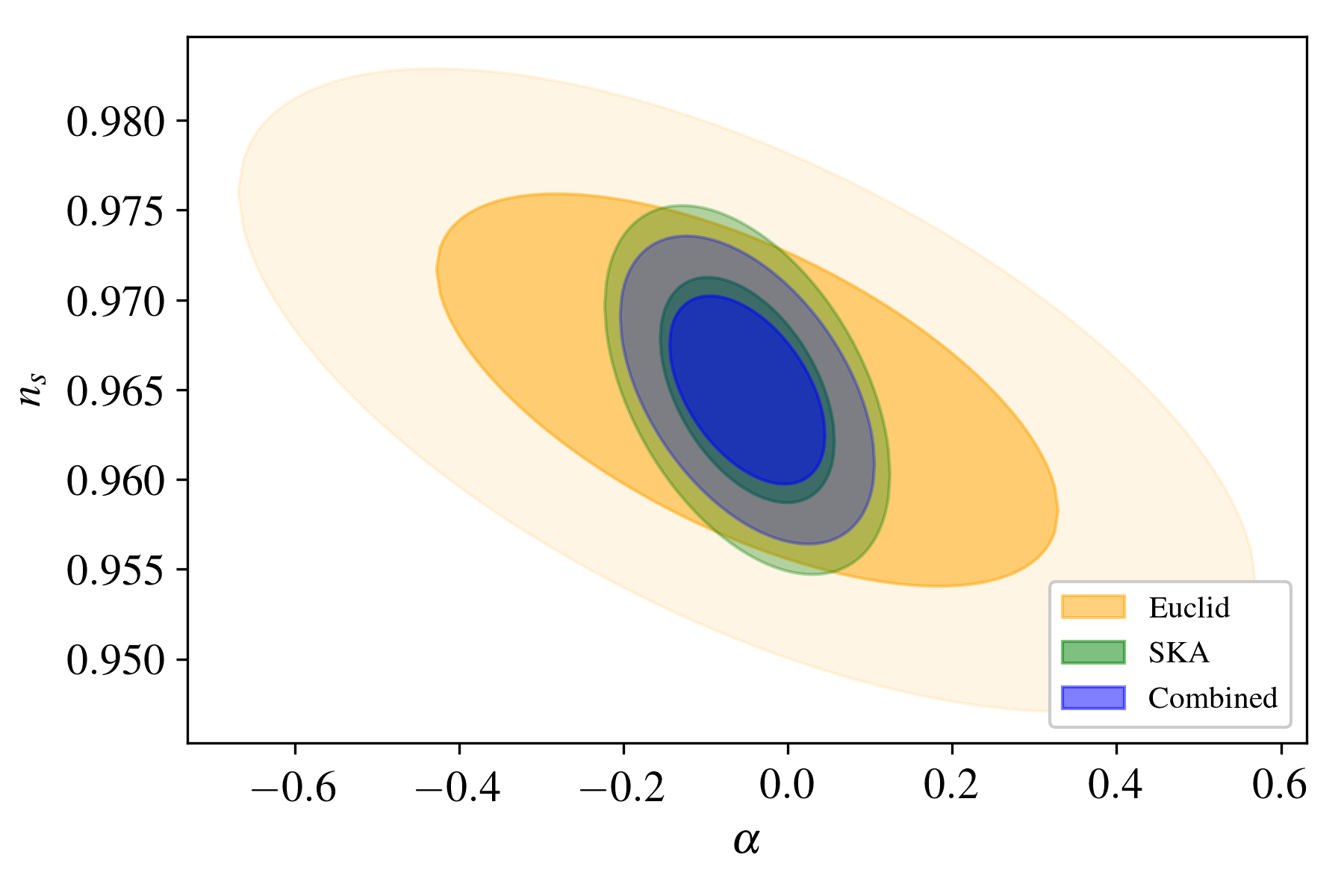}
		}
	\caption{\label{fig:m3_fm_3x}
		Confidence regions, $1\sigma$ and $2\sigma$, for selected parameters vs $\alpha$ for the disformal model~III. 
		We present the independent SKA and Euclid results, and also the combined one.
		The contours are obtained marginalizing over all the parameters except the ones being plotted.}
\end{figure}

On TABLE~\ref{tab:results_FM} we present the marginalized 1$\sigma$ C.L. errors on the fiducial parameters for all three models, while the two-dimensional contour plots for models~I, II and~III are presented in Figs.~\ref{fig:m1_fm_3x},~\ref{fig:m2_fm_3x_v2} and~\ref{fig:m3_fm_3x}, respectively.
Note that, even though we did not include any prior to the value of $\alpha$ for the disformal model~III, it must be kept in mind that the condition $\alpha < \phid^2 / (2\rhoc)$, discussed in section~\ref{sec:ex_pert_m3}, must be satisfied. 
For each fiducial model, the forecasts for SKA and Euclid are combined, marginalizing over the biases and summing the resulting Fisher matrices.

\begin{table}[ht!]
\caption{\label{tab:results_FM} 
	Results for the Fisher analysis}
\begin{ruledtabular}
\begin{tabular}{clcccccc}
\multicolumn{1}{l}{Model} & Survey & \multicolumn{1}{l}{$10^2 \times \sigma (h)$} & \multicolumn{1}{l}{$10^3 \times \sigma (\Omega_{\rm b} h^2)$} & \multicolumn{1}{l}{$10^3 \times \sigma (\Omega_\phi)$} & \multicolumn{1}{l}{$10^2 \times \sigma (\alpha)$} & \multicolumn{1}{l}{$10^2 \times \sigma (10^9 A_s)$} & \multicolumn{1}{l}{$10^3 \times \sigma (n_s)$} \\
\hline
\multirow{3}{*}{I}       
& Euclid 	&	1.46		& 	1.36		& 2.38	& 	0.50		& 	8.86		&	1.08	\\
& SKA2   	&	0.60  	& 	0.57		& 1.18	&	0.26		&	4.14		&	0.53	\\
& Combined	&	0.55  	& 	0.52		& 1.06	& 	0.22		& 	3.61		&	0.46	\\
\hline
\multirow{3}{*}{II}       
& Euclid		&	0.84		&	0.88		& 9.30 	& 	10.46	&	4.79 	& 	0.65\\
& SKA2		&	0.53		&	0.57		& 5.35	& 	5.62		&	2.88		&	0.45	\\
& Combined	&	0.44		&	0.47		& 4.43	& 	4.69		&	2.40		&	0.36	\\
\hline
\multirow{3}{*}{III}      
& Euclid		&	0.78		&	0.85	&	2.41		&	24.93 	&	3.83		&	0.72	\\
& SKA2  		&	0.47		&	0.50	&	1.24		&	6.99		&	2.59		&	0.41	\\
& Combined 	& 	0.40		&	0.42	& 	1.07		&	6.23		& 	2.13		&	0.35		
\end{tabular}
\end{ruledtabular}
\end{table}

Note that the forecasted constraints from the SKA-like survey are more stringent than those from Euclid-like one. 
The main driver of this result is the fact that the redshift window of SKA2 is larger than Euclid's, the former being able to measure galaxies at lower redshifts, and we assumed the same spacing for the redshift bins. 
Hence, the predicted errors measured by SKA are expected to be smaller.

Generally speaking, the effects of the non-minimal coupling on the observed power spectrum occur on the redshift-distortion factor, growth function, and the power spectrum (baryonic wiggle's position and shape). 
In the previous sections, it was shown that in model~I the interaction changes the evolution of both background and linear perturbations at all epochs, while in model~II the modification arises only at late times. 
On the other hand, in model~III, the coupling does not affect the background evolution but changes the behavior of linear perturbations at late times.

Since, at the background level, the main effect of the interaction is the energy transfer from CDM to the DE scalar field, the coupling causes a degeneracy between $\alpha$ and $\Omega_\phi$. 
Despite the degeneracy, both surveys would be able to constraint $\alpha$ and $\Omega_\phi$ for model~I better than for models~II and~III. 
The reason is that in model~II, the coupling affects the evolution of background and perturbed quantities only at late times, effectively being less important than in model~I. 
In model III the coupling only changes the growth function at late times, and also, since for a given $\alpha$ the negative $\Delta \fm$ compensates the growth, the redshift-distortion factor $\beta$ is not as sensitive to the non-minimal coupling as the conformal models. 
However, the precision with which all the other parameters could be measured by both surveys is higher for models~II and~III, since in model~I the modification at higher redshift leads to indetermination on these parameters.

The constraints on all parameters except $\alpha$ and $\Omega_\phi$ from SKA and Euclid are roughly of the same order of magnitude for all fiducial cosmologies considered here.
In particular, for models~II and~III, the constraints from SKA and Euclid are of the same order of magnitude for all parameters except the coupling constant.
The constraints from Euclid are much less stringent, since the modifications due to the coupling take place at low redshifts, outside its observational window.
Note also that the error on $\Omega_\phi$ for model~III is comparable to the result for model~I - even though the impact of the interaction is more important in the latter, the degeneracy between $\Omega_\phi$ and $\alpha$ hinders the determination of $\Omega_\phi$.

The combined analysis helps in reducing the degeneracy and gives slightly better constraints.
For all the fiducial cosmologies, the two-dimensional contour plots for $(\alpha, h)$ and $(\alpha, \Omega_{\rm b} h^2)$ have roughly the same shape. 
For models~II and~III, since the modifications due to the non-minimal coupling are delayed in comparison to model~I, these parameters show little to no correlation.

From the Fisher matrix analysis, we see that experiments with specifications similar to both surveys could determine if there is a non-minimal coupling in the dark sector for the conformal model~I with our choice of fiducial cosmology.
The coupling constant could be determined to the percentage level, significantly improving the current constraints \cite{Ade:2015rim}.
For model~II, the interaction could be probed to the $1\sigma$ level.
On the other hand, for model~III, the forecasted surveys do not have enough precision to determine the existence of the interaction, rendering these models observationally equivalent to the standard quintessence scenario. 
However, the combination of other cosmological probes, such as weak lensing, could help in breaking the degeneracy.


\section{Conclusion \label{sec:conclusion}}


In this paper, we extended the treatment of Einstein's gravity with a non-minimally coupled DM component recently presented in~\cite{Kimura:2017fnq}, generalizing it from the canonical scalar field Lagrangian to the general K-essence one.
In our setup, the coupling between scalar field and DM is designed through the effective metric with conformal and disformal factors, whereas, to evade solar system constraints, baryon and radiation couple minimally to metric.
Effectively, such dependence leads to an energy transfer between DM and DE.

At the homogeneous level, the interaction can be understood as if the mass of the DM particle depends on the value of the scalar field. 
As for linear perturbations, the interaction mediated by a scalar field through disformal and conformal transformations changes both the perturbed continuity and Euler equations for CDM, while baryons and radiation follow the standard equations.
We showed that the modification is important even at sub-horizon scales (in the quasi-static limit) and the overall effect can be understood in terms of an effective Hubble friction and effective gravitational coupling. 
As a result, the growth of perturbation differs from the standard uncoupled case, and the difference can be, in principle, observationally tested.

In the presence of an interaction, since the modified Kaiser formula provides information only about the effective linear growth rate, determined using the galaxy peculiar velocity field, single-redshift RSD measurements are not able to measure the actual growth. 
However, multiple-redshift RSD measurements, or a combination of different types of observations, can help in breaking the degeneracy.
Therefore, when considering extensions of the $\Lambda$CDM paradigm, one must carefully take such effects into account.
These statements are true not only for our setup but also for any cosmological model including DM interacting with DE, as long as the continuity equation for the total matter is modified.

In this paper, by considering three concrete cosmological models (two with purely conformal couplings and one with a disformal coupling), we studied the cosmological evolution of both background quantities and linear perturbations. 
The well-known coupled quintessence model, here denoted by conformal model~I, does not admit a tracker solution, and the presence of the coupling changes the evolution of matter linear density even at early times due to the modification of the effective gravitational coupling.
Then, we proposed an improved model, denoted by conformal model~II, in which the tracker behavior is preserved even in the coupled case, and the effects of the coupling on the evolution of matter linear density become important only at late times.
For both conformal models I and II, the continuity equation for CDM follows the standard one, and thus the coupling effect in the effective linear growth rate is roughly the square of the coupling constant $\alpha$.
We also proposed a purely disformal coupled model denoted by model~III. 
Thanks to a particular choice of the form of the disformal factor in this model, the energy transfer at the background level does not occur, therefore the background cosmological evolution is identical to the one in quintessence model.
Furthermore, an additional contribution due to the coupling in the effective gravitational constant is suppressed at early times, and the coupling effect on the effective linear growth rate, due to the modification of the continuity equation for CDM, is roughly of the same order as the coupling constant $\alpha$.
In all three models explored, the DE-DM coupling leads to more clustering, because of the attractive nature of the scalar fifth-force. 
Considering the conformal models I and II, the difference between the effective and actual linear growth rates are positive, such that RSD measurements would indicate a higher growth rate than the true one.
On the other hand, for the disformal model~III, the effective linear growth rate is smaller than the actual one, suppressing the redshift-distortion effect.
Hence, RSD measurements have less constraining power regarding the parameters of model~III when compared to models~I and~II.

As for the prospect of future probes measuring the DE-DM coupling, by employing the Fisher matrix analysis, we showed that assuming either of the conformal models, the absence of coupling could be probed within $1\sigma$, with the coupling constant being measured with a percent-level precision.
On the other hand, for the disformal model~III, RSD measurements alone would not be able to determine the coupling constant with sufficient accuracy.
It is expected that a combination of RSD, weak lensing, and CMB measurements would give better constraints on the coupling constant, and we hope to come back to this issue in the future.


\acknowledgments 
We thank Teruaki Suyama for many useful comments
and discussions.  This work was supported in part by JSPS Grant-in-Aid
for Scientific Research Nos.~ 17K14276 (R.K.), 18K18764 (M.Y.),
17K14304, and 19H01891 (D.Y.), MEXT KAKENHI
Grant-in-Aid for Scientific Research on Innovative Areas Nos.~15H05888,
18H04579 (M.Y.), and 18H04356 (S.Y.). This work was also partially supported by the
Mitsubishi Foundation (M.Y.), JSPS and NRF under the Japan-Korea Basic
Scientific Cooperation Program (M.Y. and S.Y.), and JSPS Bilateral
Open Partnership Joint Research Projects (F.C., R.K., and, M.Y).



\appendix 


\section{Metric transformation \label{app:metric_transformation} }


In this appendix, we summarize the expressions needed in deriving the equations of motion. 

The disformal metric transformation~\cite{Bekenstein:1992pj} is defined as
\ba
\gb_{\mu\nu}= A(\phi, X) g_{\mu\nu} + B(\phi, X) \phi_\mu \phi_\nu \, ,
\ea
where $A$ and $B$ are called, respectively, conformal and disformal functions.
In order to have a well-behaved 'new' metric, the transformation must preserve the Lorentzian signature and the causal structure (since the disformal factor changes the light-cones), 
and existence and non-singularity of the inverse transformation must be guaranteed.
These conditions, translated in terms of the conformal and disformal functions, can be summarized as
\begin{equation}
A - 2B X > 0 \,, 
\qquad 
A \left( A - 2XB \right) \left( A - A_X X + 2  B_X X^2 \right) \neq 0,
\end{equation}
for all values of $\phi$ and $X$.

The inverse of the barred metric $\gb_{\mu\nu}$ is given by 
\ba
\gb^{\mu\nu}= \frac{1}{A} \( g^{\mu\nu} - \frac{B}{A-2BX} \phi^\mu \phi^\nu \),
\ea
and the square root of the determinant of $\gb_{\mu\nu}$ is related to the one of $g_{\mu\nu}$ as
\ba
\frac{\sqrt{-\gb}}{\sqrt{-g}} = A^2 \sqrt{1- \frac{2BX}{A}}.
\ea
One can easily derive the following relations between $g_{\mu\nu}$ and $\gb_{\mu\nu}$:
\ba
\phib^\mu &=& \gb^{\mu\nu}\phib_\nu = \frac{1}{A-2BX} \phi^\mu ,\\
\Xb &=& -\frac{1}{2} \gb^{\mu\nu} \phib_\mu \phib_\nu = \frac{X}{A-2BX} , \\
\frac{\p \gb_{\mu\nu}}{\p g_{\alpha\beta}} &=&
A \delta^{~\alpha}_\mu \delta^{~\beta}_\nu 
+ \frac{1}{2} \(A_X g_{\mu\nu} + B_X \phi_\mu \phi_\nu \)\phi^\alpha\phi^\beta\,,
\\
\frac{\p g_{\mu\nu} }{ \p \gb_{\alpha\beta}} &=&
\frac{1}{A}\[
\delta^{~\alpha}_\mu \delta^{~\beta}_\nu 
- \frac{1}{2} C \(A_X g_{\mu\nu} + B_X \phi_\mu \phi_\nu \)\phi^\alpha\phi^\beta
\]\,,
\\
\frac{ \p \gb_{\alpha\beta} }{ \p \phi_\mu} &=&
B (\delta^\mu_{~\alpha} \phi_\beta +\delta^\mu_{~\beta} \phi_\alpha )
-A_X g_{\alpha\beta} \phi^\mu -B_X \phi_\alpha\phi_\beta \phi^\mu \,,
\\
C &=& \frac{1}{A-A_X X+2B_X X^2}  \,.
\ea

Also, one can define
\ba
 {\bar T}_{\rm (c)}^{\mu\nu} = 
 	\frac{2}{\sqrt{-\gb}} 
 	\frac{\delta (\sqrt{-\gb} \Ldm) }{ \delta \gb_{\mu\nu}} \,,
\ea
corresponding to the energy-momentum tensor in the frame in which dark matter is minimally coupled to gravity.
In such case, the pressure of dark matter in the old ($g_{\mu \nu}$) and new ($\gb_{\mu \nu}$) frames are related through
\ba
\Tdmd^{ij} 
	= \sqrt{ \frac{\gb}{g} } \left[ A \bar{T}^{ij} + \frac{1}{2} (A_X g_{\alpha \beta}+B_X \phi_\alpha \phi_\beta) \bar{T}_{\rm (c)}^{\alpha \beta} \phi^i \phi^j  \right] \,,
\ea
Thus the pressureless property remains the same up to the linear order in a FLRW background.


\section{Scale dependence of the mass term in the quasi-static limit \label{sec:k_independence}}


In Eq.~\eqref{eq:eom_phi_qs} we have defined the scale dependent factor $\A$.
For the canonical scalar field, Eq.~\eqref{eq:lagrangian_quintessence}, we have $\A_1 = 1$ and  $m_{\phi}^{2} = V_{\phi \phi}$, such that the scale dependence in the quasi-static limit appears as 
\begin{equation}
	\mathcal{A}(t, k) = \mathcal{A}_1 + \frac{a^2}{k^2} m_\phi^2 = 1 + \frac{a^2H^2}{k^2} \frac{V_{\phi \phi}}{H^2}.
\end{equation}
For sub-horizon scales, $k \gg aH$, as long as the ratio $V_{\phi \phi}/H^2$ is not too large, we can safely ignore the scale dependence.

\begin{figure}[!ht]
	\subfloat[\label{fig:m_over_H_sqrd_m1}]{
		\includegraphics[width=0.4\textwidth]{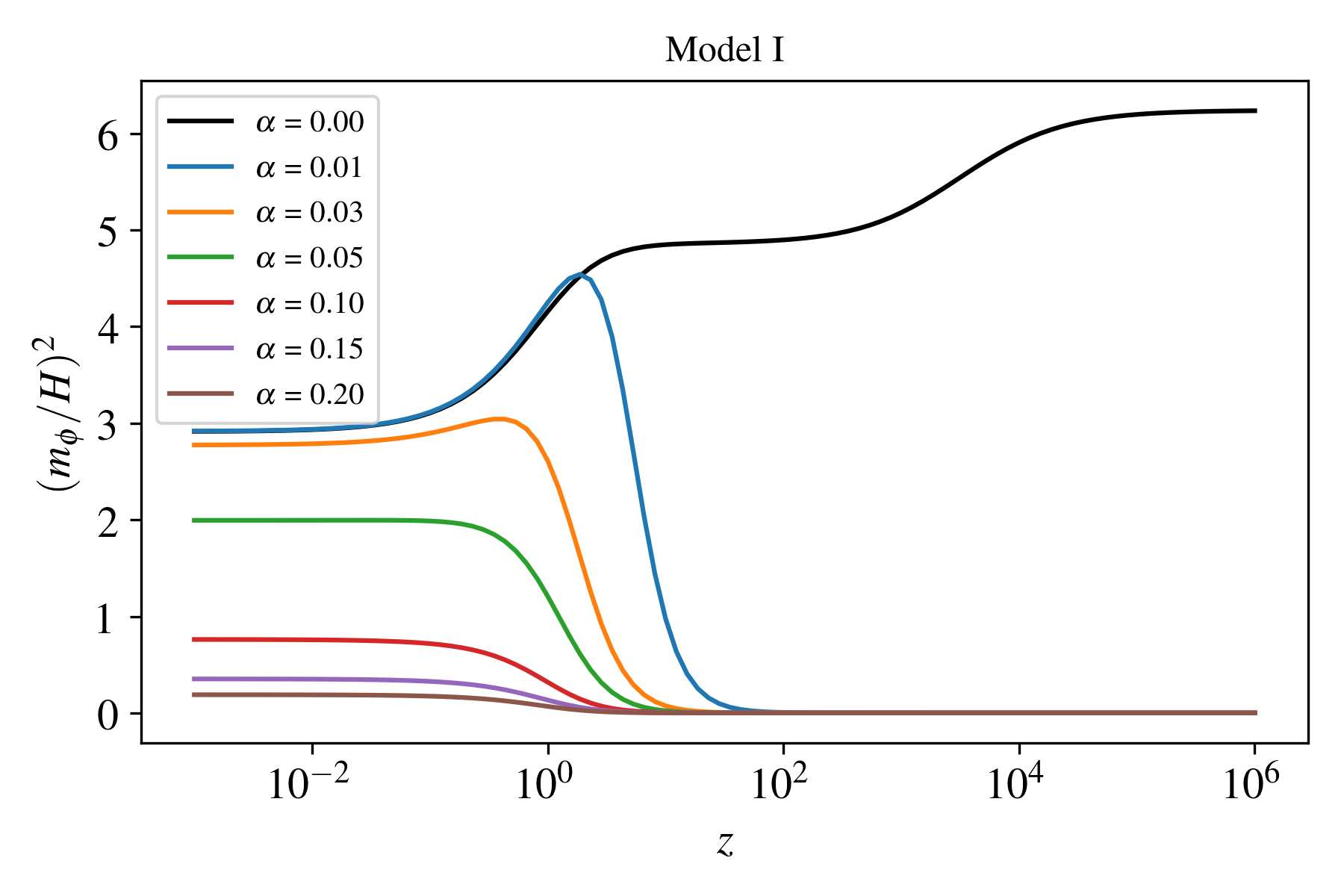}
		}
	\qquad
	\subfloat[\label{fig:m_over_H_sqrd_m2}]	{
		\includegraphics[width=0.4\textwidth]{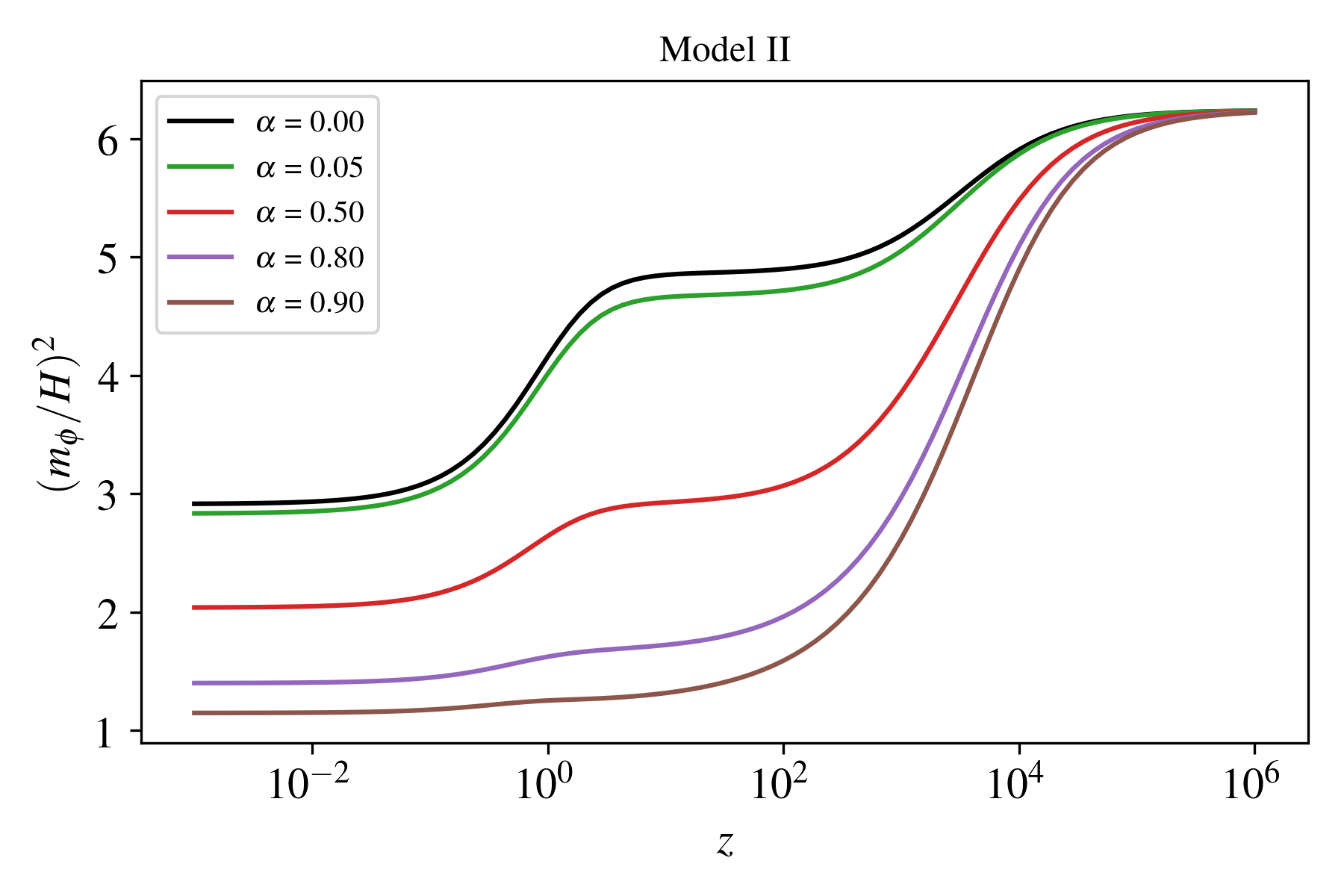}
		}
	\caption{\label{fig:m_over_H_sqrd}
		Ratio between the $m_\phi^2$ and $H^2$ in model~I (left panel (a)) and model~II (right panel (b)), as function of redshift for different values of the coupling constant.}
	\end{figure}

In Fig.~\ref{fig:m_over_H_sqrd} we present the evolution of the ratio between the scalar mass, $m_\phi$, and the Hubble parameter. 
In the uncoupled case, as well as for the conformal model~II, the mass term is roughly the same order as the Hubble parameter during most part of the cosmic evolution.
On the other hand, for model~I the mass term is much smaller than $H$ at early times, becoming of the same other around the present epoch. 
Hence, In both cases, the mass term can be safely ignored when the quasi-static limit is considered.


\section{Dispersion relation \label{sec:dispersion_relation} }


In this Appendix, we investigate the dispersion relation of scalar perturbations. 
First of all, let us consider the conformal model~I and II. In both cases, the coefficients \eqref{R1} and \eqref{R2} vanish, and therefore the gradient term in the perturbed equation for the scalar field \eqref{eq:eom_delta_phi} remains the same one as in quintessence case, implying the sound speed of the scalar perturbation is unity. Thus the sound horizon of scalar perturbation in both conformal models coincides with the cosmological horizon. 
On the other hand, the disformal model~III has non-vanishing coefficients $R_1$ and $R_2$, which modifies dispersion relation as we will see below.

The equation of motion for the perturbation of the canonical scalar field is 
\begin{equation} \label{eq:deltaphi_eom1}
	\d\phidd + 3 H \d \phid + \left( \frac{k^2}{a^2} + V_{\phi \phi} \right) \d\phi - \phid \left( \dot{\Phi} + 3 \dot{\Psi} \right) + 2 V_{\phi} \Phi = - 2 Q_0 \Phi - \d Q.
\end{equation}
In particular, for the disformal model~III we have, 
\begin{equation} \label{eq:coupling_model2}
	Q_{0} = 0, 
	\qquad 
	\d Q = 2 \alpha \frac{ \rhoc }{\phid^{2}} \frac{k^{2}}{a^{2}} \left( \phid v_{\rm c} - \d\phi \right), 
\end{equation}
and since there is no anisotropic stress, $\Phi = \Psi$. Hence, Eq.\eqref{eq:deltaphi_eom1} becomes
\begin{equation} \label{eq:deltaphi_eom_m2}
	\d\phidd + 3 H \d \phid 
		+ \left[ V_{\phi \phi} + \left(1 - 2 \alpha \frac{\rhoc }{ \phid^{2} } \right) \frac{k^2}{a^2} \right] \d\phi 
		- 4 \phid \dot{\Phi} 
		+ 2 V_{\phi} \Phi 
		+ 2 \alpha \frac{ \rhoc }{ \phid } \frac{k^{2}}{a^{2}} v_{\rm c}
		=0.
\end{equation}

To compute the dispersion relation we need to express $v_{\rm c}$ in terms of $\d\phi$, $\Psi$, and its time derivatives. 
But first, note that for model~III the Euler equations, \eqref{eq:euler_baryons} and \eqref{eq:euler_cdm}, lead to $\dot{v}_{\rm b} = \dot{v}_{\rm c}$, which means that since there is no coupling at background level, the velocity potentials will evolve in the same way. 
They will differ at most by a constant, $v_{b} = v_{c} + \textrm{constant}$. 
In what follows, we set this constant to zero. 
With that in mind, we use the $(0,i)$ component of Einstein equation, Eq.~\eqref{eq:einstein_0i}, to express the velocity as
\begin{equation} \label{eq:vc_m2}
 	(\rhob + \rhoc) v_{\rm c} = 2 \mpl^{2} \left( \dot{\Psi} + H \Psi \right) - \phid \d\phi, 
\end{equation} 
Plugging this result into Eq.~\eqref{eq:deltaphi_eom_m2}, we arrive at
\begin{equation}\label{eq:deltaphi_eom3}
	\d\phidd 
		+ \beta_1 \d\phid 
		+ \left(\beta_2 + \gamma_2 \frac{k^{2}}{a^{2}} \right) \d\phi 
		+ \left(\beta_3 + \gamma_3 \frac{k^{2}}{a^{2}} \right) \dot{\Psi}
		+ \left(\beta_4 + \gamma_4 \frac{k^{2}}{a^{2}} \right) \Psi 
		=0,
\end{equation}
where,
\begin{equation}
	\beta_1 = 3H, \quad \beta_2 = V_{\phi \phi}, \quad \beta_3 = -4 \phid, \quad \beta_4 = 2V_{\phi},
\end{equation}
\begin{equation} \label{eq:gamma_coef}
	\gamma_2 = 1 - 2\alpha \left( \frac{\rhoc}{\phid^{2}} + \frac{\rhoc}{\rhob + \rhoc} \right), \quad
	\gamma_3 = 4 \alpha \frac{\mpl^2 \rhoc }{\phid (\rhob + \rhoc)}, \quad
	\gamma_4 = 4 \alpha \frac{\mpl^2 H \rhoc }{\phid (\rhob + \rhoc)}.
\end{equation}
To close the system, we use the $(i,i)$ component of the Einstein equation, \eqref{eq:einstein_ii}, 
\begin{equation} \label{eq:potential_eom}
	\ddot{\Psi} + \alpha_1 \dot{\Psi} + \alpha_2 \Psi + \alpha_3 \d\phid + \alpha_4 \d\phi = 0, 
\end{equation}
where the coefficients are given by
\begin{equation} \label{eq:alpha_coef}
	\alpha_1 = 4H, \quad \alpha_2 = 2\dot{H} + 3H^{2} + \frac{\phid^{2}}{2\mpl^{2}}, \quad \alpha_3 = -\frac{\phid}{2\mpl^{2}}, \quad \alpha_4 = \frac{V_{\phi}}{2\mpl^{2}}.
\end{equation}
Assuming the time dependence $\d \phi = e^{i\omega t}\delta\phi_0$ and $\Psi = e^{i\omega t}\Psi_0$, the system formed by Eqs.~\eqref{eq:deltaphi_eom3} and \eqref{eq:potential_eom} admits solution if 
\begin{equation}
\mdet{-\omega^2 + i \alpha_1 \omega + \alpha 2 & i \alpha_3 \omega + \alpha_4 \\ 
	i \left( \beta_3 + \gamma_3 \frac{k^2}{a^2} \right) \omega + \beta_4 + \gamma_4 \frac{k^2}{a^2} & -\omega^2 + i \beta_1 \omega + \beta_2  + \gamma_2 \frac{k^2}{a^2}} = 0 \,.
\end{equation}
For sub-horizon scales, $k^{-1} \ll (aH)^{-1}$, the coefficients $\alpha_i$, $\beta_i$ and $\gamma_i$ are roughly constant within one Hubble time, and the large-$k$ limit of the above expression yields, 
\begin{equation}
	\omega^2 \simeq (\gamma_2 - \alpha_3 \gamma_3)k^2.
\end{equation}
Hence, the sound velocity (group velocity) is given by
\begin{equation}
	c_s^{2} = \frac{\omega^{2}}{k^{2}}  \simeq   \gamma_{2} - \alpha_3 \gamma_3 = 1 - 2\alpha \frac{\rhoc}{\phid^{2}}.
\end{equation}
As expected, the sound speed for the uncoupled canonical scalar field is equal to unity. 
However, that is changed by the disformal coupling between dark matter and dark energy. 
The fact that $c_{s}^{2}$ must be positive leads to an upper bound for the coupling constant, 
\begin{equation}
	\alpha < \frac{\phid^2}{2\rhoc}.
\end{equation}
Furthermore, since during matter epoch $\Omega_{\rm c} \ll \Omega_\phi$, we can approximate $\phid^2/\rhoc \approx 0$, and a safer bound against instabilities is $\alpha < 0$.

The quasi-static approximation is valid for modes inside the sound horizon of scalar perturbations. 
For uncoupled quintessence, since the sound speed is equal to unity, the quasi-static limit is valid for all modes well inside the Hubble horizon. 
In the general case, one can define a sound horizon scale $k_{s}^{-1} = c_{s}(aH)^{-1}$, such that the quasi-static approximation is valid for modes satisfying $k \gg k_{s}$. 
For instance, in model~III we require 
\begin{equation}
	k \gg \frac{aH}{\sqrt{1-2\alpha \frac{\rhoc}{\phid^{2}}}}\,.
\end{equation}


\bibliography{references}

\end{document}